\documentclass[pdflatex]{sn-jnl}

\jyear{2021}%

\theoremstyle{thmstyleone}%
\theoremstyle{thmstyletwo}%
\newtheorem{example}{Example}%

\theoremstyle{thmstylethree}%

\newcommand{\AG}{\textcolor{red}}
\newcommand{\YX}{\textcolor{blue}}

\usepackage{amsmath}
\allowdisplaybreaks
\usepackage{amssymb}
\usepackage{commath}
\usepackage{mathtools}
\usepackage{bbm}
\usepackage{bm}

\newcommand{\vect}[1]{\bm{#1}}

\DeclareMathOperator{\diag}{diag}
\DeclareMathOperator*{\argmin}{arg\,min} 
\DeclareMathOperator*{\argmax}{arg\,max} 
\DeclareMathOperator{\col}{col}
\DeclareMathOperator{\Cov}{Cov}

\DeclareMathOperator{\JHBL}{JHBL}

\usepackage[normalem]{ulem}

\usepackage{enumitem}
\setlist[enumerate]{leftmargin=.5in}
\setlist[itemize]{leftmargin=.5in}

\usepackage{color}
\usepackage{graphics,graphicx}
\usepackage[small]{caption}
\usepackage{subcaption}

\begin{document}

\title[Sequential edge detection]{Sequential edge detection using joint hierarchical Bayesian learning}


\author[1]{\fnm{Yao} \sur{Xiao}}\email{yao.xiao.gr@dartmouth.edu}

\author*[1]{\fnm{Anne} \sur{Gelb}}\email{Anne.E.Gelb@Dartmouth.edu}

\author[2]{\fnm{Guohui} \sur{Song}}\email{GSong@ODU.edu}

\affil*[1]{\orgdiv{Department of Mathematics}, \orgname{Dartmouth College}, \orgaddress{\city{Hanover}, \postcode{03755}, \state{NH}, \country{USA}}}

\affil[2]{\orgdiv{Department of Mathematics and Statistics}, \orgname{Old Dominion University}, \orgaddress{\city{Norfolk}, \postcode{23529}, \state{VA}, \country{USA}}}


\abstract{This paper introduces a new sparse Bayesian learning (SBL) algorithm that jointly recovers a temporal sequence of edge maps from noisy and under-sampled Fourier data.  The new method is cast in a Bayesian framework and uses a prior that simultaneously incorporates {\em intra-}image information to promote sparsity in each individual edge map with {\em inter-}image  information to promote similarities in any unchanged regions.  By treating both the edges as well as the similarity between adjacent images as random variables, there is no need to separately form regions of change. Thus we avoid both additional computational cost as well as any information loss resulting from pre-processing the image.  Our numerical examples demonstrate that our new method compares favorably with more standard SBL approaches.}

\keywords{\YX{sequential edge detection, hierarchical Bayesian learning, Fourier data}}


\pacs[MSC Classification]{\YX{15A29, 62F15, 65F22, 65K10, 68U10}}

\maketitle


\section{Introduction}
\label{sec:intro}
In applications such as environment monitoring \cite{rogers1996predicting,azzali2000mapping} and surveillance \cite{othman2012wireless,wimalajeewa2017application},  temporal sequences of images are compared to determine changes within the physical region of interest.  When data are acquired indirectly, each image in the temporal sequence is typically first recovered individually, before any comparisons are made.   However, if the scene contains rigid bodies, it is often enough to infer important surveillance and monitoring information from the corresponding time sequenced {\em edge maps}, 
\cite{gelb2017detecting,GS2017,SVGR,gelb2019reducing}. This is useful because recovering an edge map may be more accurate as well as less costly than recovering a full scale image.    

Sequential edge maps can be used more generally in image recovery processes. 
 For example, the weights in  weighted $\ell_1$ regularization methods are designed to be inversely proportional to the strength of the signal in the sparse domain (see e.g.~\cite{candes2008enhancing}). 
 Accurate edge maps can help to determine these weights off-line and in advance, as was demonstrated in \cite{adcock2019jointsparsity,gelb2019reducing,scarnati2019accelerated}.  Segmentation, classification and object-based change detection (see e.g.~\cite{xiao2022sequential}) are also examples of procedures where edge information can improve the quality of the results.  
 
The primary motivation in this investigation comes from spotlight synthetic aperture radar (SAR), where the phase history data may be considered as Fourier data, \cite{jakowatz2012spotlight}.\footnote{We note that SAR images are complex-valued and that retaining phase information is important for downstream processing.  While only real-valued images are included in this study, our approach is not inherently limited. In future investigations we will consider the modifications needed for complex-valued signal recovery.} We therefore seek to recover a sequence of edge maps from multiple observations of noisy and under-sampled Fourier measurements at different time instances, while noting that our underlying methodology can be applied to other types of measurements with some modifications.  

Several algorithms have been designed to recover the edges in a real-valued signal from (multiple) under-sampled and noisy Fourier measurements  at a {\em single instance in time}, even in situations where missing bands of Fourier data would make it impossible to recover the underlying image, \cite{GS2017,SVGR,viswanathan2012iterative}.    Many techniques have also been developed to recover the magnitude of a SAR image using a sequence of sub-aperture data acquisitions, again at a single instance in time, \cite{Cetin2005,ccetin2014sparsity,moses2004wide,stojanovic2008joint}. In this case the images are first {\em separately} recovered, each corresponding to a single data acquisition. The final estimate at each pixel is then calculated as the weighted average of the recovered images. However, processing the individual measurements to recover a set of images before combining the joint information can lead to additional information loss, especially  in the case of noisy and incomplete data collections, \cite{archibald2016image,scarnati2019accelerated,wasserman2015image,xiao2022sequential}.  By contrast, exploiting redundant information without having to first recover separate images leads to more accurate image recovery, \cite{scarnati2019accelerated,xiao2022sequential}.  Recovery algorithms that incorporate  information from multi-measurement vectors are often referred to as MMV recovery algorithms, \cite{adcock2019jointsparsity,chen2006theoretical,cotter2005sparse}.

In developing methods to recover temporal sequences of images from noisy and under-sampled Fourier measurements, changes such as translation and rotation that are expected to occur between the sequential data collections must also be accounted for. Thus none of the techniques mentioned in the preceding paragraph is suitable. 
By contrast, the sequential image recovery developed in \cite{xiao2022sequential} employs a compressive sensing (CS) framework with regularization terms designed to both account for sparsity in the appropriate transform domain as well as to promote similarities in regions for which {\em no} change has been detected.  Following the ideas in \cite{scarnati2019accelerated}, the edge maps are generated {\em without} first recovering the images so that additional information is not lost due to processing. 
In this way inter-image temporal correlations between adjacent measurements can be better exploited. Although demonstrated to yield improved accuracy when compared to algorithms which separately recover each image in the sequence, that is, without using inter-image information, the method in \cite{xiao2022sequential} is multi-step and relies on several hand-tuned parameters.  Specifically, after the edge maps for each individual image are generated from the observed data,  rigid objects must be constructed to determine the changed and unchanged regions between each adjacent image (so-called change masks), which would then in turn form the regularization term.  This step typically requires some type of clustering algorithm, leading to both a loss of information due to the extra processing and a lack of robustness due to the additional hand-tuning of parameters. Finally, there is an assumption that a moving object is not contained within another (background) object.  For example, one would have to remove the skull in a magnetic resonance image (MRI) brain scan in order to determine changes within.

Such concerns were initially addressed in \cite{xiao2022sequential1}.  There, the method in \cite{xiao2022sequential} was recast in a Bayesian framework, with an {\em intra-}image prior used to promote sparsity in the edges of each image, and a second {\em inter-}image prior used to promote the similarity of the unchanged-regions.  By treating the similarity between adjacent images as a random variable, forming the rigid boundaries as an intermediate step was no longer required. 

In this investigation we consider sequential edge map recovery (as opposed to sequential image recovery) which allows us to further simplify the approach introduced in \cite{xiao2022sequential1} by integrating both intra- and inter-image information into one prior.    That is, we introduce a prior that simultaneously promotes intra-image sparsity and inter-image similarity. To this end, we note that the classic sparse Bayesian learning (SBL) \cite{tipping2001sparse,wipf2011latent,zhang2011sparse,chen2016simultaneous,zhang2021empirical} requires a shared support of all the collected measurements to approximate edges. Such an assumption will clearly be violated when change occurs between sequential data collections. To compensate,  our proposed method introduces a new set of hyper-parameters  so that information outside the shared support is not considered in the joint estimation of the edge values. Our numerical examples demonstrate that by constructing the SBL algorithm in this way we are able to account for changes in each sequenced edge map. Such information can then be used in downstream processing as warranted by the particular application.

This rest of this paper is organized as follows.  Section \ref{sec:preliminaries} provides the necessary background for our new method, which is introduced in Section \ref{sec:newmodel}.  Numerical examples for both one- and two-dimensional signals are considered in Section \ref{sec:numresultsSBL}.  We also demonstrate how our method for recovering sequential edge maps can then be used for recovering sequential images.  Concluding remarks are given in \ref{sec:discussion_ongoing}.


\section{Preliminaries}
\label{sec:preliminaries}

Let $f_{j} \colon [-\pi,\pi]\to\mathbb R$ be a sequence of one-dimensional piecewise smooth functions at different times $j=1,\dots,J$. The corresponding Fourier samples are given as
\begin{align}
\label{eq:fourcoeff}
    \hat f_{j,l}  =\frac{1}{2\pi} \int_{-\pi}^\pi f_{j} (s)e^{-i ls}ds, \quad -\frac{n}{2}\leq l\leq \frac{n}{2}-1, \quad j=1,\dots,J.
\end{align}
We note that our approach is directly extendable to two-dimensional images, which will be demonstrated via the numerical experiments in Section \ref{sec:numresultsSBL}.

Suppose the corresponding forward model for the observations of $f_{j}$ is given by 
\begin{equation}\label{eq:forward}
    \vect g_{j} =F_{j} \vect f_{j} , \quad j=1,\dots, J,
\end{equation}
where $\vect f_{j} $ is the discrete approximation of $f_{j} $ at $n$ uniform grid points on $[-\pi,\pi]$ and {$F_{j} \in\mathbb C^{n\times n}$} is the discrete Fourier transform forward operator. In our numerical experiments, each $F_{j} $ is missing an arbitrarily chosen band-width of frequencies, indicating that the data sequence is compromised in some way, as well as being under-sampled. Details regarding the missing band-width will be made explicit in Section \ref{sec:numresultsSBL}.
We further assume that each of the $J$ observations are corrupted with additive independent and identically distributed (iid) zero-mean Gaussian noise with unknown variance $\sigma^2$ given by
\begin{equation}\label{eq:likelihood_noise}
\boldsymbol\epsilon_{j} \sim\mathcal{N}\left(0,\beta^{-1}I_n\right), \quad j=1,\dots,J,
\end{equation}
where $\beta^{-1} =\sigma^{2}$. 
The forward model \eqref{eq:forward} therefore becomes
\begin{align}\label{eq:forward_noisy} 
\tilde{\vect{g}}_{j} =F_{j} \vect f_{j} +\boldsymbol\epsilon_{j} , \quad j=1,\dots, J.
\end{align}
We seek to recover the sequential edge maps corresponding to the observations in \eqref{eq:forward_noisy} using a Bayesian framework. 
As already noted, the edge maps may be useful on their own or in downstream processing. Examples in Sections \ref{sec:img_rec}  demonstrate how the edge maps can be used effectively for sequential image recovery.

\subsection{Edge Detection from Fourier data}
\label{sec:edge detection}
Detecting the edges of a piecewise smooth signal or image from a finite number of Fourier samples, \eqref{eq:fourcoeff}, is a well-studied problem.  Due to its simple linear construction, here we employ  the concentration factor (CF) method, initially developed in \cite{gelb1999detection}. Below we briefly describe the CF method in the context of edge map recovery. For ease of notation we drop the $j$ and write $f = f_{j} $ since what follows applies to any $j = 1,\dots, J$.

Consider the piecewise analytic function $f$ which has $K$ simple discontinuities located at $\{\xi_k\}_{k=1}^\mathcal{K}$ in $(-\pi,\pi]$. The corresponding jump function is defined as 
\begin{equation}\label{eq:jumpFunc}
    [f](s)=f(s^+)-f(s^-),
\end{equation}
where $f(s^+)$ and $f(s^-)$ represent the right- and left-hand side limit of $f$ at location $s$. 
We then define the  signal $\vect x \in \mathbb{R}^n$ as the jump function of  $f$ evaluated at $n$ uniform grid points in $(-\pi,\pi]$, given by ${s_\nu} = -\pi + \frac{2\pi \nu}{n}$.   Specifically, 
\begin{equation}
\label{eq:xvector}
{\vect x} = \{x_{\nu}\}_{\nu = 1}^n = \{[f](s_\nu)\}_{\nu = 1}^n.
\end{equation}
Observe that ${\bf x}$ is sparse, as it is zero everywhere except for indices corresponding to points in the domain where there is an edge, or jump discontinuity, in the function $f$. 
An equivalent expression of \eqref{eq:xvector} is given by 
\begin{align*}
    x_\nu = \sum^{K}_{k=1}[f](\xi_k)I_{\xi_k}(s_{\nu}), \quad {\nu = 1,\dots,n,}
\end{align*}
where $I_{\xi_k}(s_{\nu})$ is the indicator function defined as
\begin{equation*}
    I_{\xi_k}(s{_\nu}) =\begin{cases}
    1, &s{_\nu} = \xi_k\\
    0,&\text{otherwise}.
\end{cases}
\end{equation*}

Given the set of Fourier coefficients $\{\hat f_l\}_{l=-n/2}^{n/2-1}$ of each $\vect f$ in the temporal sequence, the CF edge detection method is formulated as 
\begin{align}\label{eq:1d_CF}
    S^\sigma_n f(s) = \sum_{l=-n/2}^{n/2-1} \tau(l) \hat f_l e^{i ls},
\end{align}
where
$$ \tau(l) = i \,\text{sgn}(l) \sigma\left(\frac{2\abs{l}}{n}\right),\quad l = -n/2,\dots,n/2-1.$$
Here $\sigma(\eta)$, $\eta \in [0,1]$, is an admissible concentration factor that guarantees the convergence of \eqref{eq:1d_CF}  to  $[f](s)$, \cite{gelb1999detection,gelb2000detection}.\footnote{Loosely speaking, $\tau(l)$ is a band-pass filter that amplifies more of the high-frequency coefficients which typically contain information about the edges of the underlying function $f$. We also note that $\tau(l)$ can be constructed for {\em discrete} Fourier measurements. Hence in this regard the method developed in our investigation applies to other forms of measurement data, as well as multiple data sources, so long as they can be stored as (discrete) Fourier samples.} As the number of discontinuities in two dimension is infinite, the corresponding extension of \eqref{eq:1d_CF} becomes undefined. Parameterization of the corresponding edge curve and the rotation of the concentration factor are incorporated into the CF technique to circumvent this issue for two-dimensional case, see \cite{adcock2019jointsparsity,xiao2022sequential}.

Given all $n$ Fourier coefficients in \eqref{eq:fourcoeff} for each $j = 1,\dots, J$, it is possible to recover the sequence of edge maps for moderate levels of noise.   It is also possible to incorporate inter-image information from the temporal sequence of data sets to improve each individual edge map recovery, \cite{xiao2022sequential}.  As already noted, however, this additional step requires more processing and therefore more hand-tuning of  parameters. Moreover, as the SNR decreases and as fewer coefficients in each data set become available (for example, if a band of Fourier data is not usable), correlating information from the sequenced data becomes more challenging.  Thus we are motivated to use a sparse Bayesian learning approach to incorporate the temporal inter-image information into the process of edge map recovery, which we now describe.

\subsection{Sparse Bayesian Learning}
\label{sec:SBL}
We start by applying the CF method in \eqref{eq:1d_CF} on  $\tilde{\vect g}_{j} $ in \eqref{eq:forward_noisy} to generate $j = 1,\dots,J$ approximations to $\vect x_{j} $ in \eqref{eq:xvector} as 
\begin{align}\label{eq:edge_est}
\begin{split}
    \vect y_{j}&=E_{j}\left(F_{j}\vect f_{j}+\boldsymbol\epsilon_{j}\right),\\
    &= E_{j}\vect{g}_{j}+E_{j}\boldsymbol\epsilon_{j},\\
    &=\tilde{\vect x}_{j}+\tilde{\boldsymbol\epsilon}_{j},
    \quad j=1,\dots, J.
\end{split}
\end{align}
Here $E_{j}\in {\mathbb C^{n\times n}}$ is the (diagonal) edge detection operator with entries $E_{j}(l,l) = \tau(l-(n/2+1))$, $l = 1,\dots,n$,
and $\tilde{\vect x}_{j}$ is an approximation to the sparse signal ${\vect x}_{j}$. We note that each matrix $E_{j}$ is designed specifically for the case when bands of Fourier data are unavailable, see \cite{viswanathan2012iterative}.

We then stack all measurements into a vector and obtain the new model as
\begin{equation}\label{eq:forward_m}
    Y= X+\mathcal E,
\end{equation}
where $Y=\col\left([\vect y_{1},\dots,\vect y_{J}]^T\right) \in \mathbb R^{nJ}$ and  $X=\col\left([\vect x_{1},\dots,\vect x_{J}]^T\right)\in\mathbb R^{nJ}$.
That is, there are $n$ arrays of length $J$ in $X$ and $Y$ where $Y_i$ denotes the $i^{th}$ array of  measurements at location $i$, and $X_i$ the corresponding solution.  

Compressive sensing (CS) methods \cite{candes2006stable,candes2006robust,candes2006near,donoho2006compressed,candes2008enhancing,langer2017automated} are often used to solve problems in the form of \eqref{eq:forward_m}.  In this regard the classical unconstrained optimization problem used to recover the underlying signal is given by 
\begin{equation}\label{eq:cs_model}
    \argmin_{X}(\norm{Y-X}^2_2+\lambda \norm{X}_1),
\end{equation}
where the first term is used to promote data fidelity and the second (regularization) term encourages sparsity in the solution.    As is explained in the classical CS literature, the $\ell_1$ norm  in \eqref{eq:cs_model}  serves as a surrogate for $\norm{\cdot}_0$, since the  $0$ ``norm'' (pseudo-norm) yields an intractable problem \cite{wipf2004minimization,candes2006stable}.  The regularization parameter $\lambda$ balances the contribution between the terms, with smaller $\lambda$ implying high quality data and vice versa.  Importantly, \eqref{eq:cs_model} does not account for the correlated information  available from the $J$ data sets. 

Methods to exploit the joint sparsity attainable from the multiple measurement vectors (MMVs) have been subsequently developed in \cite{cotter2005sparse,chen2006theoretical,eldar2009robust,zheng2012subspace,deng2013group,singh2016weighted}, and somewhat relatedly, additional refinement to \eqref{eq:cs_model} can be made by employing {\em weighted} $\ell_1$ or $\ell_2$ regularization, \cite{candes2008enhancing,churchill2018edge,gelb2019reducing,scarnati2019accelerated,ren2020imaging}.  Both approaches have the effect of more heavily penalizing sparse regions in the sparse domain of the solution than in locations of support. Ultimately these techniques require information regarding the  support locations in the solution's sparse domain, which may not be readily accessible  when the measurement data are heavily corrupted.

As an alternative, \eqref{eq:forward_m} can be viewed from a {\em sparse Bayesian learning} (SBL) perspective.  By choosing uninformative hyper-priors, no advanced knowledge about the support in the sparse domain is required. Since SBL serves as the foundation of the proposed algorithm in our investigation,  we briefly review it below.

The inverse problem \eqref{eq:forward_m} can be formulated in a hierarchical Bayesian framework  by extending $X$, $Y$ and a collective parameter $\boldsymbol\theta$ into random variables (see e.g.~\cite{calvetti2007introduction,kaipio2006statistical}). We use the following density functions to describe the relationships among $X$, $Y$ and $\boldsymbol\theta$:
\begin{itemize}
\item The {\em prior} $p(X\vert\boldsymbol\theta)$ is the probability distribution of $X$ conditioned on $\boldsymbol\theta$.
\item The {\em hyper-prior} $p(\boldsymbol\theta)$ is the probability distribution of $\theta$.
\item The {\em likelihood} $p(Y\vert X,\boldsymbol\theta)$ is the probability distribution of $Y$ conditioned on $X$ and $\boldsymbol\theta$.
\item The {\em posterior} $p(X,\boldsymbol\theta \vert Y)$ is the joint probability distribution of $X$ and parameter $\boldsymbol\theta$ conditioned on $Y$.
\end{itemize} 
Our goal is to recover the posterior distribution, which by Bayes' theorem is given by
\begin{equation}\label{eq:bayes}
p(X,\boldsymbol\theta\vert Y)=\frac{p(Y\vert X)p(X\vert\boldsymbol\theta)p(\boldsymbol\theta)}{p(Y)}.
\end{equation}
In particular, $\boldsymbol\theta$ is not pre-determined a {\em priori} but rather as a part of the Bayesian inference. A main challenge in the above formulation is the computation of the marginal distribution \(p(Y) \),  usually an intractable high-dimensional integral. As it is impractical to compute the posterior \(p(X,\vect{\theta}\vert Y) \) directly from \eqref{eq:bayes}, we instead seek its approximation. Specifically, we first employ the empirical Bayes approach to obtain a point estimate \(\hat{\vect{\theta}}  \) and then compute the conditional posterior \(p(X\vert\hat{\vect{\theta}} , Y) \) as an approximation of the joint posterior.   A point estimate of \(X \) can also be realized as the {\em maximum a posteriori} (MAP) estimate of the conditional posterior, given by
\begin{equation}
    X^\ast=\argmax_{X}p(X, \hat{\vect{\theta}} \vert Y).
    \label{eq:BayesMAP}
\end{equation}
We note that \eqref{eq:cs_model} is equivalent to finding a point estimate approximation to \eqref{eq:BayesMAP} when using a Laplace prior with a pre-determined hyper-parameter $\lambda$.

Sparse Bayesian learning (SBL) is a catch-all phrase for a class of algorithms designed to calculate the hyper-parameter estimate \(\hat{\vect{\theta}}\) and the corresponding conditional posterior distribution when considering a hierarchical prior. More generally there are a range of techniques for solving inverse problems using the Bayesian approach that focus in particular on (application dependent) prior and hyper-prior estimation, \cite{mackay1992bayesian,mackay1999comparison,mohammad1996full,mohammad1996joint,molina1999bayesian}.  For example, the hierarchical Gaussian prior (HGP) approach guarantees a closed-form posterior distribution \cite{bardsley2012mcmc,calvetti2019hierachical,calvetti2020hybrid,calvetti2020sparsity,
calvetti2020sparse}. Indeed, many SBL algorithms choose conjugate priors \cite{tipping2001sparse, wipf2004sparse, wipf2007empirical,zhang2011sparse, wipf2011latent,chen2016simultaneous}. In this framework the hyper-parameters are often approximated by the Expectation Maximization (EM) \cite{dempster1977maximum} or the evidence maximization approach \cite{mackay1992bayesian}.
In some cases the hyper-parameters for the sparse prior are determined empirically from the given data \cite{wipf2007empirical,pereira2015empirical,zhang2021empirical}.  Finally, we note that (joint recovery) SBL is designed for stationary support in the sparse domain, which is pointedly {\em not} our assumption in this investigation.

In Section \ref{sec:newmodel} we propose a new technique that provides more accurate and efficient MAP estimates of the temporal sequence of solution posteriors. Our approach exploits the temporal correlation between the neighboring data sets in the sequence by introducing a new set of hyper-parameters which are then incorporated into the  algorithm developed in \cite{tipping2001sparse}.\footnote{To be clear, the algorithm in \cite{tipping2001sparse} considers a stationary sparsity profile.  The magnitudes at the jump locations are allowed to vary in this setting, however, which is also the assumed case for MMV methods in \cite{wipf2007empirical,zhang2011sparse}.}

\subsection{Hierarchical Bayesian Framework}
We first specify each of the terms used in \eqref{eq:bayes} and subsequent MAP estimate \eqref{eq:BayesMAP}.   To derive the proposed algorithm, we work through the standard SBL framework and start with the case where ${\boldsymbol x}$ is a stationary signal for which we have $J \ge 1$ sets of observable data.  We relax this assumption to accommodate for signal sequence with non-stationary support in Section \ref{sec:newmodel}. 

\subsubsection{The likelihood}
The likelihood function describes the relationship of the solution $X$, the observation noise $\mathcal{E}$, and the data sets $Y$. When considering individual parts of the temporal sequence,
for each pair of $\vect{y}_{j}$ and $\vect{x}_{j}$, $j=1,\dots,J$ in \eqref{eq:edge_est}, we assume the zero-mean, i.i.d Gaussian distributed noise $\boldsymbol \epsilon_{j}\in\mathbb{R}^{n}$ in \eqref{eq:likelihood_noise} with the precision parameter $\beta>0$. 
This assumption leads to the following likelihood function:
\begin{align}\label{eq:likelihood}
    p(Y\vert X, \beta) = (2\pi)^{-\frac{nJ}{2}}\beta^{\frac{nJ}{2}}\exp{\left\{-\frac{\beta}{2}\norm{Y-X}^2_2\right\}}.
\end{align}
In many applications the maximum likelihood estimate (MLE) is used to obtain a point estimate for the solution. However, overfitting  can be an issue, especially in low SNR environments.  This issue often motivates using the Bayesian approach, in which an appropriate prior distribution on the solution is imposed. 

\subsubsection{The prior (stationary case)} \label{sec:prior}
The desired solution $X$ in \eqref{eq:forward_m} contains the discrete jump approximation of the underlying signal $f_j$. Thus it should be sparse. 
As already noted there are plenty of potential prior distributions that promote sparsity, including the Laplace prior \cite{figueiredo2007majorization}, the hyper-Laplacian prior \cite{levin2007image, krishnan2009fast}, and the mixture-of-Gaussian prior \cite{fergus2006removing}. Since explicit formulas are available for conjugate priors, here we consider a Gaussian prior distribution for each $X_i$ in \eqref{eq:BayesMAP}, conditioned on the hyper-parameters $s_i$ given by
\begin{equation}\label{eq:prior_i_classic}
p(X_{i}\vert s_i) \sim \mathcal N(0,s_i^{-1}I_J), \quad i=1,\dots,n.
\end{equation}
In particular, the hyper-parameter $\boldsymbol s =(s_1,\dots,s_n)$ assumes a {\em stationary} sparsity profile across the temporal sequence and the value of pixels are i.i.d.~to each other.  In this case the elements in each array $X_i$ are the same. 
The value of the hyper-parameter forces the sparsity of the posterior by concentrating most of the probability at \(X_{i} \approx 0 \) when \(s_{i} ^{-1} \approx 0\). In \cite{tipping2001sparse,wipf2007empirical}, the recovery of the sparse signal (in the deterministic case) amounts to determining the $i = 1,\dots, n$  entries in $\boldsymbol x$ that correspond to large $s_i^{-1}$.

\subsection{The hyper-prior (stationary case)}
\label{sec:hyper_prior}
From the discussion in Section \ref{sec:prior}, it is clear that to promote sparsity of the solution $X$ in the conditional Gaussian distributed prior \eqref{eq:prior}, the entries of the hyper-parameter $\boldsymbol s$ should be able to vary significantly. This can be achieved by using an uninformative hyper-prior as the density function of $\boldsymbol s$ and then treating each $s_1,\dots, s_n$ as a random variable. Following similar discussions in \cite{tipping2001sparse,wipf2007empirical,glaubitz2022generalized} here we employ the gamma distribution for each $s_i$, $i=1,\dots, n$, in \eqref{eq:prior_i} given by
\begin{equation}
    \label{eq:hyperhyper1}
p(s_i)=\Gamma(s_i\vert a_s,b_s) = \frac{b_s^{a_s}}{\Gamma(a_s)}s_i^{a_s-1}e^{-b_ss_i},
\end{equation}
where $a_s,b_s>0$ are the shape and rate parameters of the gamma distribution \cite{artin2015gamma}. In particular, the gamma distribution is uninformative because it has mean $a_s/b_s\to\infty$ and variance $a_s/b_s^2\to\infty$ if $a_s<\infty$ and $b_s\to 0$.   
By choosing $a_s=1$ and $b_s=10^{-4}$ \cite{tipping2001sparse,bardsley2012mcmc,glaubitz2022generalized}, the hyper-parameters $s_1,\dots,s_n$ will be able to vary as needed depending on the observed data sets.  We can conveniently assume the same uninformative hyper-prior for the hyper-parameter $\beta$ in \eqref{eq:likelihood}
\begin{equation}\label{eq:hyper-prior-beta}
p(\beta^{-1}) = \Gamma(\beta\vert a_\beta,b_\beta),
\end{equation}
where we analogously choose $a_\beta=1$ and $b_\beta=10^{-4}$ for simplicity.

Although our formulation generates more parameters to compute, from a methodological perspective, an accurate inference approximation may prevent over-fitting \cite{neal2012bayesian}. This hierarchical formulation of the prior distribution falls under the category of {\em automatic relevance determination} (ARD) prior \cite{mackay1996bayesian,neal2012bayesian}. Based on the evidence from the data sets, the uninformative hyper-priors allow the posterior function of the solution $X$ to concentrate at large values of $\boldsymbol s$ \cite{tipping2001sparse}. The corresponding values of $X$ at those locations must have values near $0$, which are deemed as ``irrelevant'',  i.e.~as not contributing to the data sets.


\section{Exploiting inter-image information}
\label{sec:newmodel}
The Gaussian distributed prior conditioned on $\boldsymbol s$ in \eqref{eq:prior_i_classic} assumes a stationary sparsity profile across the temporal sequence of observations. Such an assumption fails whenever the sparsity profile is not stationary.  For example, if an object within a scene moves from one time frame to another, we can analogously consider the case where $x_{j,i} \ne 0$,  for some $i \in [0,n] , j \in [1,J]$, but that $x_{j^\prime, i} = 0$ for some $j^\prime \ne j$.  In the stationary framework the estimate of $s_i^{-1}$ will be ``averaged out''  over {\em all} the $J$ recoveries.  In particular if $j = 1$ and subsequently $j' = 2,\dots,J$,  then (for large enough $J$) $s_i^{-1} \approx 0$, and correspondingly yields $x_{1,i} \approx 0$. Figure \ref{fig:rec_SBL} illustrates this behavior for $J = 6$, where the translating support locations are lost due to the incorrect assumption regarding stationary support.  

We address this issue by introducing new hyper-parameters $\boldsymbol q=(q_1,\dots,q_n)$  in the prior covariance matrix to account for potential changes between neighboring signals, that is, the temporal correlation, at each location. This will allow appropriate moderation of the strength of $\boldsymbol s$ on the conditioned prior distribution of $X$.  More specifically we adapt the Gaussian prior distribution of $X_i$ in \eqref{eq:prior_i_classic}  to be conditioned on the hyper-parameters $s_i$ and $q_i$ as 
\begin{equation}\label{eq:prior_i}
p(X_{i}\vert s_i,q_i) \sim \mathcal N(0,[\Sigma_{i}(\vect{s},\vect{q})]^{-1}), \quad i=1,\dots,n.
\end{equation}
where
\begin{equation}\label{eq:prior_cov_i}
    \Sigma_{i}(\vect{s},\vect{q})=\begin{pmatrix}
    s_i  & q_i  		& 		  		&\\
    q_i &  \ddots  & 	\ddots  &\\
          &  \ddots & \ddots   & q_i \\
          &        		& q_i 		& s_i
    \end{pmatrix}
    \quad\in\mathbb R^{J\times J},
\end{equation}
We note that we construct the prior covariance matrix  as a tri-diagonal  matrix for convenience, and other banded or sparse matrices may also be appropriate. Importantly, values from measurements at the same location are {\em not} assumed to be  i.i.d, as in \cite{wipf2007empirical}, but are instead correlated. This is a departure from standard SBL approaches described in Section \ref{sec:prior} used for stationary observations. 

With \eqref{eq:prior_cov_i} in hand, we now employ the conditional Gaussian prior density distribution over $X$ given by \cite{tipping2001sparse,wipf2007empirical,glaubitz2022generalized}
\begin{equation}\label{eq:prior}
p(X\vert \boldsymbol s,\boldsymbol q)=(2\pi)^{-\frac{nJ}{2}} \abs{\Sigma(\vect{s},\vect{q})}^{-\frac{1}{2}}\exp{\left\{-\frac{1}{2}X^T\Sigma^{-1}(\vect{s},\vect{q}) X\right\}},
\end{equation}
where $\Sigma(\vect{s},\vect{q}) $ is the block diagonal matrix given by
\begin{align}\label{eq:Sigma-s-q}
    \Sigma(\vect{s},\vect{q}) =\diag(\Sigma_{1}(\vect{s},\vect{q}),\dots,\Sigma_{n}(\vect{s},\vect{q})).
\end{align}
 The same uninformative hyper-prior as used before in \eqref{eq:hyperhyper1} is adopted for the hyper-parameters \(\boldsymbol s \) and $\boldsymbol q$: 
 \begin{align}\label{eq:hyper-hyper-prior}
    p(\boldsymbol s)= \prod^n_{i=1}\Gamma(s_{i} \vert a, b),\quad 
    p(\boldsymbol q) = \prod^n_{i=1}\Gamma(q_{i} \vert a, b).
\end{align}
where $a = 1$ and $b = 10^{-4}$.

\subsection{Inference with joint hierarchical Bayesian learning (JBHT)}
We first compute the MAP estimate \(\hat{\vect{\theta}}\) of all hyper-parameters \(\vect{\theta} = (\vect{s},\vect{q},\beta) \) as defined by their density functions in \eqref{eq:hyper-prior-beta} and \eqref{eq:hyper-hyper-prior} by maximizing the posterior of \(\vect{\theta} \), that is,
\begin{align}\label{eq:MAP-theta}
    \hat{\vect{\theta}}=\mathop{\arg\max}\limits_{\vect{\theta}}p(\vect{\theta}\vert Y).  
\end{align}
We then derive the conditional posterior \(p(X\vert\hat{\vect{\theta}}, Y)\) and a point estimate of \(X\) by maximizing the conditional posterior. In what follows we describe how this can be accomplished.

We start with the computation of \(\hat{\vect{\theta}}\) in \eqref{eq:MAP-theta}. Using Bayes' theorem \AG{\eqref{eq:bayes}}, 
we rewrite \eqref{eq:MAP-theta} as 
\begin{align*}
    \hat{\vect{\theta}}=\mathop{\arg\min}\limits_{\vect{\theta}}[-\ln p(Y\vert\vect{\theta}) - \ln p(\vect{\theta})] = \mathop{\arg\min}\limits_{\vect{\theta}} [-\ln p(Y\vert\vect{\theta})-\ln  p(\vect{s}) -\ln p(\vect{q}) -\ln p(\beta)].
\end{align*}
We then plug the likelihood function \eqref{eq:likelihood} and the prior \eqref{eq:prior} into 
\begin{align*}
    p(Y\vert\vect{\theta}) = \int p(Y\vert X, \beta) p(X\vert\vect{s},\vect{q})dX,
\end{align*}
and then obtain from a standard derivation (see e.g.~\cite{tipping2001sparse}) 
\begin{align*}
    p(Y\vert\vect{\theta}) = (2\pi)^{-\frac{nJ}{2}}\abs{\beta^{-1}I+\Sigma(\vect{s},\vect{q})}^{-\frac{1}{2}}\exp{\left\{ -\frac{1}{2}Y^T(\beta^{-1}I+\Sigma(\vect{s},\vect{q}))^{-1}Y \right\}}.
\end{align*}
We now combine this expression with the hyper-priors \eqref{eq:hyper-hyper-prior} and \eqref{eq:hyper-prior-beta} to obtain 
\begin{align*}
    \hat{\vect{\theta}} = \mathop{\arg\min}\limits_{\vect{\theta} = (\vect{s},\vect{q},\beta)} \mathcal{L}(\vect{\theta})\coloneqq &\ln \left\vert \beta^{-1}I+\Sigma(\vect{s},\vect{q}) \right\vert + Y^T(\beta^{-1}I+\Sigma(\vect{s},\vect{q}))^{-1}Y\\
    & - 2b \left(\sum_{i=1}^{nJ}s_i + \sum_{i=1}^{nJ}q_i + \beta\right),
\end{align*}
where \(b=10^{-4}\) based on previous choices of $b_s$ and $b_\beta$. 

Following a similar approach from \cite{tipping2001sparse}, we are now able to construct a solution to this minimization problem. Specifically, we derive iterative algorithms based on the partial derivatives with respect to the log of each parameter, which are computed as
\begin{align}\label{eq:L-partials}
    \begin{aligned}
        \frac{\partial\mathcal{L}}{\partial{\ln s_i}} &= 2 - s_i\left[\frac{1}{J}\norm{[\vect{\mu}(\vect{\theta})]_i}^2+[\Lambda(\vect{\theta})]_{i,i}  -[\Sigma(\vect{s},\vect{q})]_{i,i}  + 2b \right],\\
        \frac{\partial\mathcal{L}}{\partial{\ln q_i}} &= 2 - 2q_{i} \left[\frac{1}{J}[\vect{\mu}(\vect{\theta})]_i^T [\vect{\mu}(\vect{\theta})]_{i+1}+[\Lambda(\vect{\theta})]_{i,i+1}-[\Sigma(\vect{s},\vect{q})]_{i,i+1}+b\right],\\
        \frac{\partial\mathcal{L}}{\partial{\ln \beta}} &= (N+2) - \beta \left[\text{tr}(\Lambda(\vect{\theta}))+\norm{Y-\vect{\mu}(\vect{\theta})}^2+2b\right],
    \end{aligned}
\end{align}
where 
\begin{align}\label{eq:mu-Lambda}
    \vect{\mu}(\vect{\theta}) = \beta\Lambda(\vect{\theta}) Y,\quad\quad \Lambda(\vect{\theta}) = \left(\beta I+[\Sigma(\vect{s},\vect{q})]^{-1}\right)^{-1},
\end{align}
\([\vect{\mu}(\vect{\theta})]_{i}  \) is the \(i \)-th block of \(\vect{\mu}(\vect{\theta})\), and 
\begin{align*}
    A_{i,j}: = \frac{1}{J}\sum_{k=1}^{J} A((i-1)J +k, (j-1)J +k), \quad 1\leq i,j\leq n
\end{align*}
for \(A\in \mathbb{R}^{nJ\times nJ} \). 

We are now able to formulate the minimizer \(\hat{\vect{\theta}} \)  as a fixed point of a specified operator by setting each partial derivative in \eqref{eq:L-partials} to zero and then compute the fixed point via iterative algorithm(s). In particular, we use the following iterative formulas to update \(\vect{\theta}^{(k+1)} \) based on the current value \(\vect{\theta}^{(k)} = (\vect{s}^{(k)}, \vect{q}^{(k)}, \beta^{(k)}) \):  
\begin{align}
    s_{i} ^{(k+1)} &= \frac{2}{\frac{1}{J}\norm{[\vect{\mu}(\vect{\theta}^{(k)})]_i}^2+[\Lambda(\vect{\theta}^{(k)})]_{i,i}  -[\Sigma(\vect{s}^{(k)},\vect{q}^{(k)})]_{i,i}  + 2b}, \label{eq:SBL-s} \\
    q_{i} ^{(k+1)} &= \frac{1}{\frac{1}{J}[\vect{\mu}(\vect{\theta}^{(k)})]_i^T [\vect{\mu}(\vect{\theta}^{(k)})]_{i+1}+[\Lambda(\vect{\theta}^{(k)})]_{i,i+1}-[\Sigma(\vect{s}^{(k)},\vect{q}^{(k)})]_{i,i+1}+b}, \label{eq:SBL-q} 
\end{align}
each for $1 \le i \le n$, and
\begin{align}
    \beta^{(k+1)} &= \frac{2N+2}{\text{tr}(\Lambda(\vect{\theta}^{(k)}))+\norm{Y-\vect{\mu}(\vect{\theta}^{(k)})}^2+2b}. \label{eq:SBL-beta}
\end{align}

With the  approximation \(\vect{\theta}^{\ast}\) of \(\hat{\vect{\theta}}\) in hand, we can now compute the point estimate of \(X \) by maximizing the conditional posterior as
\begin{align*}
    X^{\ast} = \mathop{\arg\max}\limits_{X} p(X\vert\vect{\theta}^{\ast}, Y). 
\end{align*}
By using the conjugacy of Gaussian distributions \cite{Gelman2015}, it follows from the Gaussian likelihood \(p(Y\vert X, \beta) \) in \eqref{eq:likelihood} and the Gaussian prior \(p(X\vert \vect{s},\vect{q}) \) in \eqref{eq:prior} that 
\begin{align}\label{eq:conditional-posterior}
    p(X\vert \vect{\theta},Y) = (2\pi)^{-\frac{nJ}{2}}\abs{\Lambda(\vect{\theta})}^{-\frac{1}{2}}\exp{\left\{ -\frac{1}{2}\left(X-\vect{\mu}(\vect{\theta})\right)^T [\Lambda(\vect{\theta})]^{-1}\left(X-\vect{\mu}(\vect{\theta})\right) \right\}},
\end{align}
where \(\vect{\mu} \) and \(\Lambda \) are defined in \eqref{eq:mu-Lambda}. From \eqref{eq:conditional-posterior} we have our estimate
\begin{align*}
    X^{\ast} = \vect{\mu}(\vect{\theta}^{\ast}).
\end{align*}
Our new JHBL approach is summarized in Algorithm \ref{alg:JHBL}.

\begin{algorithm}[h!]
    \caption{JHBL approach of estimating $X$  ($\boldsymbol x_{\JHBL}^\beta$ in numerical examples)}
    \label{alg:JHBL}
        \hspace*{\algorithmicindent} \textbf{Input:} Measurements $Y$ in \eqref{eq:forward_m}\\
        \hspace*{\algorithmicindent} \textbf{Output:} Signal estimate $X^{\ast}$.
    \begin{algorithmic}[1]
    \State{Initialize $a=1$, $b = 10^{-4}$, and \(\vect{\theta}^{(0)}=(\vect{s}^{(0)}, \vect{q}^{(0)}, \beta^{(0)}) \).}
    \Repeat
        \State{Update $\vect{s}^{(k+1)}$ by \eqref{eq:SBL-s}}
        \State{Update $\vect{q}^{(k+1)}$ by \eqref{eq:SBL-q}}
        \State{Update $\beta^{(k+1)}$ by \eqref{eq:SBL-beta}}
        \State{Update $k\to k+1$}
    \Until{convergence at \(\vect{\theta}^{\ast} = (\vect{s}^{\ast}, \vect{q}^{\ast}, \beta^{\ast}) \) }

    \State Compute \( X^{\ast} = \vect{\mu}(\vect{\theta}^{\ast}) \) as in \eqref{eq:mu-Lambda}. 
    \end{algorithmic}
\end{algorithm}

As discussed in \cite{wipf2007empirical},  it may be possible to have information regarding $\beta$, in which case it does not have to be determined. Hence we also include Algorithm \ref{alg:JHBL-fixed-beta}, which is the JHBL algorithm given such information. 

\begin{algorithm}[h!]
    \caption{JHBL approach of estimating $X$ with fixed $\beta $  ($\vect{x}_{\JHBL}$ in numerical examples)}
    \label{alg:JHBL-fixed-beta}
        \hspace*{\algorithmicindent} \textbf{Input:} Measurements $Y$ in \eqref{eq:forward_m}\\
        \hspace*{\algorithmicindent} \textbf{Output:} Signal estimate $X^{\ast}$.
    \begin{algorithmic}[1]
    \State{Initialize $a=1$, $b = 10^{-4}$, \(\beta \)  and \(\vect{\theta}^{(0)} \).}
    \Repeat
        \State{Update $\vect{s}^{(k+1)}$ by \eqref{eq:SBL-s}}
        \State{Update $\vect{q}^{(k+1)}$ by \eqref{eq:SBL-q}}
        \State{Update $k\to k+1$}
    \Until{convergence at \(\vect{\theta}^{\ast} = (\vect{s}^{\ast}, \vect{q}^{\ast}, \beta) \) }

    \State Compute \( X^{\ast} = \vect{\mu}(\vect{\theta}^{\ast}) \) as in \eqref{eq:mu-Lambda}. 
    \end{algorithmic}
\end{algorithm}

\subsection{Inference with refined JHBL}
As discussed in Section \ref{sec:hyper_prior}, the solution $X$ can have non-zero entries only corresponding to very small $\boldsymbol s$. As was also discussed, the components of $\boldsymbol s$ may be large over any interval in which there is a change in the edge location, which results from the ``averaging out'' of the information in the temporal sequence. We emphasize that whenever {\em any} $s_i$ is large, it will dominate the information provided by both the inverse prior covariance matrix $\Sigma$ in \eqref{eq:prior} and the posterior covariance matrix $\Lambda$ in \eqref{eq:conditional-posterior}. This is because the re-estimate rule of $q_i$ in \eqref{eq:SBL-q} involves both $\Sigma$ and $\Lambda$ meaning that $\boldsymbol q$ contains similar information as $\boldsymbol s$ does. 
 
A better approach would be to use an update rule for  $\boldsymbol q$, which represents the {\em temporal correlation} of the $J$ observations at each location $i = 1,\dots,n$, that is independent of $\boldsymbol s$, which corresponds to probability that an edge occurs at that location in any {\em individual} image.  Thus we propose to compute a point estimate \(X^{(k)} \) given the current hyper-parameter estimate \(\vect{\theta}^{(k)} \)
\begin{align}
    X^{(k)} = \mathop{\arg\max}\limits_{X} p(X\vert\vect{\theta}^{(k)}, Y) = \vect{\mu}(\vect{\theta}^{(k)})
\label{eqn:Xkiterate}
\end{align}   
and use the covariance matrix of the given temporal sequence to compute $q$ as
\begin{equation}\label{eq:q_update_correct}
    \vect{q}^{(k+1)} = \diag(\Cov( \vect{x}_{1}^{(k)},\dots, \vect{x}_J^{(k)})).
\end{equation}
Here $\Cov(\cdot)$ denotes the covariance matrix of the input.

Using \eqref{eq:q_update_correct} helps to mitigate the problem corresponding to large $s_i$ for any individual image having an out-sized impact on the remaining images in the sequence.  It is still the case, however, that large $s_i$ will have dominant influence in the neighborhoods of $X$ for which there is translation of a nonzero value, or edge.  
To compensate for this problem, our method must account for nonzero values of $\boldsymbol s$ in these neighborhoods so that the translating edge is identified. In this regard, we first get individual estimate of \(\vect{s} \) from each \( \vect{y}_{j} \) separately, which could be viewed as a special case of \eqref{eq:SBL-s} with \(J=1\) and \(Y=\vect{y}_{j} \) for each \(1\leq j\leq J\). Indeed, we will calculate the components of each individual estimate \(\vect{u}_{j} ^{(k+1)} \), $j = 1,\dots,J$, as 
\begin{align}\label{eq:SBL-u}
    u_{j,i}^{(k+1)} = \frac{2}{\norm{[\vect{\mu}_{j} (\vect{\theta}^{(k)})]_i}^2+[\Lambda_{j} (\vect{\theta}^{(k)})]_{i,i}  -[\Sigma_{j} (\vect{s}^{(k)},\vect{q}^{(k)})]_{i,i}  + 2b},  \quad 1\leq i\leq n, 
\end{align}  
where \( \vect{\mu}_{j} (\vect{\theta}^{(k)}) = \beta^{(k)} \Lambda_{j} (\vect{\theta}^{(k)}) \vect{y}_{j} \), \(\Lambda_{j} (\vect{\theta}^{(k)}) = (\beta I + [\Sigma_{j} (\vect{s}^{(k)},\vect{q}^{(k)})]^{-1} )^{-1}  \), and \(\Sigma_{j} (\vect{s}^{(k)},\vect{q}^{(k)}) = \operatorname{diag} (\vect{s}^{(k)}) \). We then use \AG{\eqref{eq:SBL-u}} to refine \(\vect{s}^{(k+1)}\) as
\begin{align}\label{eq:SBL-s-refine}
    s_{i} ^{(k+1)} = \mathop{\min}_{1\leq j\leq J}  \left\vert u_{j,i}^{(k+1)}  \right\vert, \quad i \in I, 
\end{align}
where $I=\{i\colon q^{(k+1)}_i \geq \vartheta\max_{i=1,\dots,n}q^{(k+1)}_i\}$ for some $\vartheta\in[0,1]$. The parameter $\vartheta$ is inherently related to the scale of the sparse signal, the SNR, and the distance between nonzero entries (resolution) in the underlying solution ${\vect x}_{j} $.  In our examples we use a rough estimate for $\vartheta$ that would be accessible from the measurements without any additional tuning. The idea in using \eqref{eq:SBL-s-refine} is to redefine $\vect{s}$ in possible regions of change, which is informed by $\vect{q}$ defined by \eqref{eq:q_update_correct}.  Since larger entries of $q_{i}$'s indicate more likely regions of change, \eqref{eq:SBL-s-refine} seeks to mitigate the impact of $s_{i}$'s in those regions. 

Observe that \eqref{eq:SBL-s-refine} does not consider the joint sparsity profile of the temporal sequence, but rather seeks to confirm whether a nonzero value (or edge)  was present at any $i \in I$ over $j = 1,\dots, J$. Specifically, by using \eqref{eq:SBL-s-refine} the solution posterior is allowed to produce nonzero values at the edges over the neighborhood where changes occur.

We present this refined JHBL approach in Algorithm \ref{alg:JHBL-refine}. 

\begin{algorithm}[h!]
    \caption{Refined JHBL approach of estimating $X$  ($\boldsymbol x_{\JHBL}^{\beta, \vect{q}}$ in numerical examples)}
    \label{alg:JHBL-refine}
        \hspace*{\algorithmicindent} \textbf{Input:} Measurements $Y$ in \eqref{eq:forward_m}\\
        \hspace*{\algorithmicindent} \textbf{Output:} Signal estimate $X^{\ast}$.
    \begin{algorithmic}[1]
    \State{Initialize $a=1$, $b = 10^{-4}$, and \(\vect{\theta}^{(0)}=(\vect{s}^{(0)}, \vect{q}^{(0)}, \beta^{(0)}) \).}
    \Repeat
        \State{Update $\vect{s}^{(k+1)}$ by \eqref{eq:SBL-s}}
        \State{Update $\vect{q}^{(k+1)}$ by \eqref{eq:q_update_correct}}
        \State{Refine $\vect{s}^{(k+1)}$ by \eqref{eq:SBL-s-refine}}
        \State{Update $\beta^{(k+1)}$ by \eqref{eq:SBL-beta}}
        \State{Update $k\to k+1$}
    \Until{convergence at \(\vect{\theta}^{\ast} = (\vect{s}^{\ast}, \vect{q}^{\ast}, \beta^{\ast}) \) }

    \State Compute \( X^{\ast} = \vect{\mu}(\vect{\theta}^{\ast}) \) as in \eqref{eq:mu-Lambda}. 
    \end{algorithmic}
\end{algorithm}

\subsection{Inference of two-dimensional images}
\label{sec:inference}
We observe that the major computational cost of Algorithm \ref{alg:JHBL-refine} is the calculation of \(\Lambda(\vect{\theta}^{(k)}) \) repeatedly at each iteration \(k \), which is the inverse of a large-scale matrix. This becomes computationally prohibitive for 2D images since the size \(n^{2} \) of a vectorized \(n\times n \) image is typically huge. In what follows we therefore describe an alternative approach that enables an efficient and robust recovery, specifically by approximating $\(\Lambda(\vect{\theta}^{(k)}) \)$.

We begin by using a diagonal approximation of \(\Sigma(\vect{s}, \vect{q}) \) in \eqref{eq:Sigma-s-q} given by
\begin{align*}
    \widetilde{\Sigma} (\vect{s}) = \operatorname{diag} (s_{1} I_J, s_{2}I_J, \ldots , s_{n} I_J).
\end{align*}
That is, we ignore the sub-diagonals \(\vect{q}\).  We note that the magnitude of \(\vect{q} \) is typically smaller than the magnitude of \(\vect{s} \). The approximations of \(\vect{\mu}(\vect{\theta}) \) and \(\Lambda(\vect{\theta}) \) are defined correspondingly as
\begin{align*}
    \widetilde{\vect{\mu}} (\vect{\theta}) = \beta \widetilde{\Lambda}(\vect{\theta}) Y, \quad \widetilde{\Lambda} (\vect{\theta})= \left(\beta I + \widetilde{\Sigma} (\vect{s})^{-1} \right)^{-1}. 
\end{align*}  
We then use these approximations in the partial derivative \eqref{eq:L-partials} to derive the corresponding iterative formulas for updating \(\vect{s} \) and \(\beta \):
\begin{align}\label{eq:SBL-2D-s}
    s_i ^{(k+1)} = \frac{J+2}{\norm{[\widetilde{\vect{\mu}}(\vect{\theta}^{(k)})]_i}^2 + 2b}
\end{align}
and
\begin{align}\label{eq:SBL-2D-beta}
    \beta ^{(k+1)} = \frac{nJ+2}{\norm{Y-\widetilde{\vect{\mu}}(\vect{\theta}^{(k)})}^2+2b}.
\end{align}
We emphasize that the above iterations could be computed much faster than \eqref{eq:SBL-s} and \eqref{eq:SBL-beta} since the approximated matrix \(\widetilde{\Lambda}  \) is diagonal. 

Analogous to  \eqref{eqn:Xkiterate}, we then have
\begin{equation}
\label{eq:Xkiterate2D}
 X^{(k)} = \widetilde{\vect{\mu}} (\vect{\theta}^{(k)}),
\end{equation}
for which we employ the same refinement techniques as \eqref{eq:q_update_correct} and \eqref{eq:SBL-s-refine} respectively  given by 
\begin{align}\label{eq:SBL-2D-q}
   \vect{q}^{(k+1)} = \diag(\Cov( \vect{x}_{1}^{(k)},\dots, \vect{x}_J^{(k)})), 
\end{align}
and 
\begin{align}\label{eq:SBL-2D-s-refine}
    s_{i} ^{(k+1)} = \mathop{\min}_{1\leq j\leq J}  \left\vert u_{j,i}^{(k+1)}  \right\vert, \quad i \in I, 
\end{align}
where \(u_{j,i}^{(k+1)}\) is similar to \eqref{eq:SBL-u} with the diagonal approximations and 
$$I=\{i\colon q^{(k+1)}_i \geq \vartheta\max_{i=1,\dots,n}q^{(k+1)}_i\}$$ for some $\vartheta\in[0,1]$.

We present this refined JHBL approach for 2D images in Algorithm \ref{alg:JHBL-refine-2D}.

\begin{algorithm}[h!]
    \caption{Refined JHBL approach of estimating 2D $X$  ($\boldsymbol x_{\JHBL}^{\beta, \vect{q}}$ in numerical examples)}
    \label{alg:JHBL-refine-2D}
        \hspace*{\algorithmicindent} \textbf{Input:} Measurements $Y$ in \eqref{eq:forward_m}\\
        \hspace*{\algorithmicindent} \textbf{Output:} Signal estimate $X^{\ast}$.
    \begin{algorithmic}[1]
    \State{Initialize $a=1$, $b = 10^{-4}$, and \(\vect{\theta}^{(0)}=(\vect{s}^{(0)}, \vect{q}^{(0)}, \beta^{(0)}) \).}
    \Repeat
        \State{Update $\vect{s}^{(k+1)}$ by \eqref{eq:SBL-2D-s}}
        \State{Update $\vect{q}^{(k+1)}$ by \eqref{eq:SBL-2D-q}}
        \State{Refine $\vect{s}^{(k+1)}$ by \eqref{eq:SBL-2D-s-refine}}
        \State{Update $\beta^{(k+1)}$ by \eqref{eq:SBL-2D-beta}}
        \State{Update $k\to k+1$}
    \Until{convergence at \(\vect{\theta}^{\ast} = (\vect{s}^{\ast}, \vect{q}^{\ast}, \beta^{\ast}) \) }

    \State Compute \( X^{\ast} = \vect{\mu}(\vect{\theta}^{\ast}) \) as in \AG{\eqref{eq:Xkiterate2D}}. 
    \end{algorithmic}
\end{algorithm}


\section{Numerical results}
\label{sec:numresultsSBL}
We now provide some one- and two-dimensional examples to demonstrate the efficacy of our method.

\subsection{One-dimensional piecewise continuous signal}
\label{sec:test_1Dsignal}
We consider the temporal sequence of $J=6$ one-dimensional piecewise smooth functions given by
\begin{example}
\label{example:falpha}
The functions $f_{j} \colon[-\pi,\pi]$ are defined as
\begin{equation}\label{eq:1d_fun}
f_{j} (\alpha_i) = \begin{cases}
1.5, & -\frac{3\pi}{4}-.5+.1j\leq \alpha_i<-\frac{\pi}{2}-.5+.1j\\
\frac{7}{4}-\frac{\alpha_i}{2}+\sin(\alpha_i-\frac{\pi}{4}), & -\frac{\pi}{4}\le \alpha_i<\frac{\pi}{8}\\
\frac{11\alpha_i}{4}-5, & \frac{3\pi}{8}\le \alpha_i<\frac{3\pi}{4}\\
0, & \text{otherwise}.
\end{cases}
\end{equation}
\end{example}

\begin{figure}[h!]
\centering
\begin{subfigure}[b]{.8\textwidth}
\includegraphics[width=\textwidth]{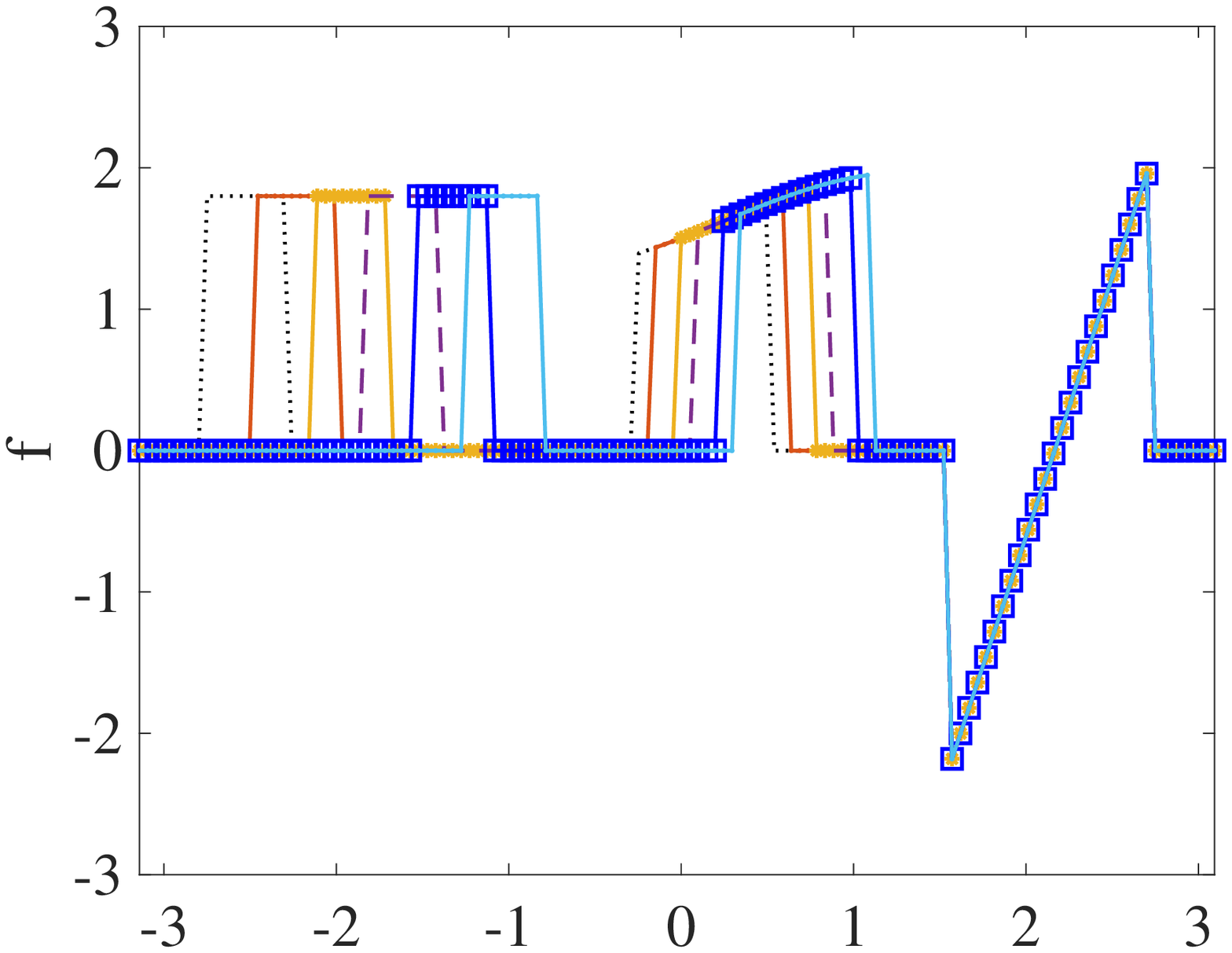}
\caption{$f_{j} $, $j=1,\dots,6$ in \eqref{eq:1d_fun}}
\end{subfigure}
\\
\begin{subfigure}[b]{.45\textwidth}
\includegraphics[width=\textwidth]{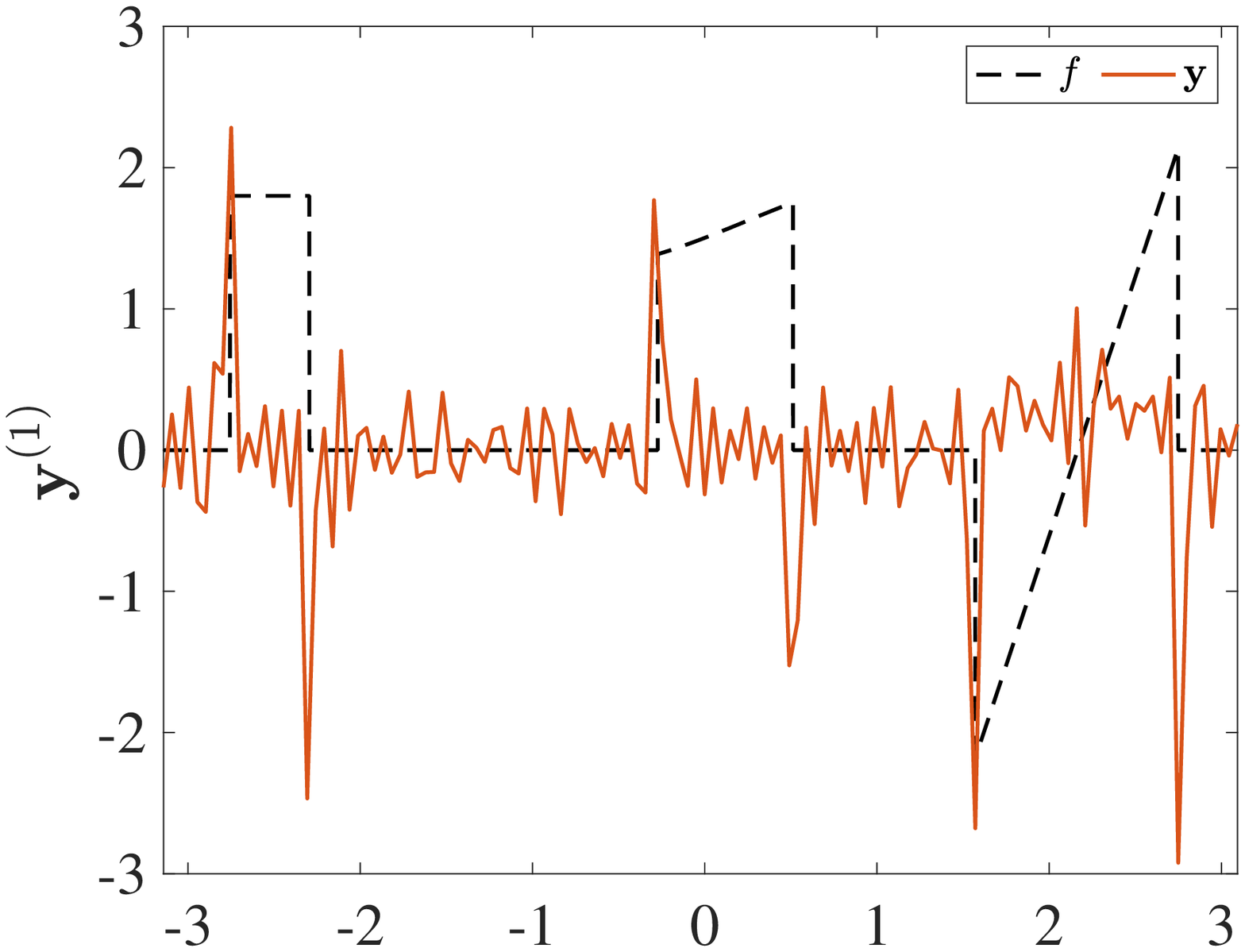}
\caption{$\vect{y}_{1} $}
\end{subfigure}
\begin{subfigure}[b]{.42\textwidth}
\includegraphics[width=\textwidth]{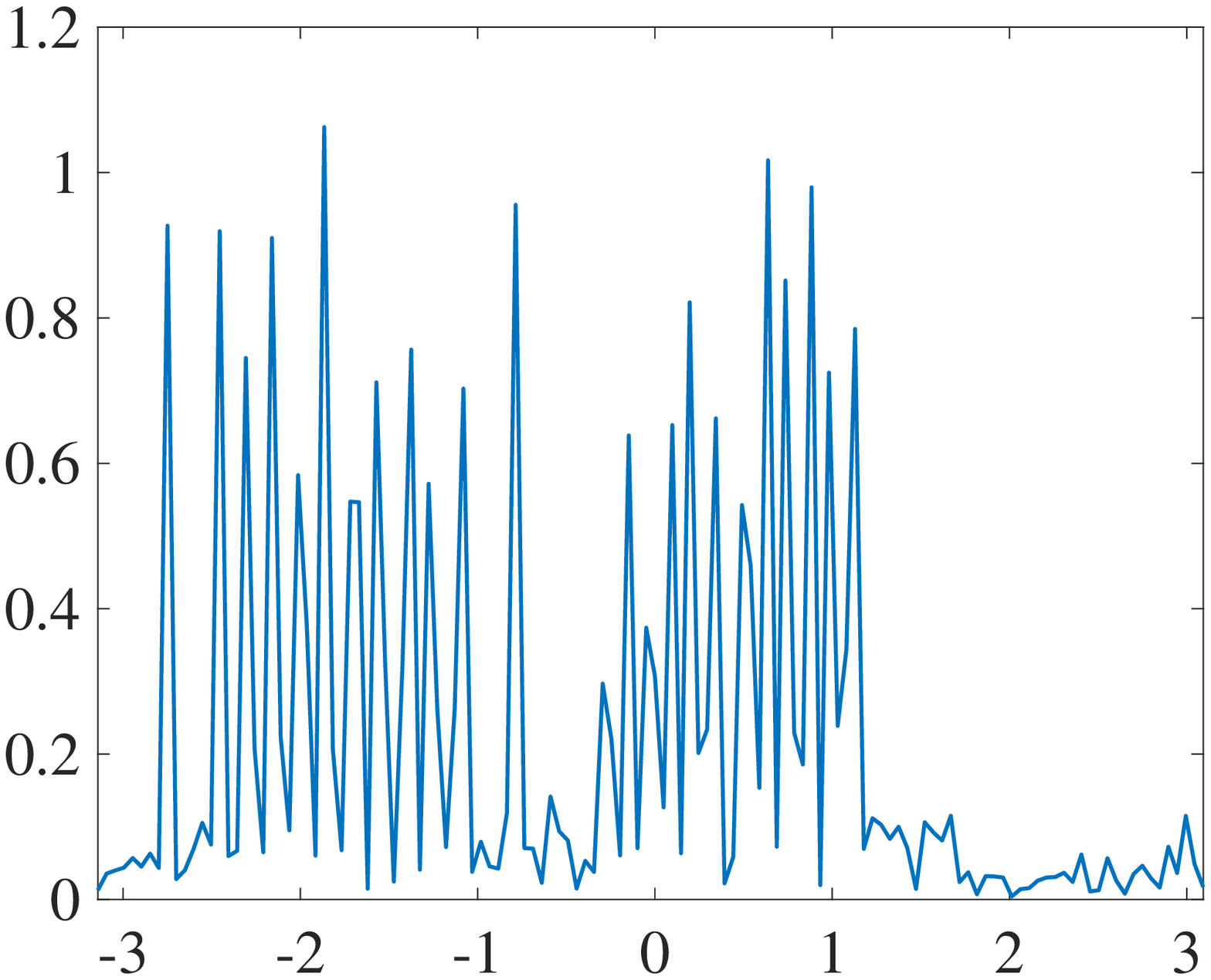}
\caption{The initialization of $\boldsymbol q$}
\end{subfigure}

\caption{(a) The sequence of functions given by Example \ref{example:falpha}. (b) $\vect{y}_{1} $ as defined in \eqref{eq:forward_m}. (c) hyper-parameter $\boldsymbol q$ initialized in \eqref{eq:q_update_correct}.  
}
\label{fig:noisy_edge}
\end{figure}

The sequence of functions given by Example \ref{example:falpha} are depicted in Figure \ref{fig:noisy_edge}(a), while Figure \ref{fig:noisy_edge}(b) shows the measurements $\vect{y}_{1} $ as given by \eqref{eq:edge_est}. We note that the similar oscillations  are likewise observed in each $\vect{y}_{j}$, $j = 2,\dots, 6$.  The Fourier measurements in \eqref{eq:fourcoeff}, as discretized by ${\boldsymbol g}_{j} $, $j = 1,\dots,6$, given in \eqref{eq:forward_noisy}, are also each missing the symmetric frequency band $\mathcal{K}_j$, $j = 1,\dots, J$, given by
\begin{equation}
\label{eq:bandzero_1d}
\mathcal{K}_j = [\pm(10j + 13),\pm(10j + 15)],
\end{equation} 
Our goal is to recover the jump function (edges) of each  $f^{(j)}$ as defined in \eqref{eq:jumpFunc} given by ${\boldsymbol x}_{j} $ in \eqref{eq:xvector}. The domain $[-\pi,\pi]$ is uniformly discretized with $n=128$ grid points such that $\alpha_i= \frac{(2 \pi)(i-1)}{n}$, $i=1,\dots,n$.    
In our experiments we consider additive i.i.d. Gaussian noise of zero-mean and $.2$-variance (signal-to-noise ratio $\text{SNR} \approx 20$) where we define the SNR specifically for \eqref{eq:forward_m} as
\begin{equation}
\label{eq:SNR}
{\text{SNR}=10\log_{10}\left(\frac{\bar{f}^2}{\sigma^2}\right)},
\end{equation}
where $\bar{f}$ is the mean value of $f$ over the observed grid points. 

From Figure \ref{fig:noisy_edge}(c) we observe that the covariance matrix of the acquired data in \eqref{eq:forward} can roughly determine the interval in which the edge locations have moved. It is not very accurate, however, as there are spurious oscillations that affect the ability to determine the true shifted jump locations.  Such a coarse initial estimate of the change region might be all that is available in a variety of applications, and indeed motivates our proposed framework here. That is, our method can still be useful even when the change region is not accurately described.

\begin{figure}[h!]
\centering
\begin{subfigure}[b]{.31\textwidth}
\includegraphics[width=\textwidth]{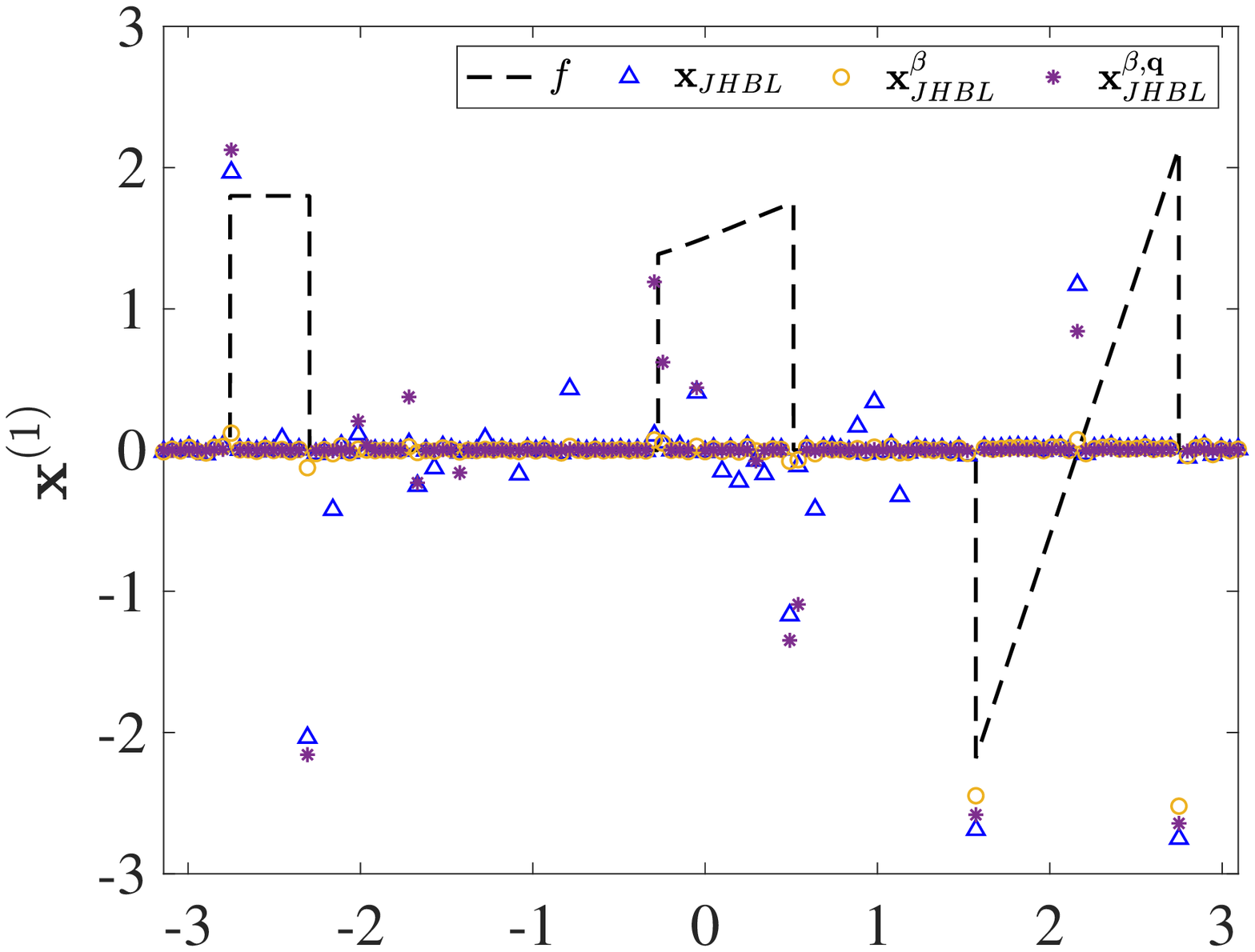}
\end{subfigure}
\begin{subfigure}[b]{.31\textwidth}
\includegraphics[width=\textwidth]{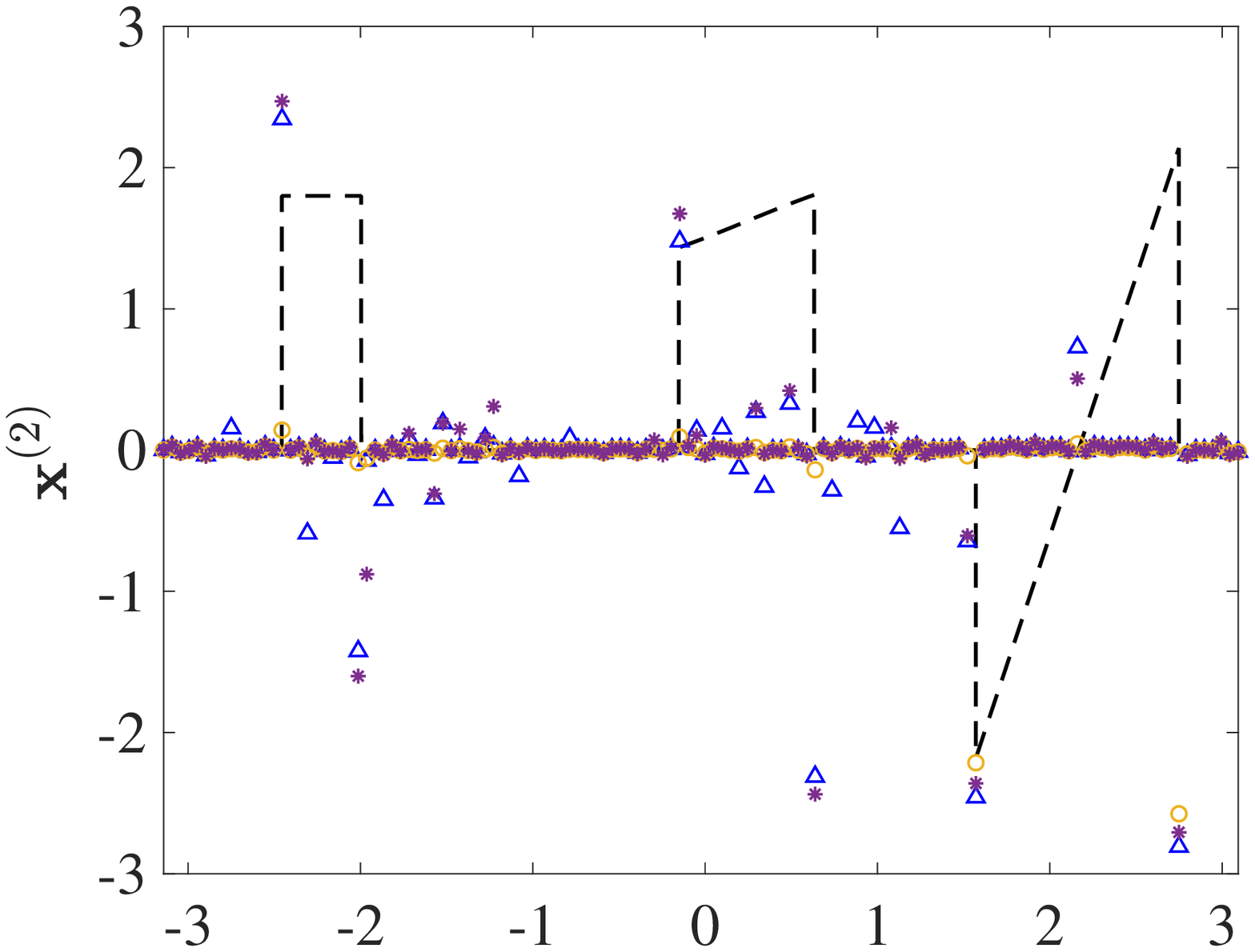}
\end{subfigure}
\begin{subfigure}[b]{.31\textwidth}
\includegraphics[width=\textwidth]{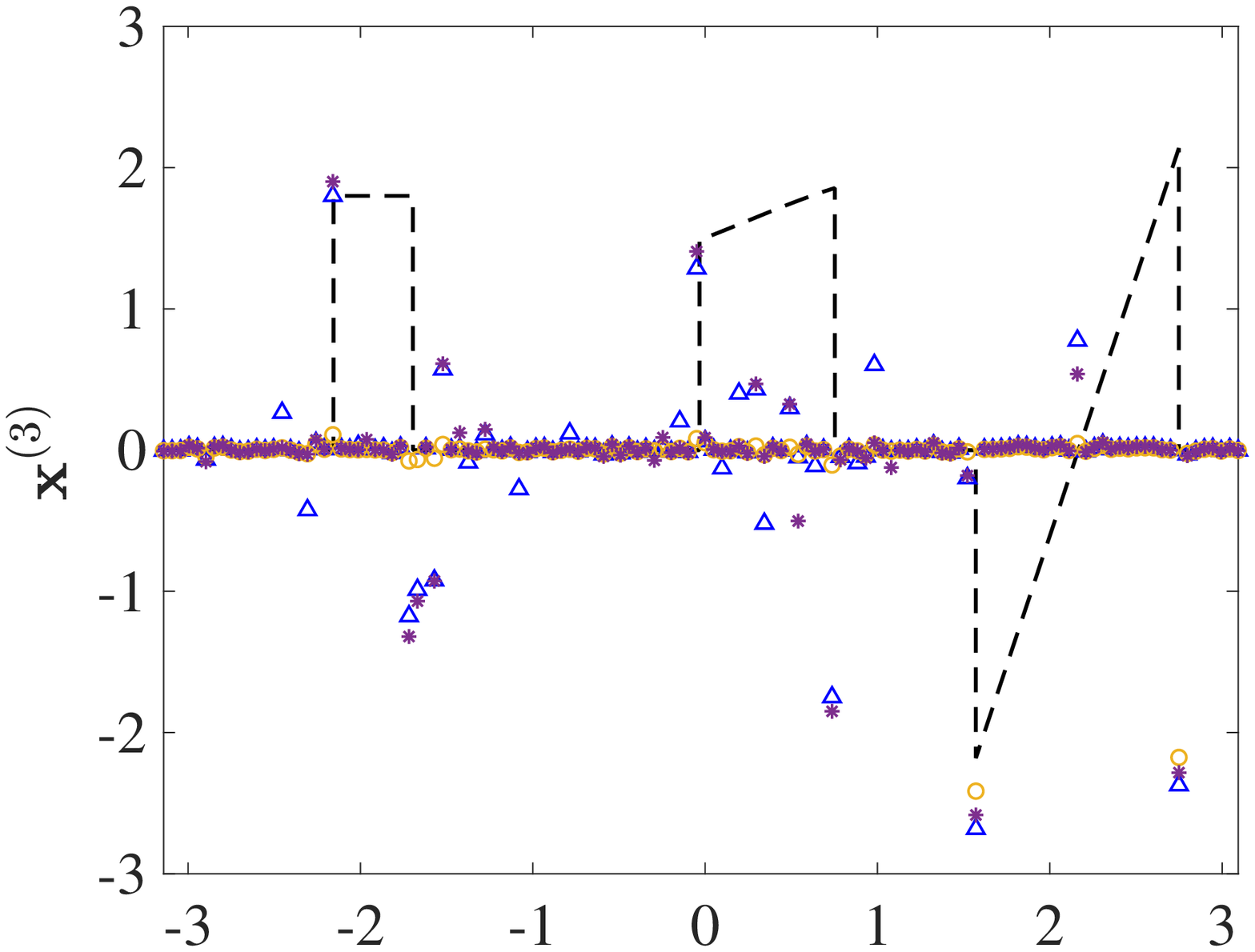}
\end{subfigure}
\\
\begin{subfigure}[b]{.31\textwidth}
\includegraphics[width=\textwidth]{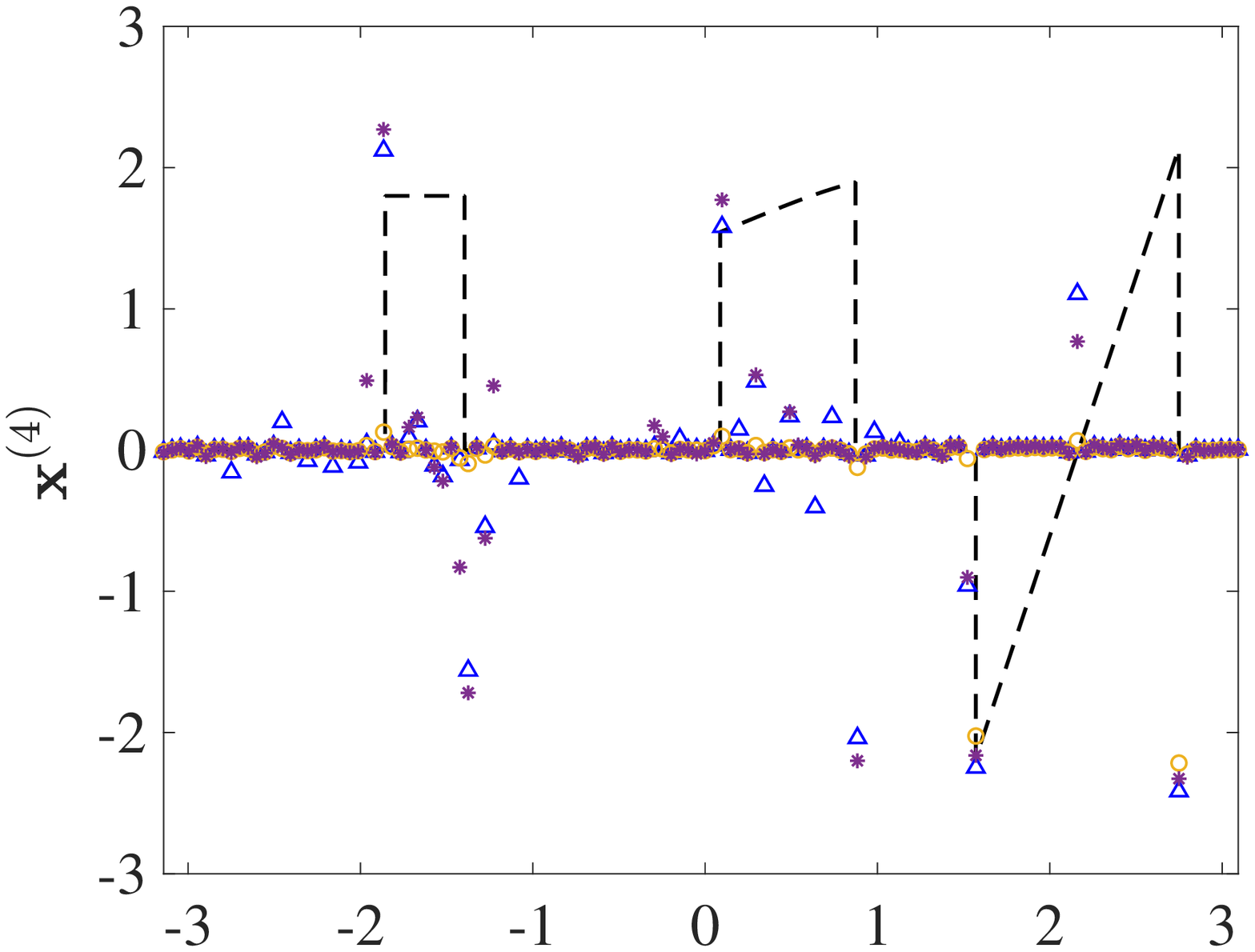}
\end{subfigure}
\begin{subfigure}[b]{.31\textwidth}
\includegraphics[width=\textwidth]{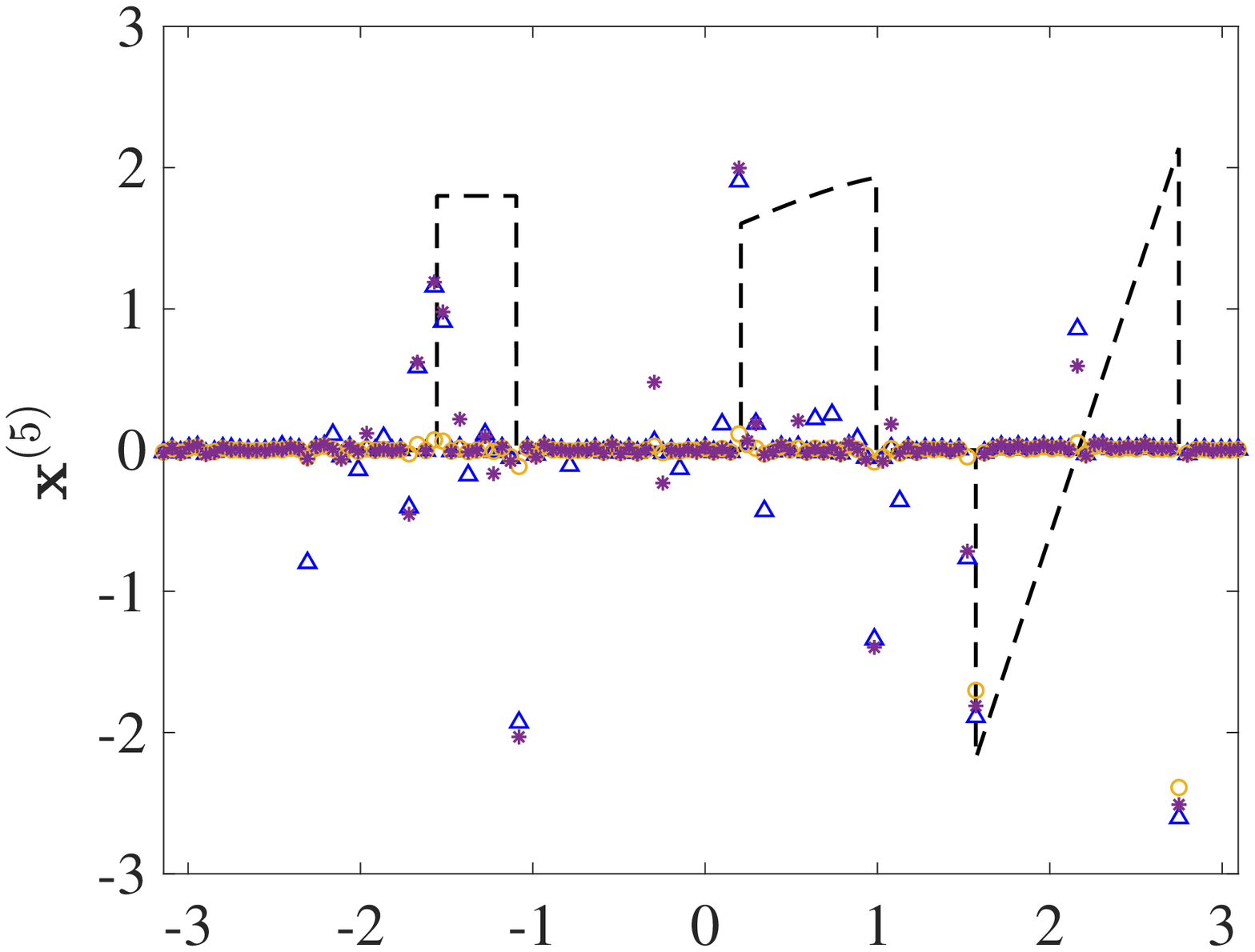}
\end{subfigure}
\begin{subfigure}[b]{.31\textwidth}
\includegraphics[width=\textwidth]{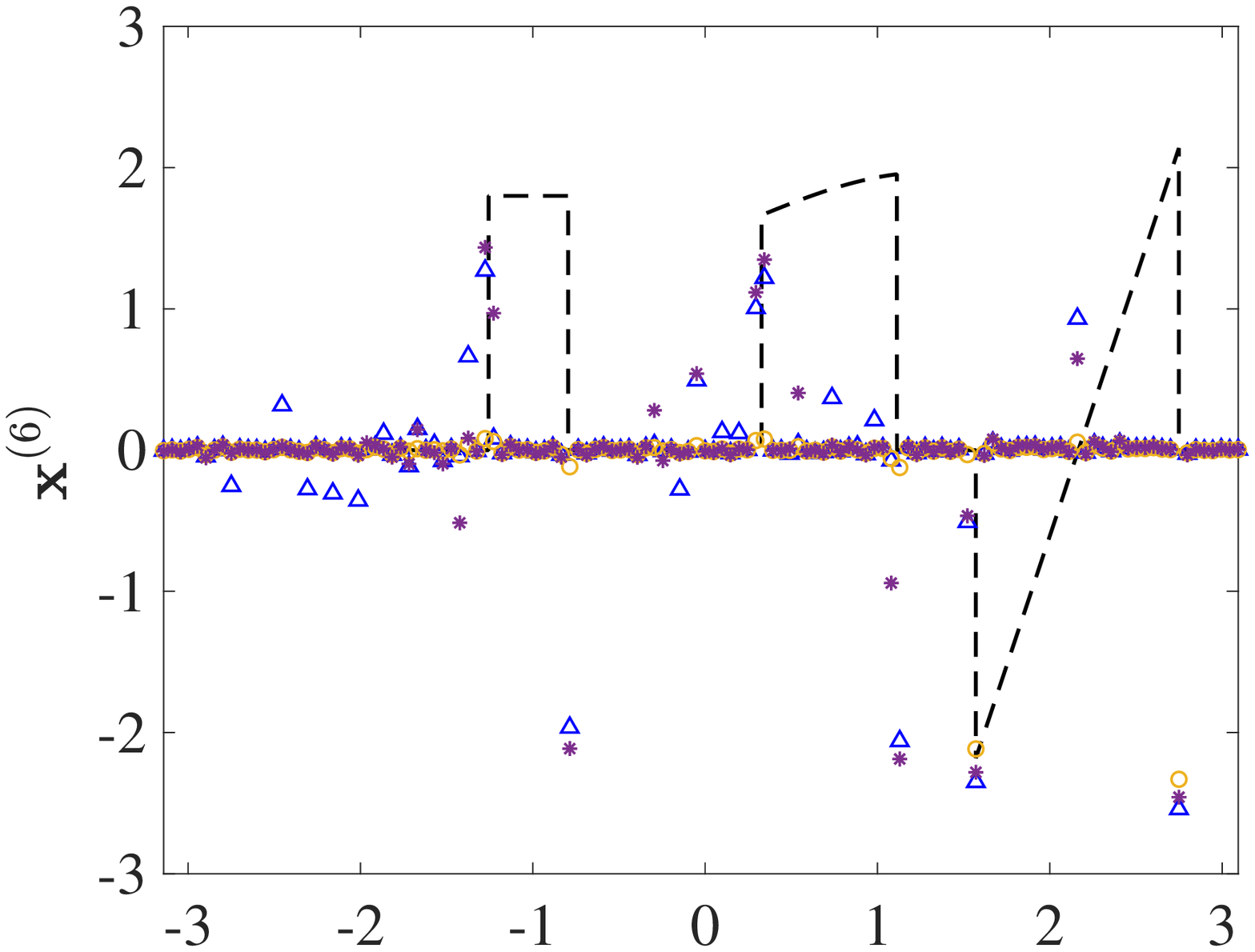}
\end{subfigure}

\caption{Sequential edge map recovery for Fourier measurements given by \eqref{eq:bandzero_1d}, $J = 6$.  The legend is provided in the top left figure.}
\label{fig:rec_SBL}
\end{figure}
Figure \ref{fig:rec_SBL} compares the results of our new method to other existing approaches that seek to recover sparse signals from a sequence of given data.  These include 
(1) $\vect{x}_{\JHBL}$,  the JHBL recovery using Algorithm \ref{alg:JHBL-fixed-beta} with $\beta$ chosen a-priori and (2) $\vect{x}_{\JHBL}^\beta$,  the JHBL recovery using Algorithm \ref{alg:JHBL} for updating $\beta$ iteratively. We denote the results from our proposed method, realized using Algorithm \ref{alg:JHBL-refine}, as $\vect{x}_{\JHBL}^{\beta, \vect{q}}$.  We note that in our examples the exact value of $\beta$ is assumed for $\vect{x}_{\JHBL}$, but not for $\vect{x}_{\JHBL}^{\beta}$ or $\vect{x}_{\JHBL}^{\beta,\vect{q}}$. As can be observed from the results in Figure \ref{fig:rec_SBL}, by explicitly including change information, as is the main feature of Algorithm \ref{alg:JHBL-refine}, we are able to more accurately capture the sequence of jump function recoveries. We used $\vartheta=0.05$ in \eqref{eq:SBL-s-refine} and note that some additional tuning may improve the results. More discussion regarding how parameter $\vartheta$ may be chosen follows \eqref{eq:SBL-s-refine}.

Finally, we note that it is possible of course to use the standard SBL to recover the individual signals separately, that is without combining information from other data in the sequence.  In this case the joint sparsity method is reduced to standard $\ell_1$ regularization (compressive sensing) given by \eqref{eq:cs_model}. While the results for this simple example are comparable, we are interested in the problem  where there is obstruction in each  data acquisition  that hampers individual recovery.  For example, in our two-dimensional examples we consider the case where there are missing bands of Fourier data in each acquisition, so that no single acquisition has enough data to accurately recover the underlying image.

\subsection{Two-dimensional data}
\label{sec:img_rec}
Our two-dimensional examples include a magnetic resonance image (MRI)  and a synthetic aperture radar (SAR) image.  As was done for the one-dimensional sparse signal  example, we also compare edge maps respectively recovered  using Algorithm \ref{alg:JHBL}, Algorithm \ref{alg:JHBL-fixed-beta}, and Algorithm \ref{alg:JHBL-refine-2D}.
In contrast to the one-dimensional signal example, here we show how the temporal edge maps are employed for the downstream process of full image recovery. 

\subsubsection*{Sequential mage recovery using edge maps}

Since it pertains to the overall usefulness of our new sequential edge map recovery procedure, we now include a brief review of some commonly employed algorithms for sequential image recovery.  Due to the sparsity in the edge domain, CS algorithms \cite{candes2006robust,candes2006stable,candes2006near,donoho2006compressed} are often employed to either separately (see e.g.~above references) or jointly  (see e.g.~\cite{adcock2019jointsparsity,cotter2005sparse,chen2006theoretical,eldar2009robust,xiao2022sequential}) recover the corresponding sequence of images.\footnote{The joint recovery methods in these publications consider cases of non-overlapping support in the sparse domain, which is consistent with the assumptions for the temporal sequence of images discussed here.}  For the standard single measurement case, the standard CS algorithm may be written as
\begin{equation}\label{eq:CS_model_image}
    \Tilde{\vect{f}}_{j}  = \argmin_{\vect{f}^\ast} \left(\frac{1}{2}\norm{F_{j} \vect{f}^\ast-\Tilde{\vect{g}}_{j} }_2^2+ \lambda\norm{\mathcal{L}\vect{f}^\ast}_1\right),
\end{equation}
where ${\mathcal L}$ is a sparsifying transform operator, designed here to promote sparsity in the edge domain.\footnote{Since our examples only include piecewise constant structures, we can simply choose ${\mathcal L}$ as a first order differencing operator, and note that high order differencing may also be used as appropriate (see e.g. \cite{archibald2016image}).} If the data acquisition  model \eqref{eq:forward_noisy} is accurate, it should be the case that any measurement $\Tilde{\vect{g}}_{j} $ carries sufficient information to recover its corresponding image so long as the regularization parameter is suitably chosen.  The performance of  CS algorithms deteriorate for seriously under-sampled data with low SNR, however, \cite{shchukina2017pitfalls,kang2019compressive}. Further, even if optimally chosen, the {\em global} impact of regularization term in \eqref{eq:CS_model_image} will make it impossible to resolve local features in these environments. 

Spatially varying regularization parameters can improve the accuracy of the CS solution. Specifically, the parameter should be constructed to more heavily penalize the solution in the true sparse regions (in the edge domain) and less so in regions of presumed support. The (re-)weighted $\ell_1$ regularization method is designed for this purpose, with the most basic form written as \cite{candes2008enhancing}
\begin{equation}\label{eq:wl1_model}
    \Tilde{\vect{f}}_{j}  = \argmin_{\vect{f}^\ast}\left(\frac{1}{2}\norm{F_{j} \vect{f}^\ast-\Tilde{\vect{g}}_{j} }_2^2+\norm{W\mathcal{L}\vect{f}^\ast}_1\right).
\end{equation}
Here the entries $\vect{w}_{j}$ of the diagonal weighting matrix $W=\diag{(\vect{w}_{j} )}$ typically are iteratively constructed, yielding expensive computational cost. Furthermore, errors in $W$ due to noise and incompleteness of acquired data will be fed into \eqref{eq:wl1_model} and the resulting error will propagate at each iteration. 

As the edge information of each part of the temporal sequence has been determined prior to the image reconstruction, following what was done in \cite{adcock2019jointsparsity,gelb2019reducing,scarnati2019accelerated}, we avoid iterating on $W$ and define a pre-computed weighting matrix as 
\begin{equation}\label{eq:weights}
    \omega^{(j)}(i) = \frac{1}{\text{vec}\left(\Tilde{\vect{x}}_{j} \right)(i)+\epsilon},
\end{equation}
where $\text{vec}(\cdot)$ is to stack input vertically as a vector and $\epsilon=10^{-4}$ is introduced to avoid zero-valued denominator. We further scale the matrix with maximum value 1 and replace \eqref{eq:weights} as
\begin{equation}
    \label{eq:weights_scaled}
    \omega^{(j)}(i) = \frac{\omega^{(j)}(i)}{\max_{i}{\abs{ \omega^{(j)}(i)}}}.
\end{equation}
We follow \cite{gelb2019reducing} and employ the Alternating Direction Method of Multipliers (ADMM) (see \cite{boyd2011distributed}), to solve the convex optimization problem in \eqref{eq:wl1_model}.

\subsubsection*{Sequential MRI}

\begin{figure}[h!]
    \centering
    \begin{subfigure}[b]{.23\textwidth}
    \includegraphics[width=\textwidth]{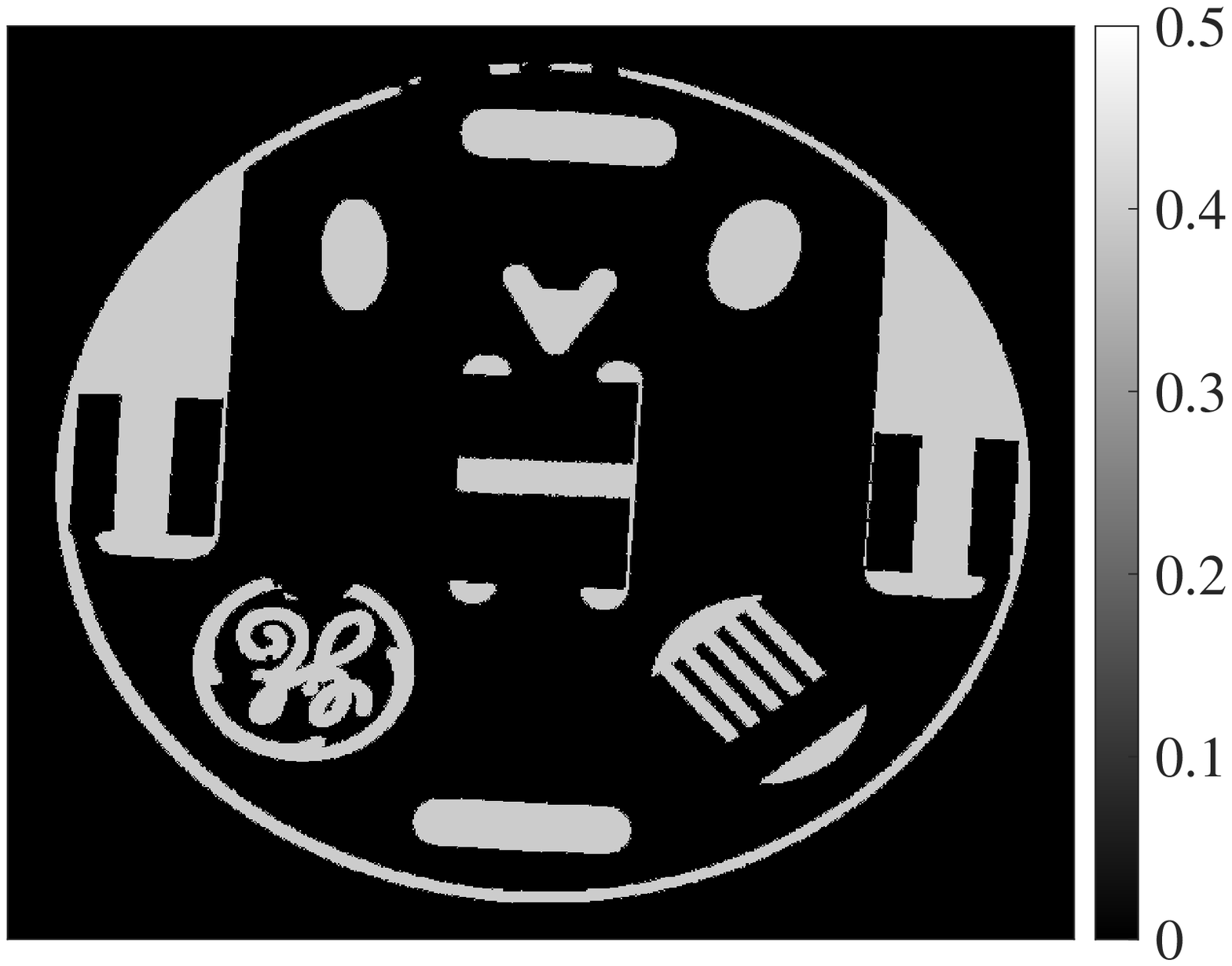}
    \caption{$\vect{f}_1$}
    \end{subfigure}
    \begin{subfigure}[b]{.23\textwidth}
    \includegraphics[width=\textwidth]{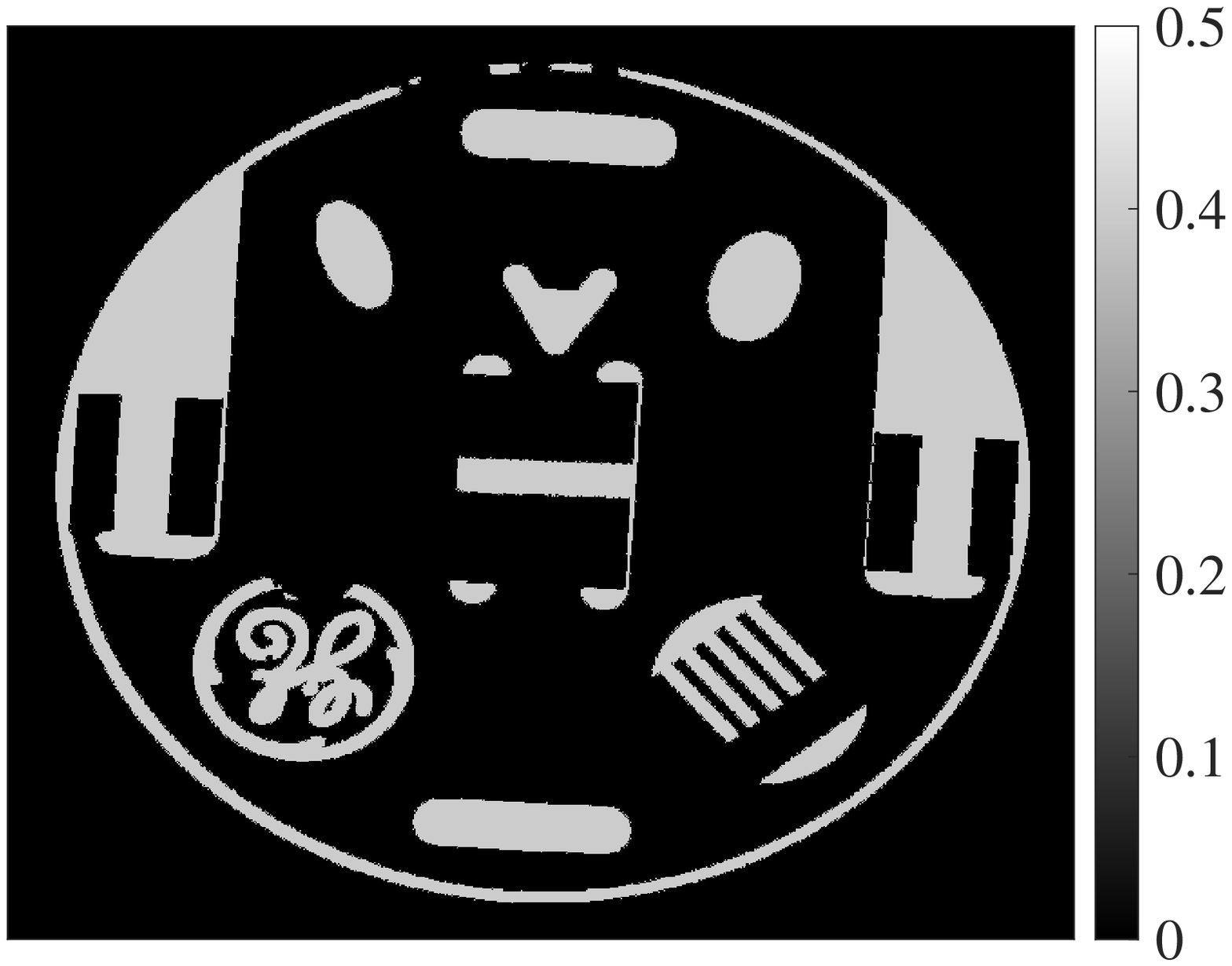}
    \caption{$\vect{f}_2$}
    \end{subfigure}
    \begin{subfigure}[b]{.23\textwidth}
    \includegraphics[width=\textwidth]{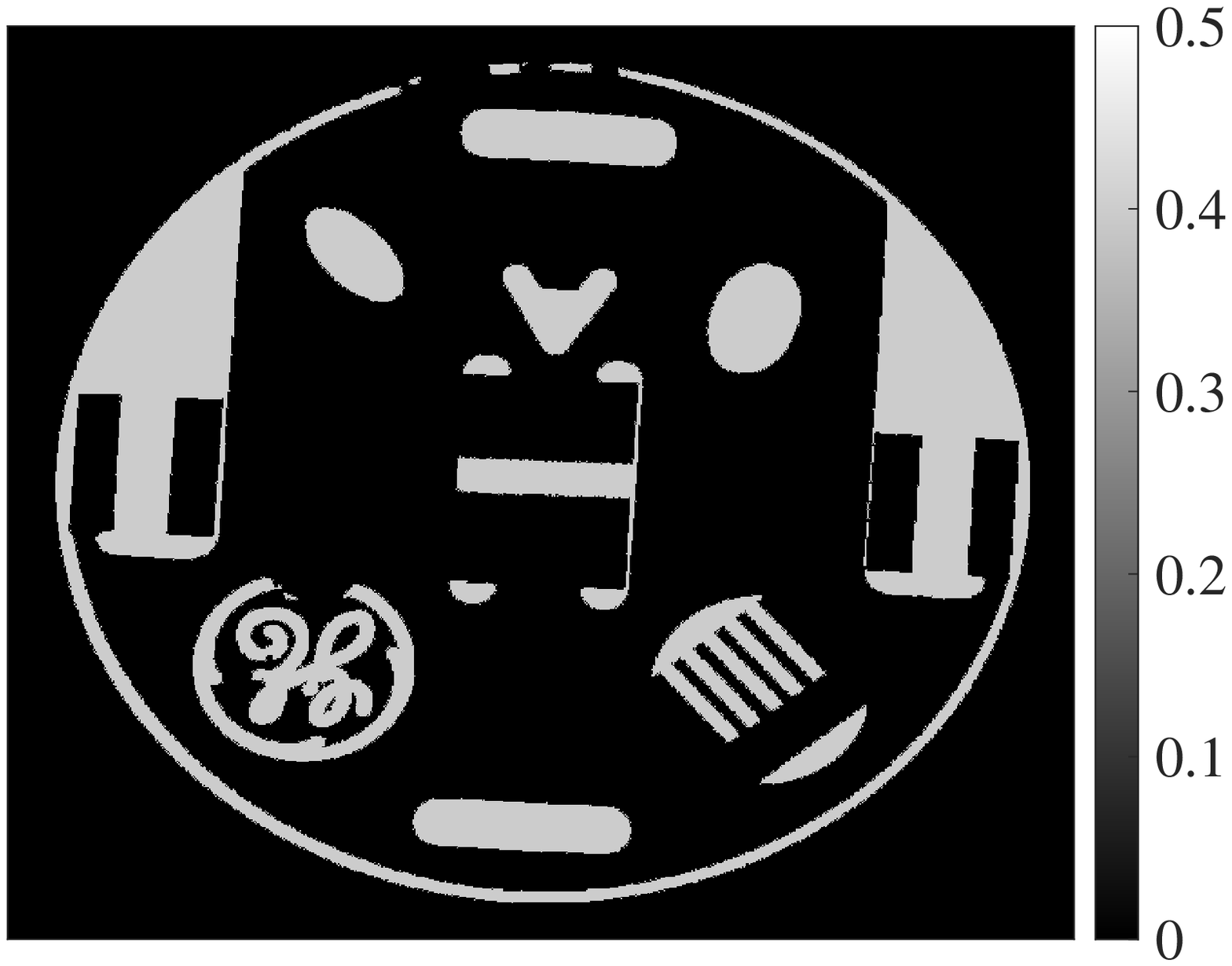}
    \caption{$\vect{f}_3$}
    \end{subfigure}
    \begin{subfigure}[b]{.23\textwidth}
    \includegraphics[width=\textwidth]{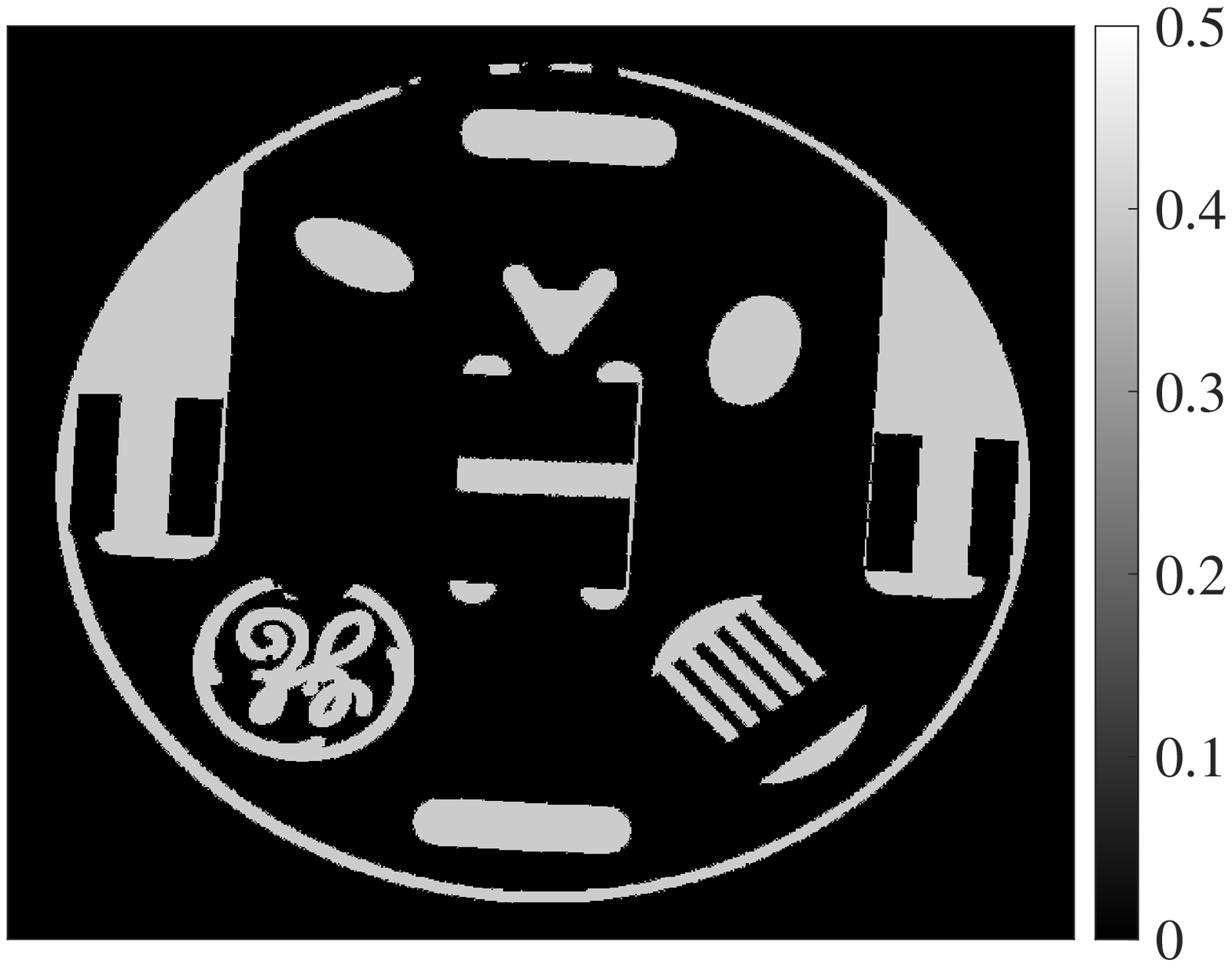}
    \caption{$\vect{f}_4$}
    \end{subfigure}
    \\
    \begin{subfigure}[b]{.23\textwidth}
    \includegraphics[width=\textwidth]{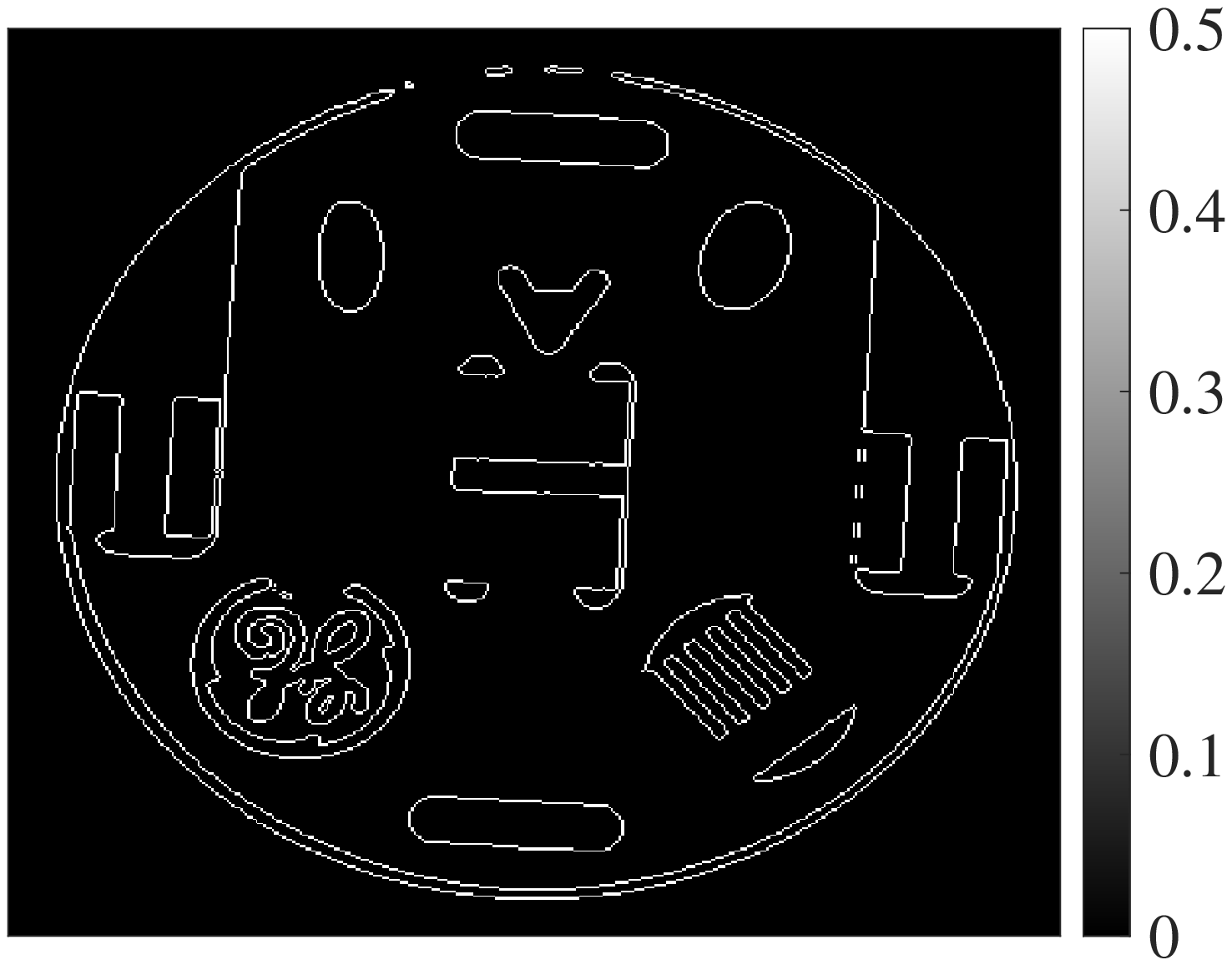}
    \caption{$\vect{x}_{1}$}
    \end{subfigure}
    \begin{subfigure}[b]{.23\textwidth}
    \includegraphics[width=\textwidth]{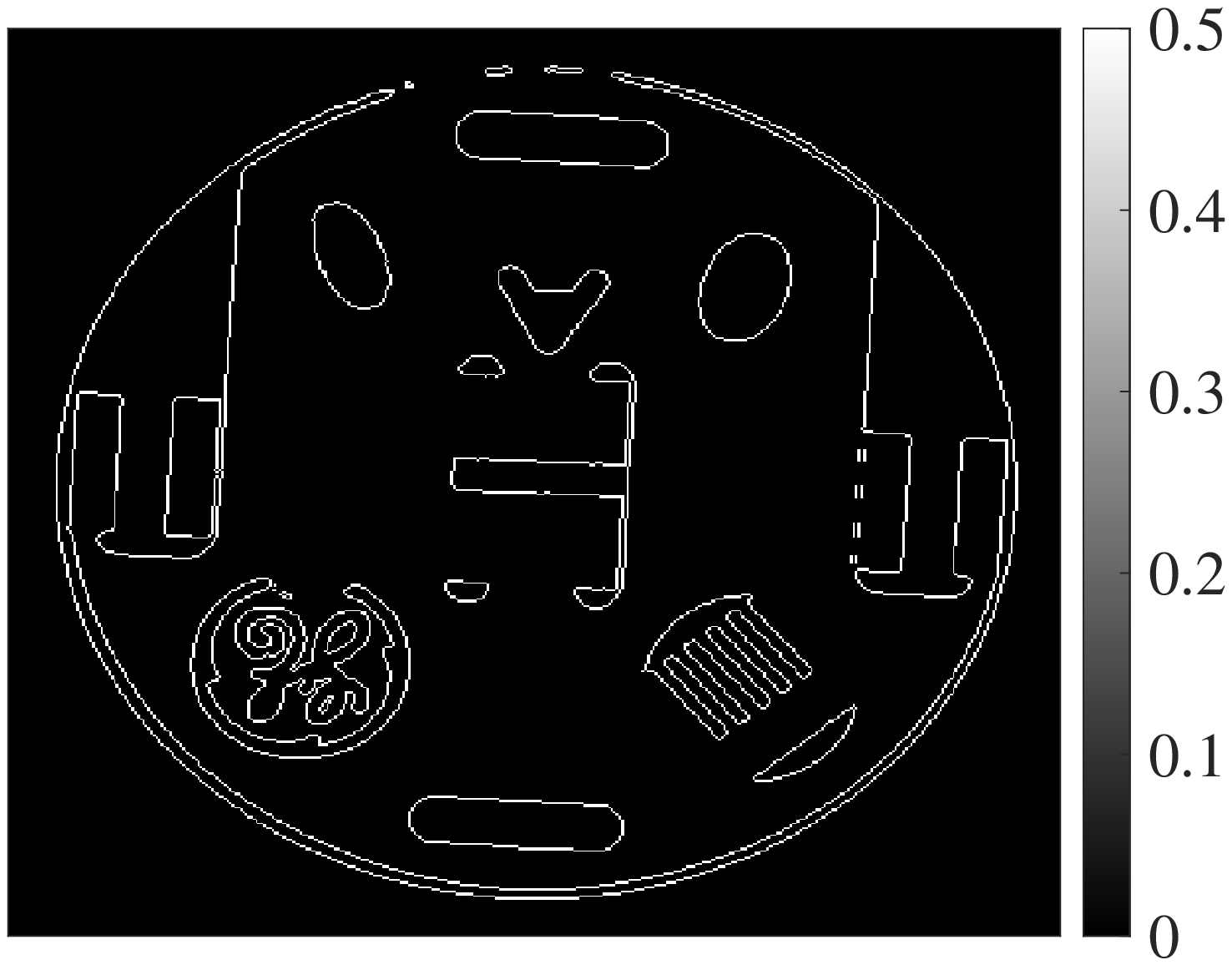}
    \caption{$\vect{x}_2$}
    \end{subfigure}
    \begin{subfigure}[b]{.23\textwidth}
    \includegraphics[width=\textwidth]{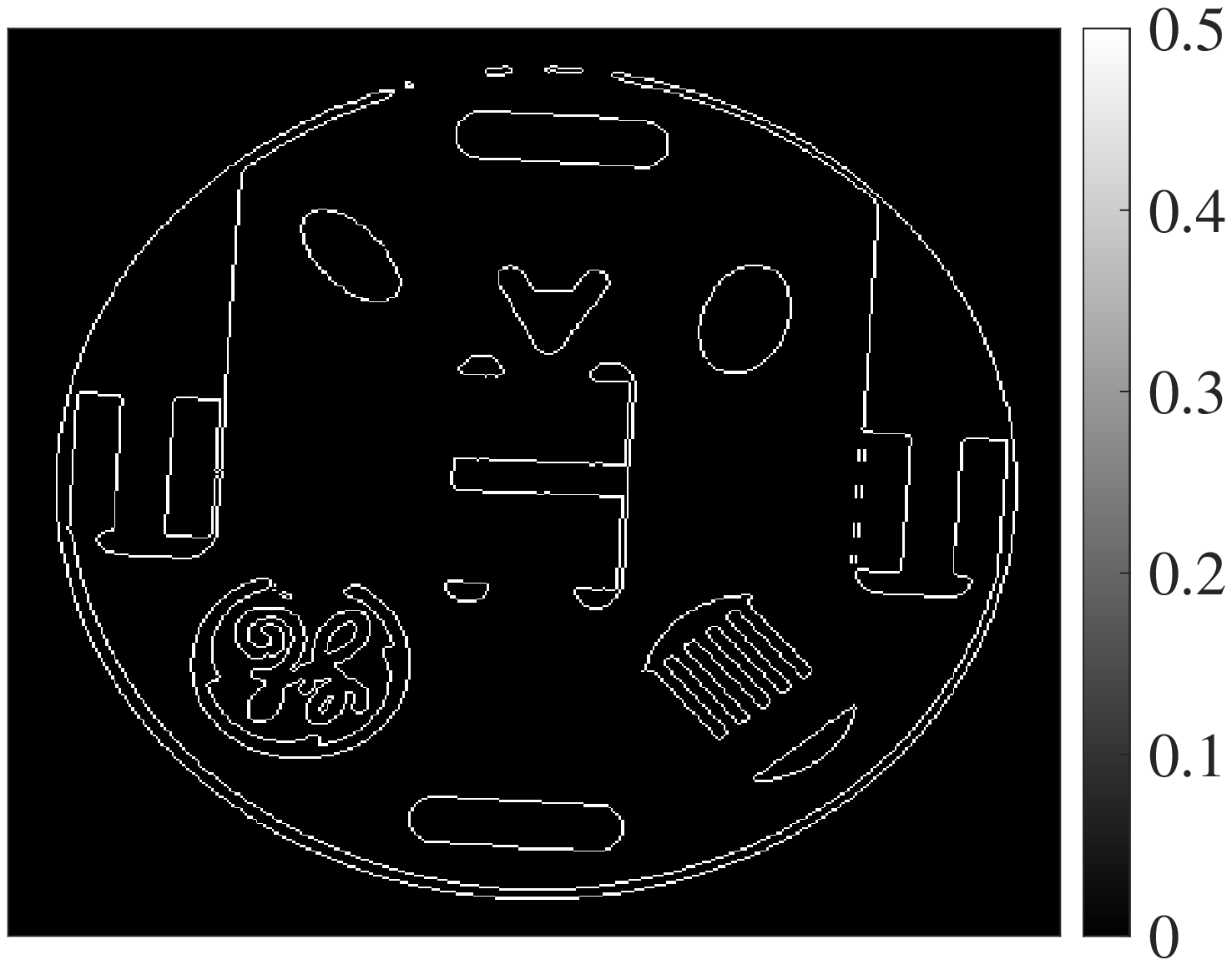}
    \caption{$\vect{x}_3$}
    \end{subfigure}
    \begin{subfigure}[b]{.23\textwidth}
    \includegraphics[width=\textwidth]{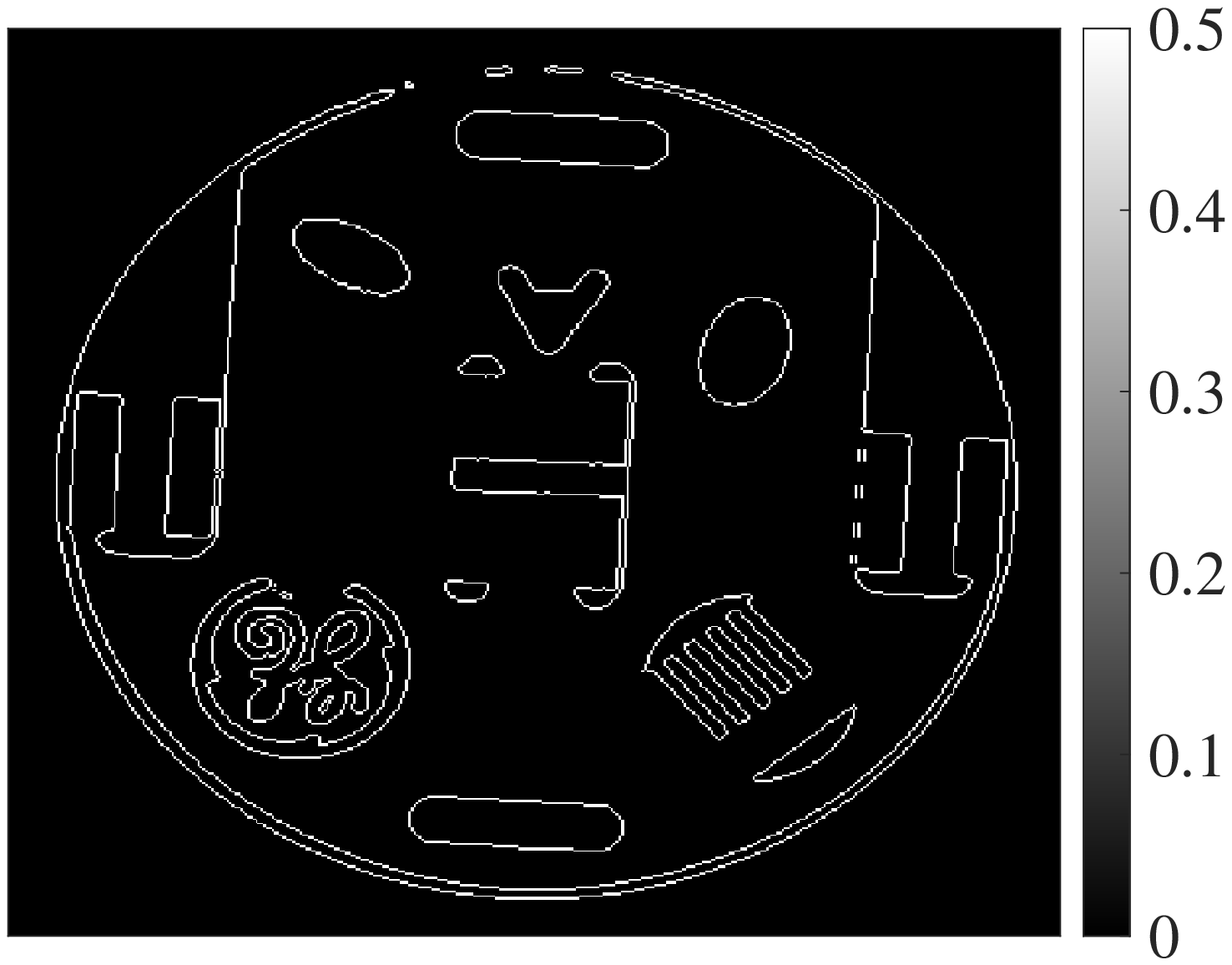}
    \caption{$\vect{x}_4$}
    \end{subfigure}
    \\
    \begin{subfigure}[b]{.23\textwidth}
    \includegraphics[width=\textwidth]{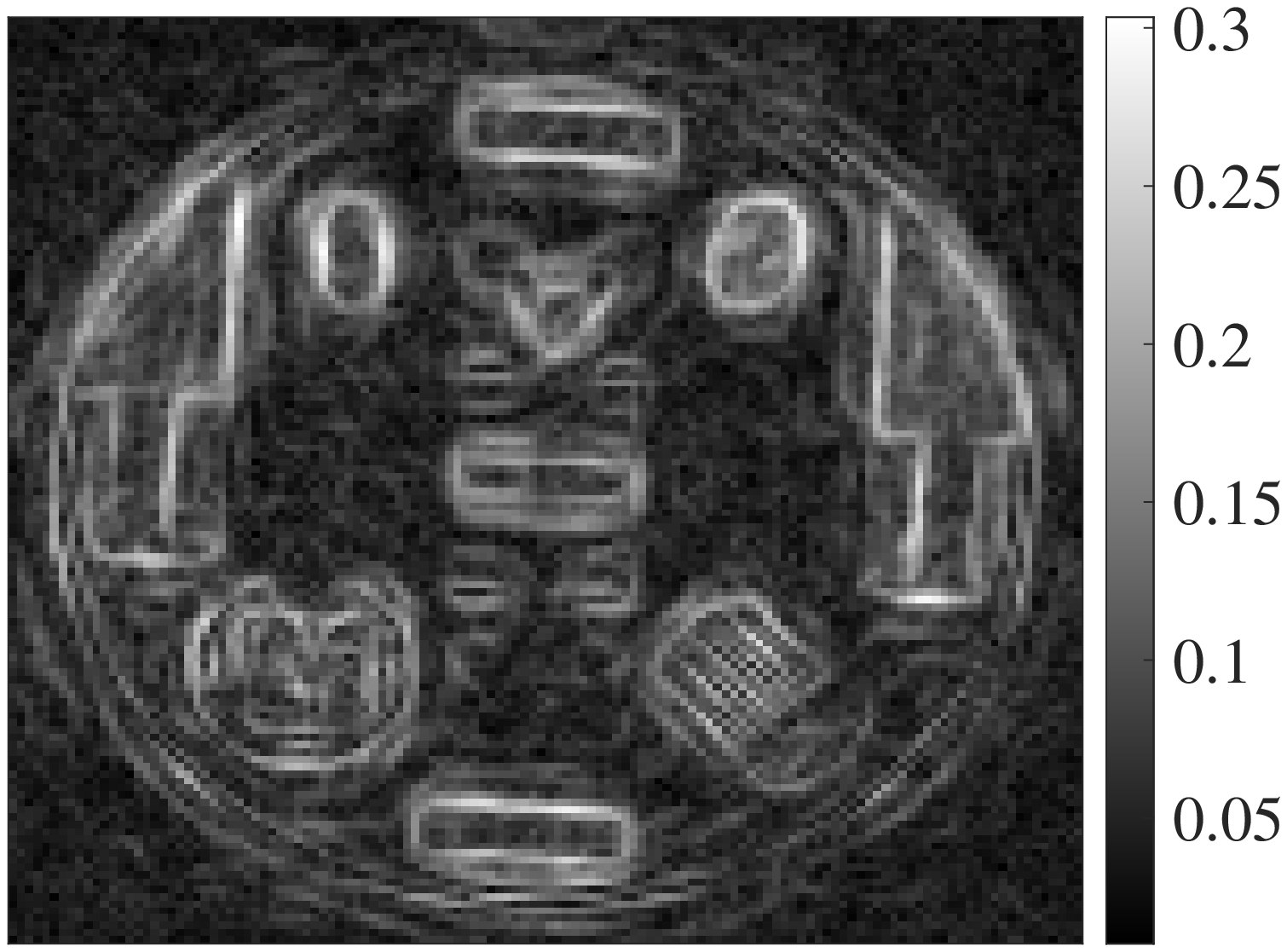}
    \caption{$\mathbf{y}_{1}$}
    \end{subfigure}
    \begin{subfigure}[b]{.23\textwidth}
    \includegraphics[width=\textwidth]{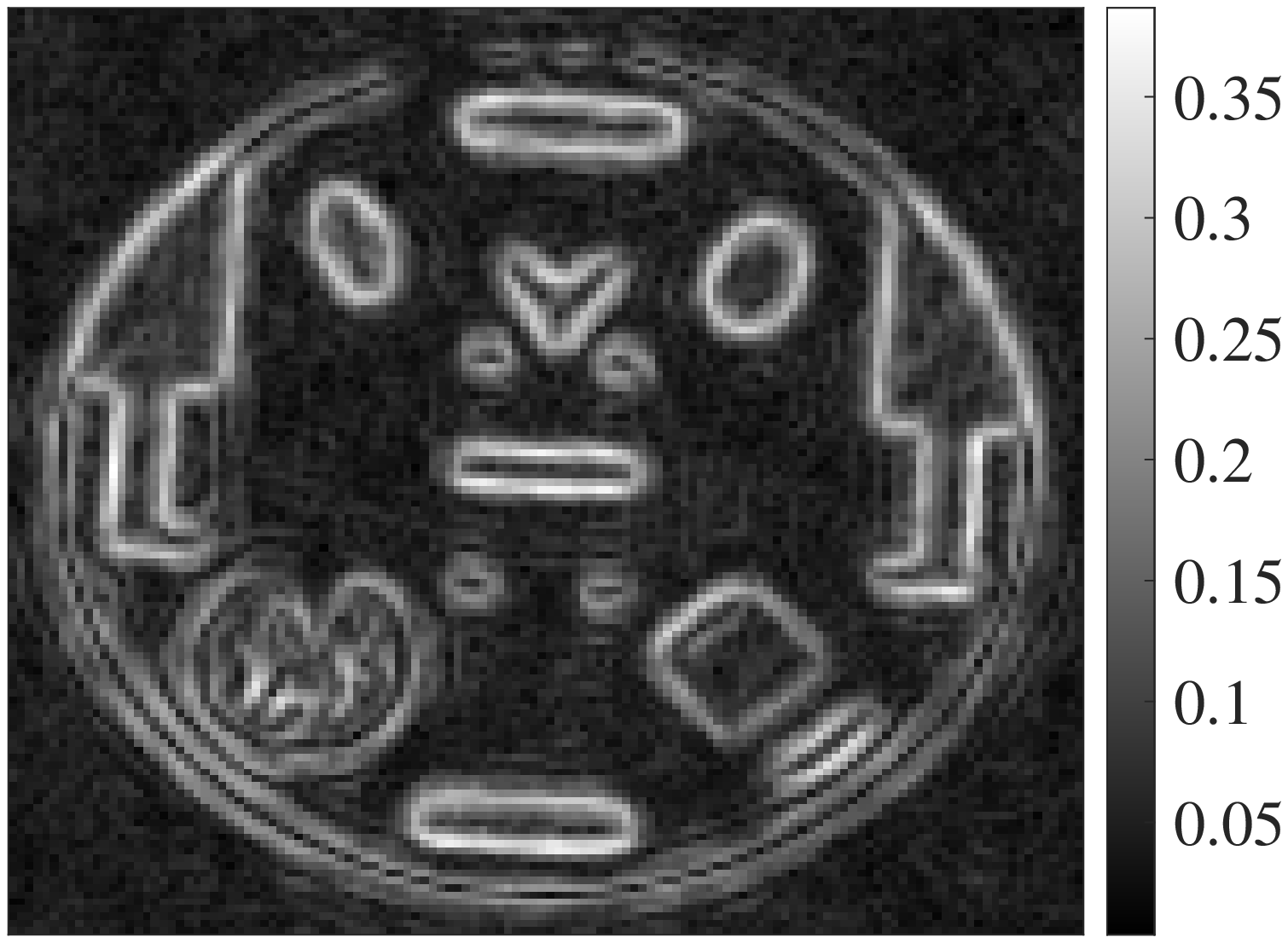}
    \caption{$\mathbf{y}_2$}
    \end{subfigure}
    \begin{subfigure}[b]{.23\textwidth}
    \includegraphics[width=\textwidth]{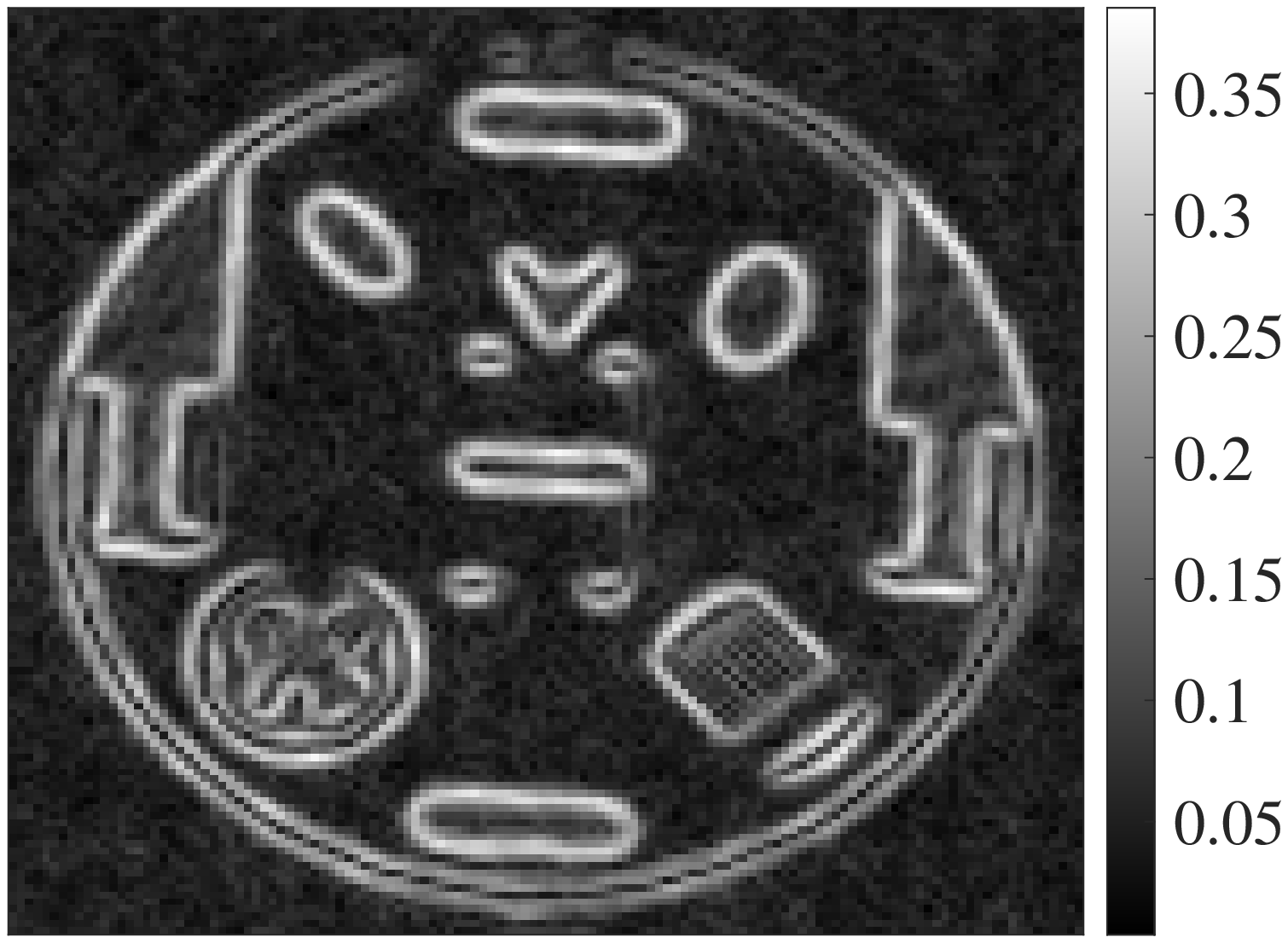}
    \caption{$\mathbf{y}_3$}
    \end{subfigure}
    \begin{subfigure}[b]{.23\textwidth}
    \includegraphics[width=\textwidth]{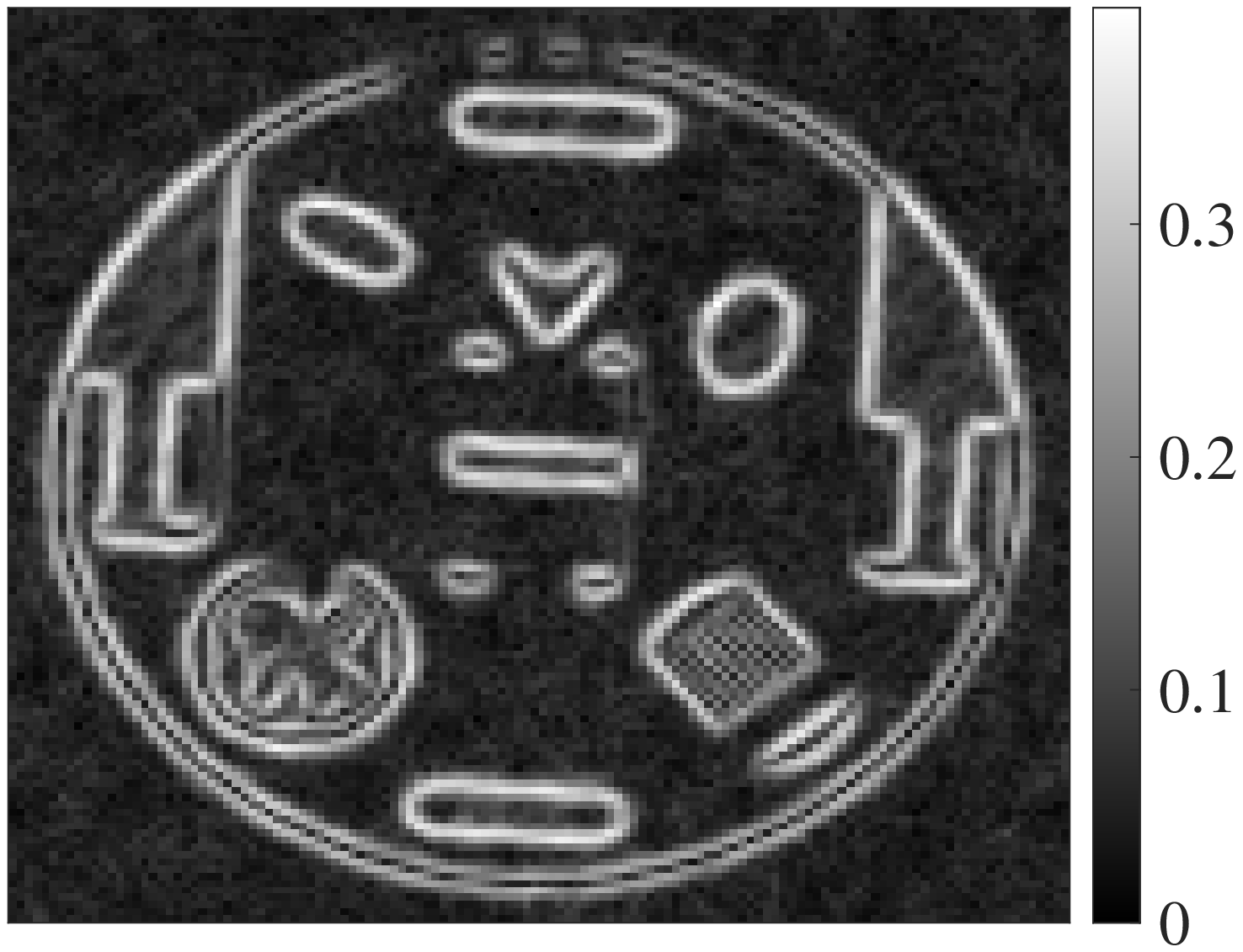}
    \caption{$\mathbf{y}_4$}
    \end{subfigure}
    
    \caption{(top row) The underlying MRI images, \cite{Lalwanietal}.  Two rotating/translating ellipses are imposed on each image.
    (middle row) The exact edge map  for each  MRI image.
    (bottom row) The edge approximations using the two-dimensional extension of the CF method \eqref{eq:1d_CF} (see e.g. \cite{xiao2022sequential}) from the under-sampled and noisy Fourier samples.  Here we use SNR $=2$ for the noise standard deviation in \eqref{eq:SNR}.} \label{fig:MRI}
\end{figure}
Figure  \ref{fig:MRI}(top) shows four sequential MRI images. Two ellipses that are rotating/translating in time  are super-imposed on each static image.   We assume we are given the corresponding Fourier data for $J = 6$ (with four shown for better visualization) sequential images,  which we simulate by taking the discrete Fourier transform of each $128\times 128$ pixelated image.  We also ``zero out'' a symmetric band $\mathcal{K}_j$, $j = 1,\dots, J$, given by
\begin{equation}
\label{eq:bandzero}
\mathcal{K}_j = [\pm(10j + 1),\pm(10(j+1))],
\end{equation}
and then add noise with {SNR $=2$}. 
The intervals for the missing data bands were somewhat arbitrarily chosen, with the idea being that in each case relevant information in each data set of the sequence  would be compromised in some way. Figure \ref{fig:MRI}(middle) shows the exact edge map at each time, while  
Figure \ref{fig:MRI}(bottom) displays the edge maps obtained using the two-dimensional expansion of the CF method in \eqref{eq:1d_CF} (see e.g.~\cite{adcock2019jointsparsity,gelb2017detecting,xiao2022sequential} for discussion on two-dimensional CF method expansion).

\begin{figure}[h!]
    \centering
    \begin{subfigure}[b]{.23\textwidth}
    \includegraphics[width=\textwidth]{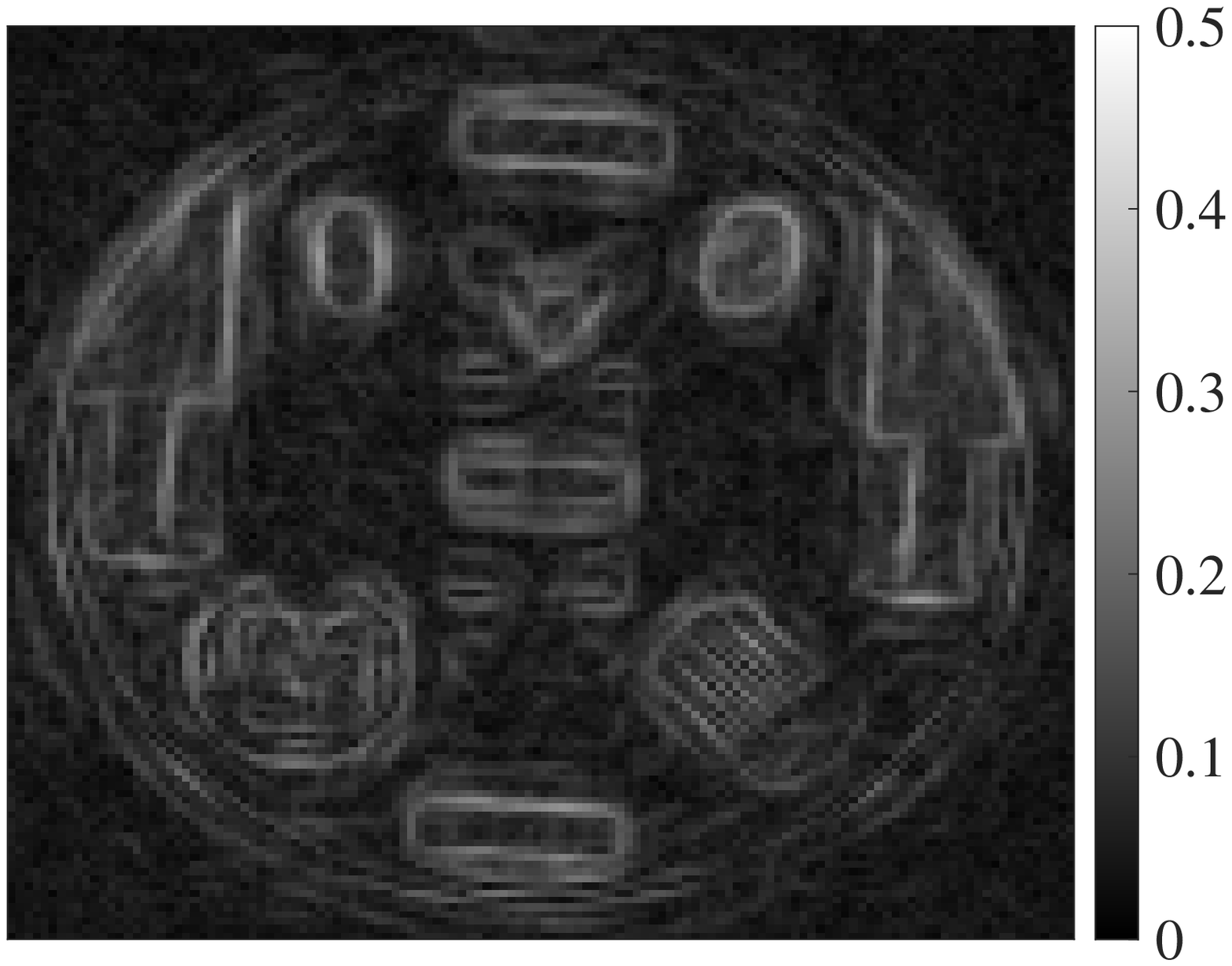}
    \caption{${\vect{x}_1}_{\JHBL}$}
    \end{subfigure}
    \begin{subfigure}[b]{.23\textwidth}
    \includegraphics[width=\textwidth]{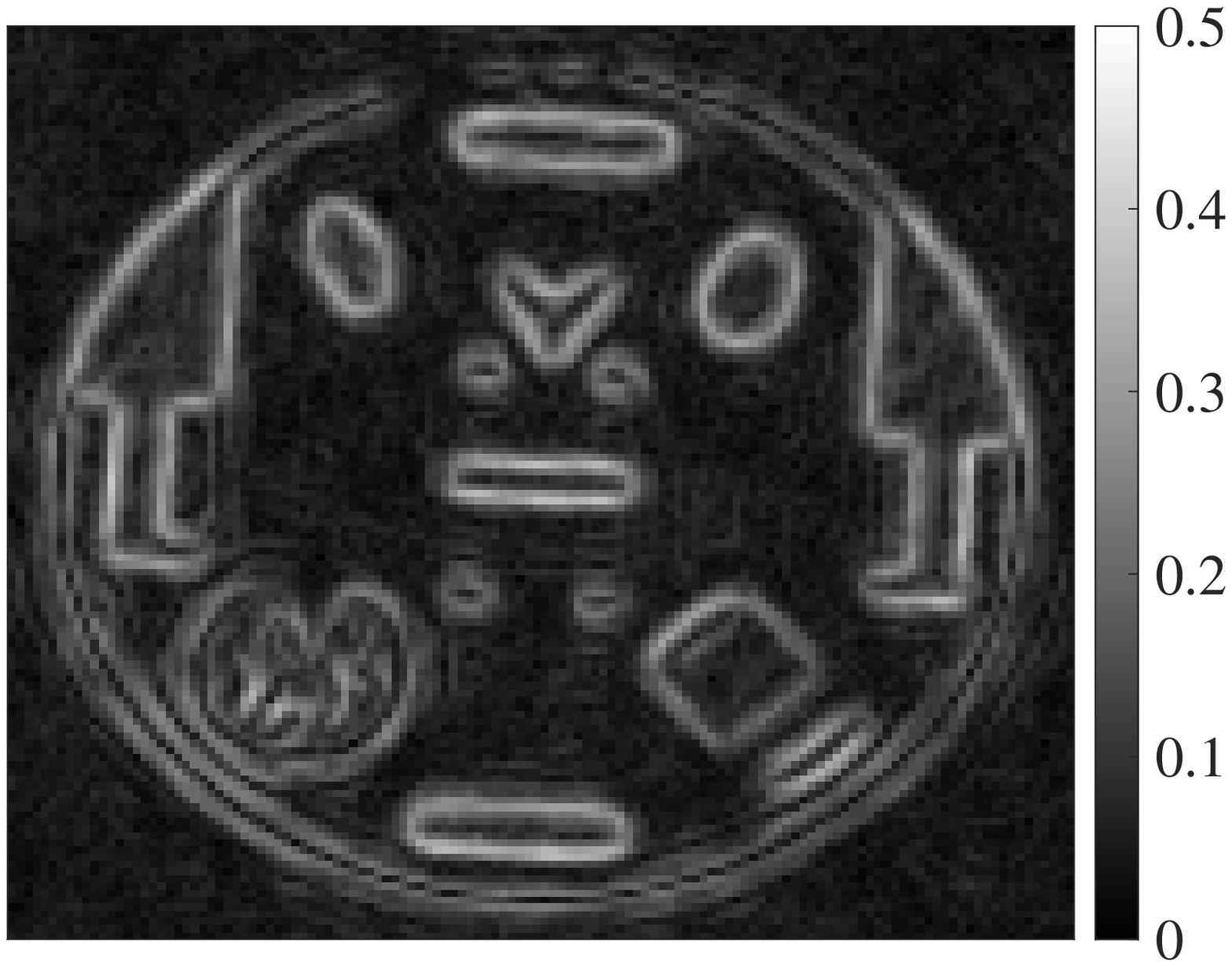}
    \caption{${\vect{x}_2}_{\JHBL}$}
    \end{subfigure}
    \begin{subfigure}[b]{.23\textwidth}
    \includegraphics[width=\textwidth]{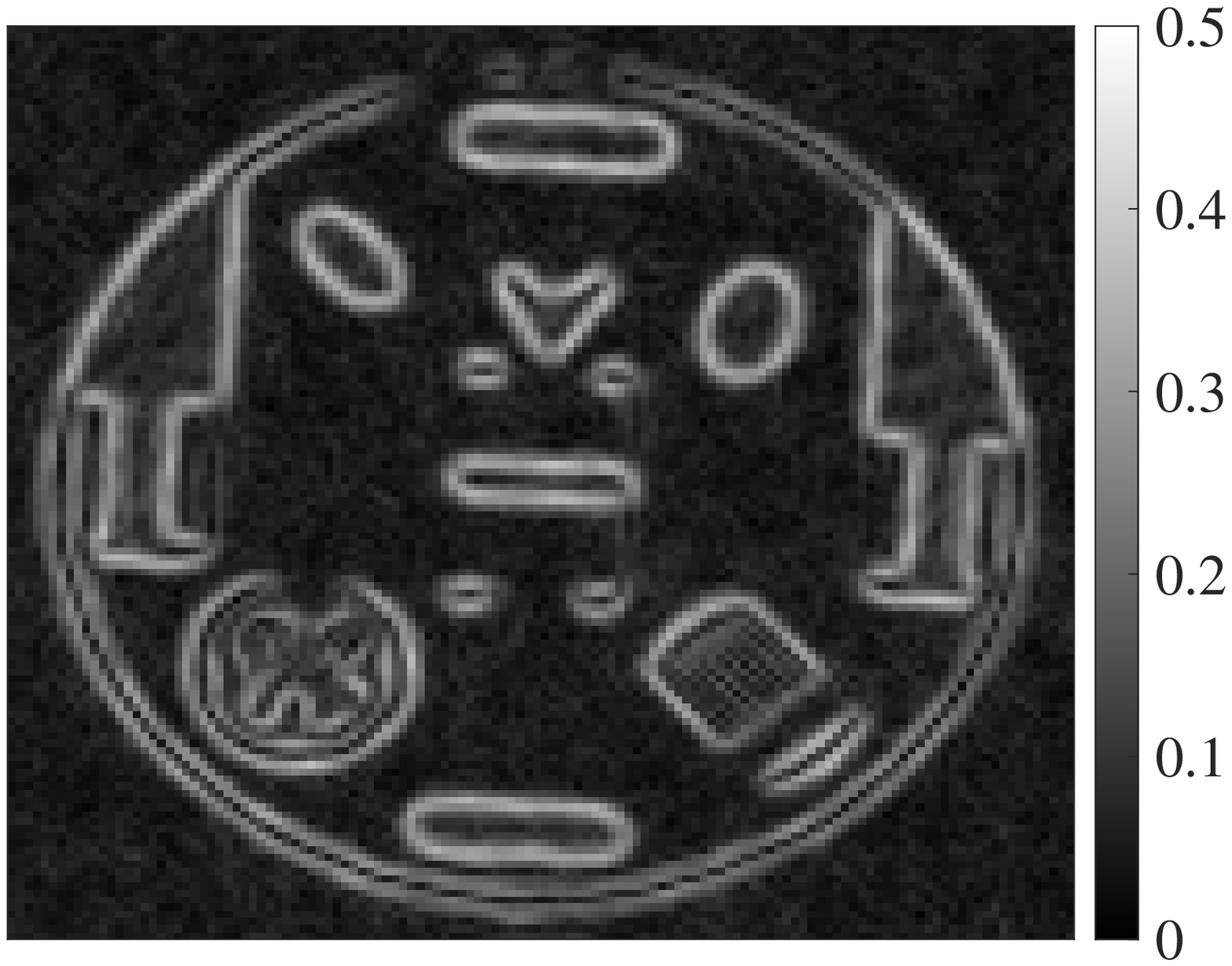}
    \caption{${\vect{x}_3}_{\JHBL}$}
    \end{subfigure}
    \begin{subfigure}[b]{.23\textwidth}
    \includegraphics[width=\textwidth]{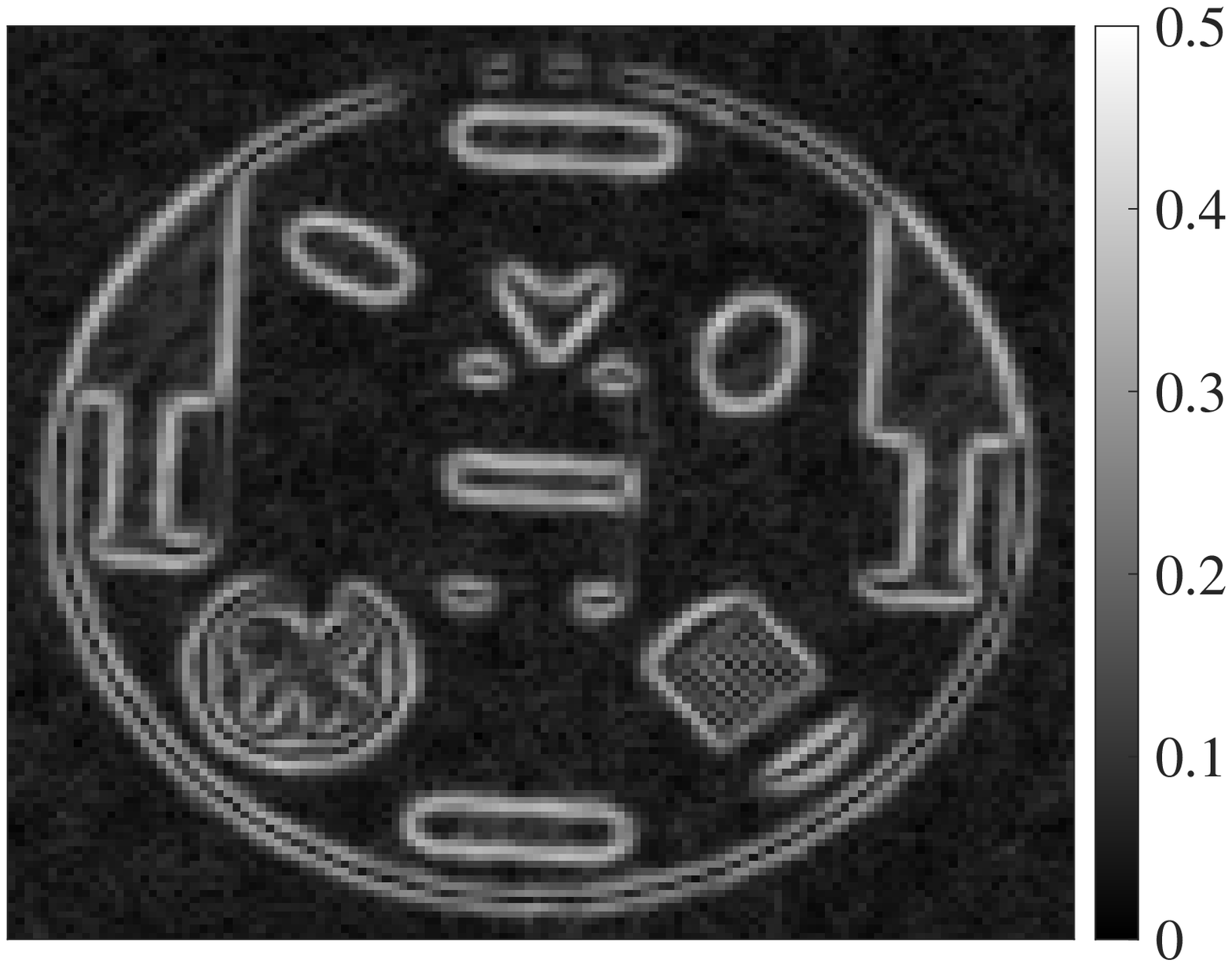}
    \caption{${\vect{x}_4}_{\JHBL}$}
    \end{subfigure}
    \\
    \begin{subfigure}[b]{.23\textwidth}
    \includegraphics[width=\textwidth]{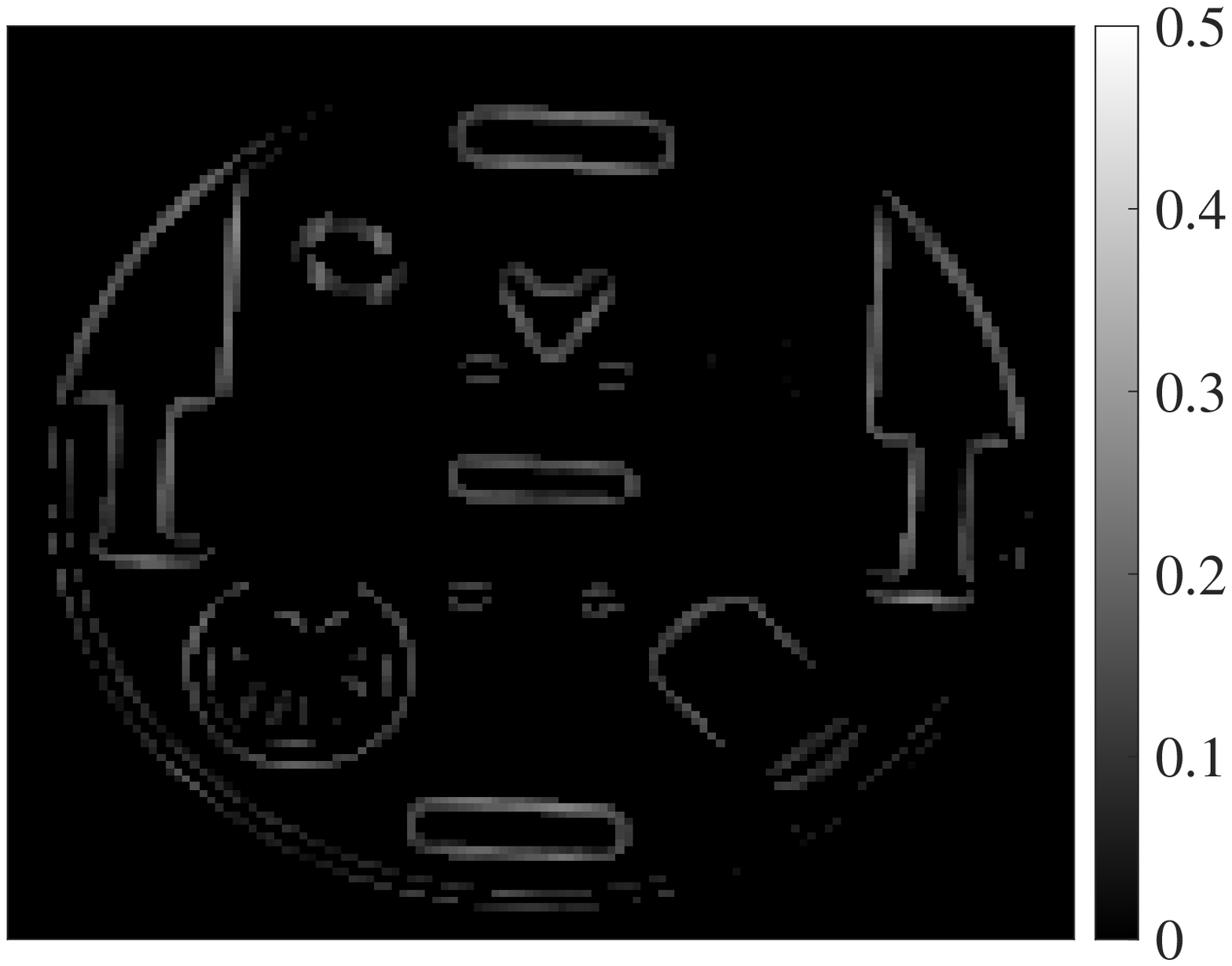}
    \caption{${\vect{x}_1}^{\beta}_{\JHBL}$}
    \end{subfigure}
    \begin{subfigure}[b]{.23\textwidth}
    \includegraphics[width=\textwidth]{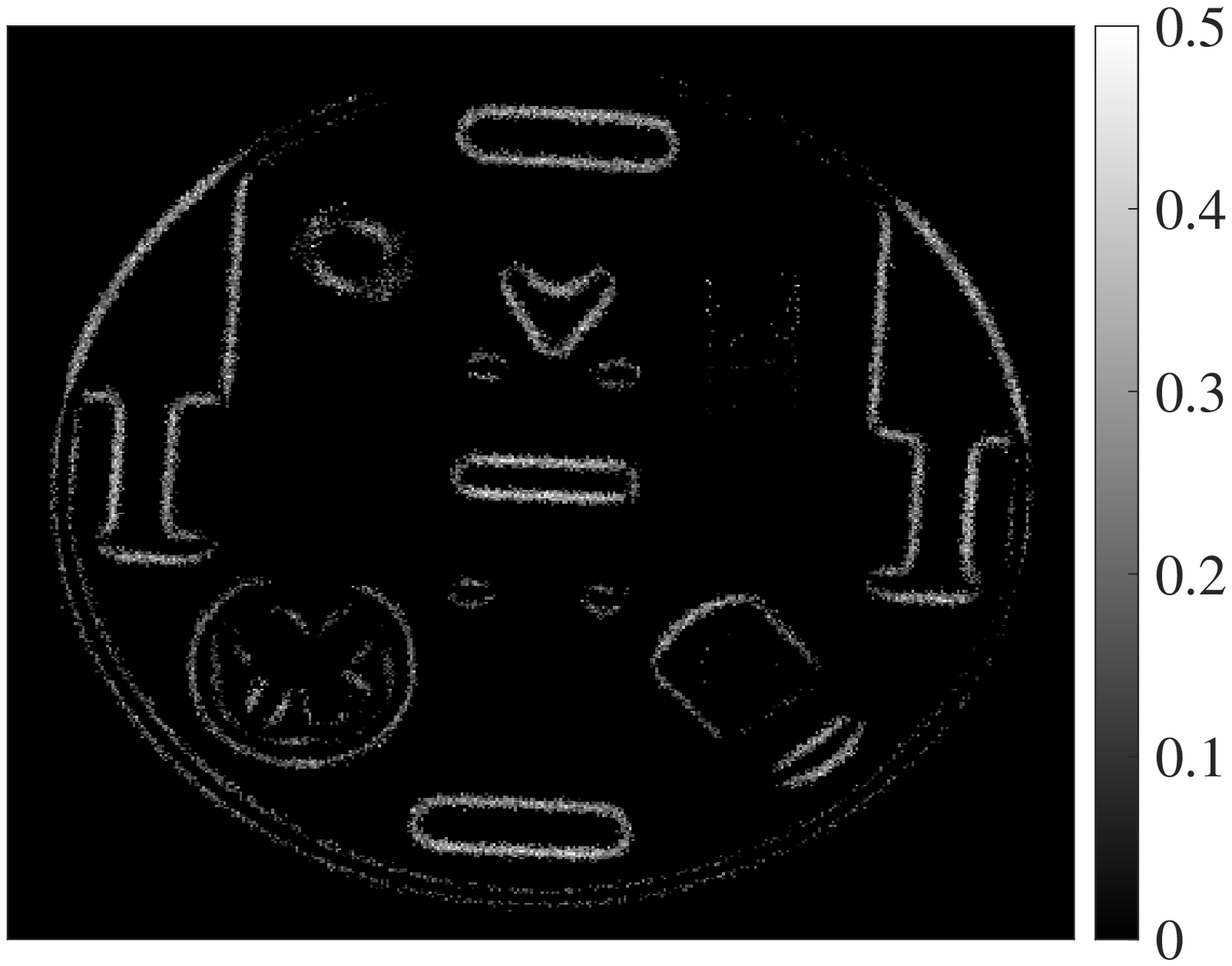}
    \caption{${\vect{x}_2}^{\beta}_{\JHBL}$}
    \end{subfigure}
    \begin{subfigure}[b]{.23\textwidth}
    \includegraphics[width=\textwidth]{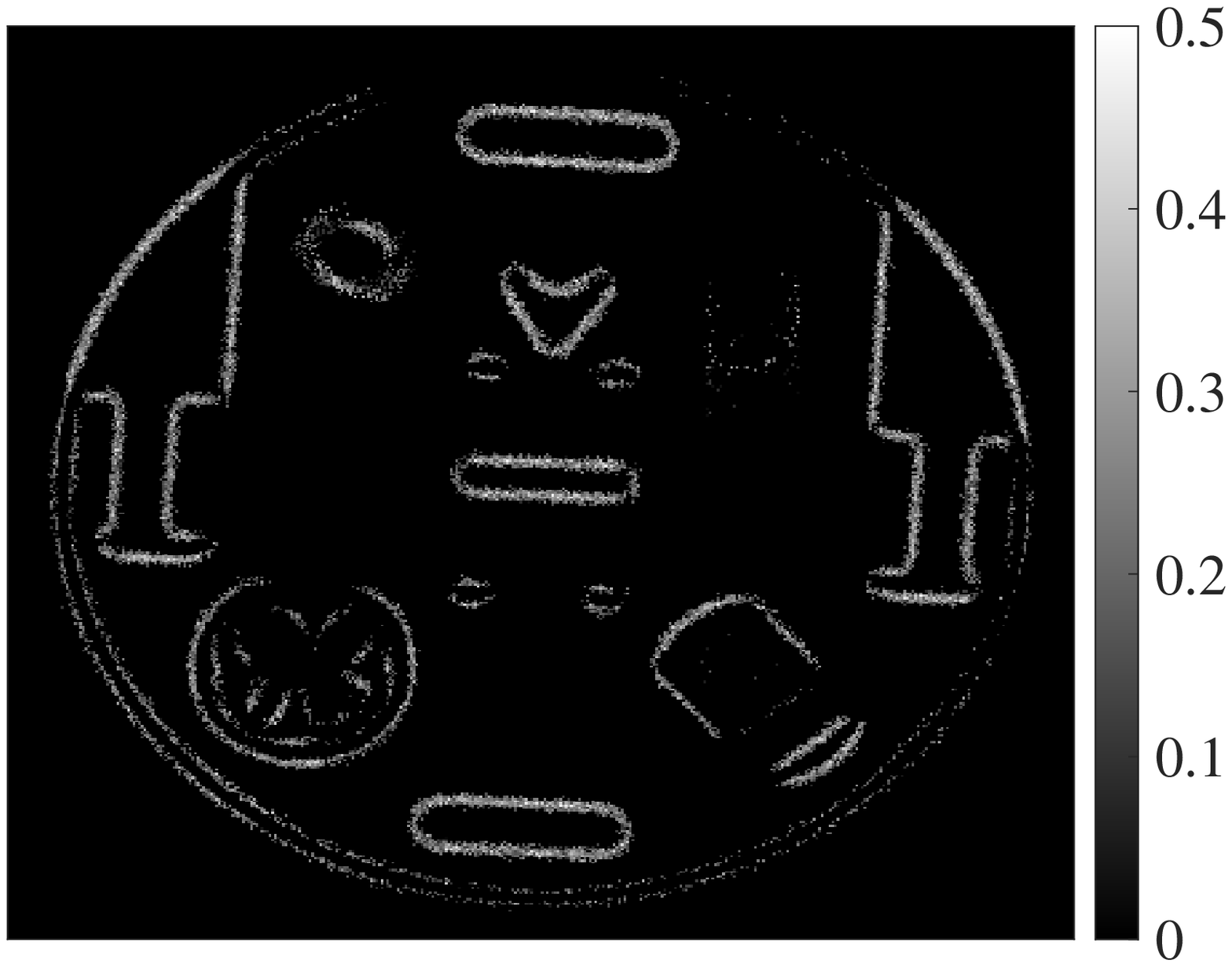}
    \caption{${\vect{x}_3}^{\beta}_{\JHBL}$}
    \end{subfigure}
    \begin{subfigure}[b]{.23\textwidth}
    \includegraphics[width=\textwidth]{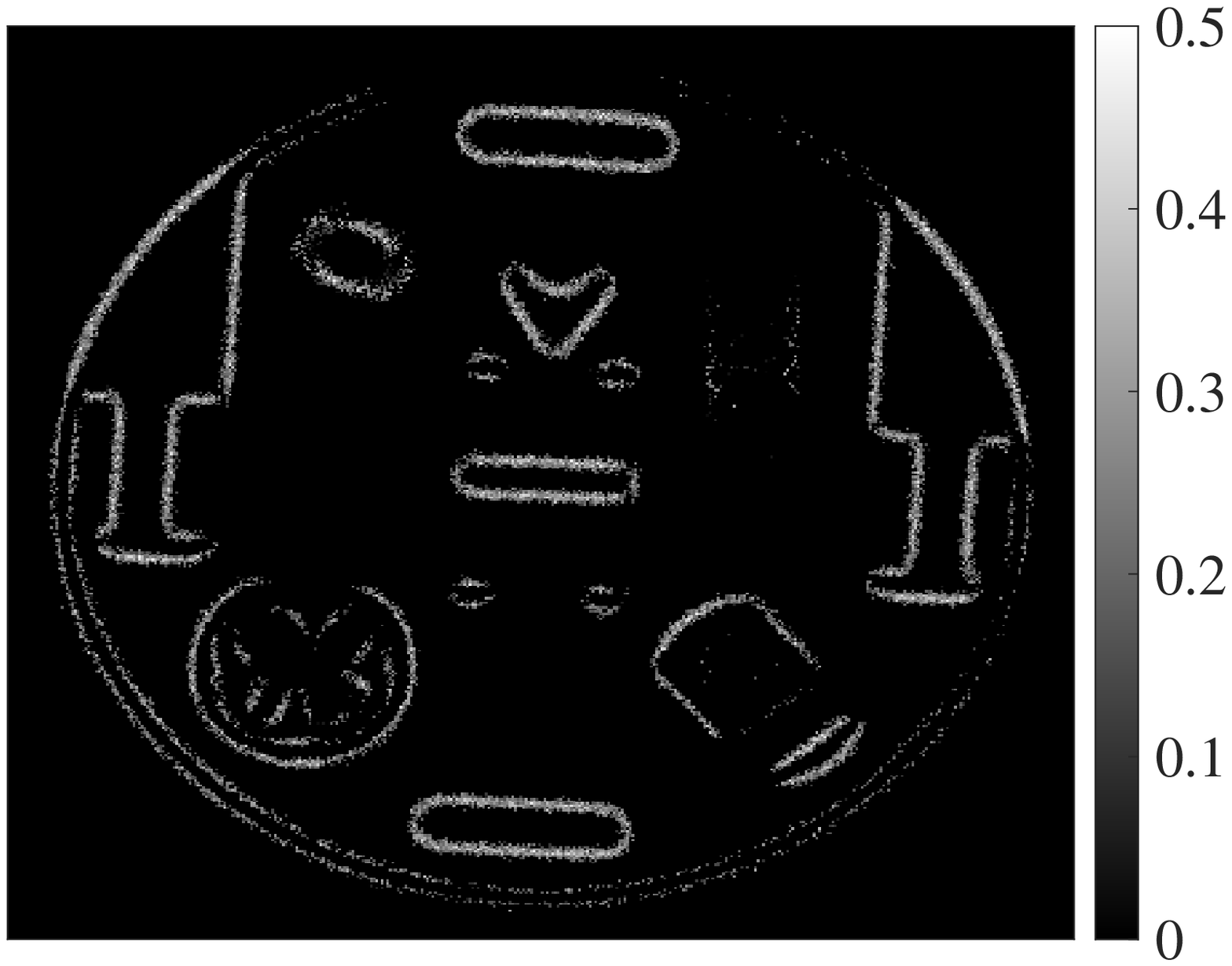}
    \caption{${\vect{x}_4}^{\beta}_{\JHBL}$}
    \end{subfigure}
    \\
    \begin{subfigure}[b]{.23\textwidth}
    \includegraphics[width=\textwidth]{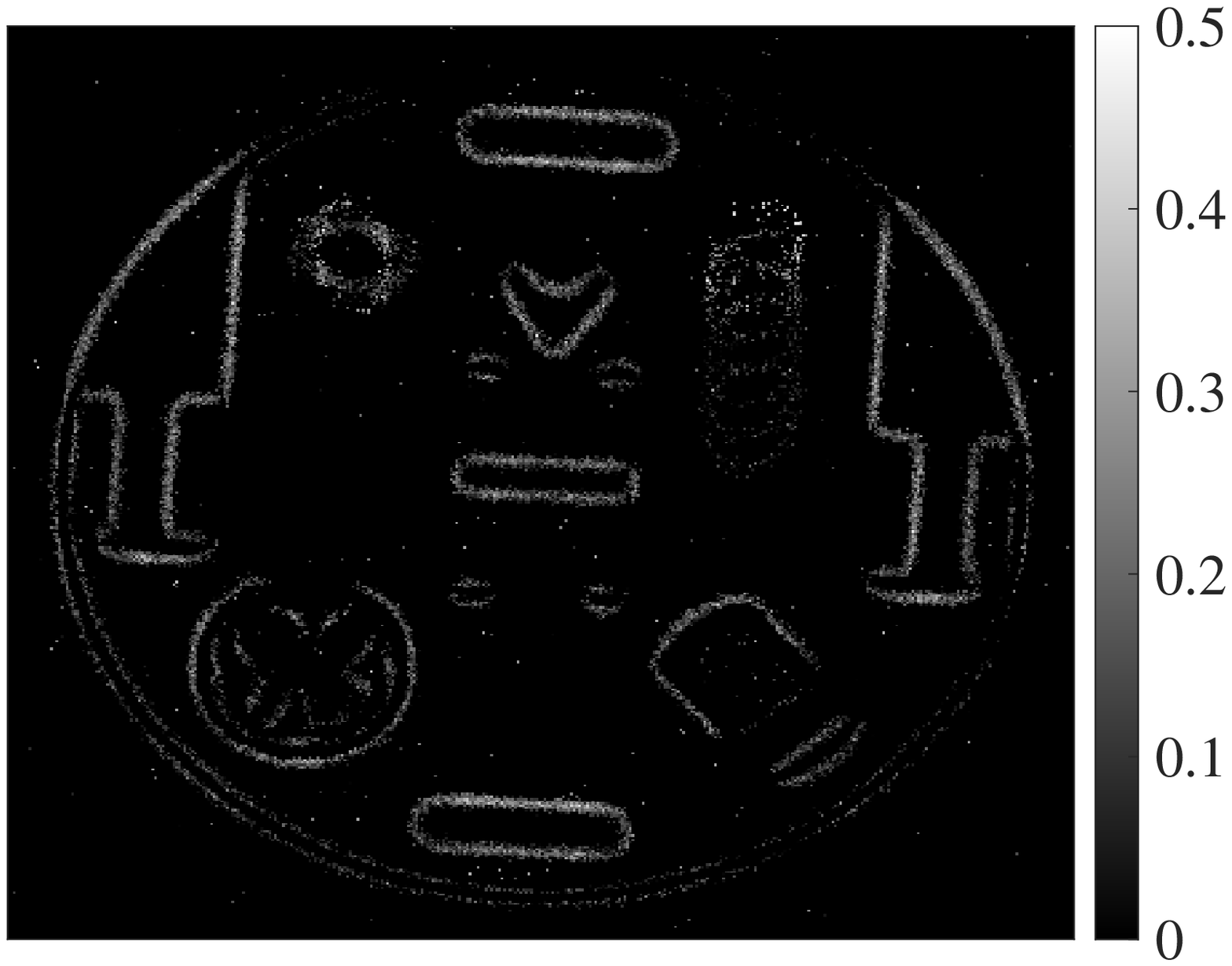}
    \caption{${\vect{x}_1}^{\beta,\boldsymbol q}_{\JHBL}$}
    \end{subfigure}
    \begin{subfigure}[b]{.23\textwidth}
    \includegraphics[width=\textwidth]{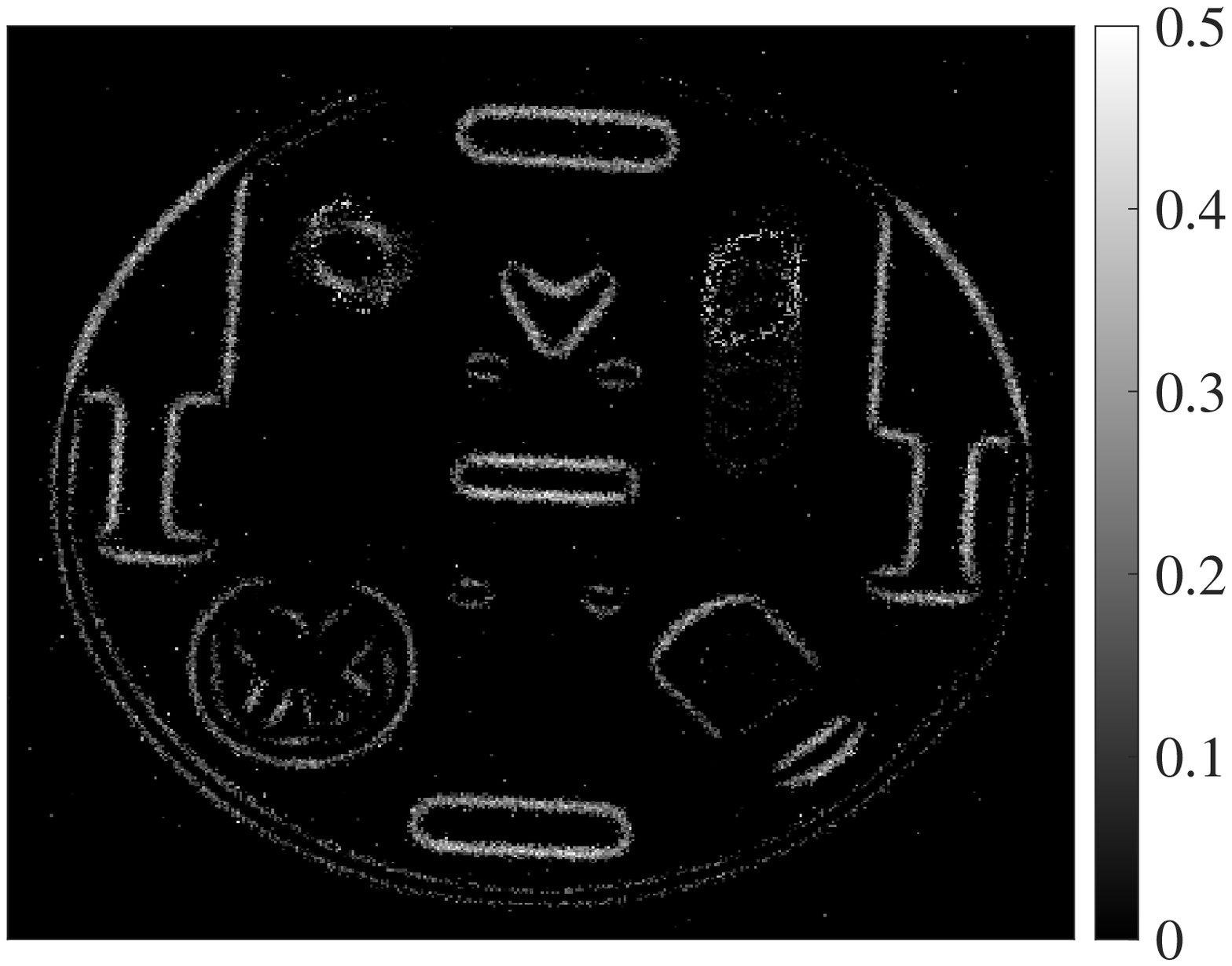}
    \caption{${\vect{x}_2}^{\beta,\boldsymbol q}_{\JHBL}$}
    \end{subfigure}
    \begin{subfigure}[b]{.23\textwidth}
    \includegraphics[width=\textwidth]{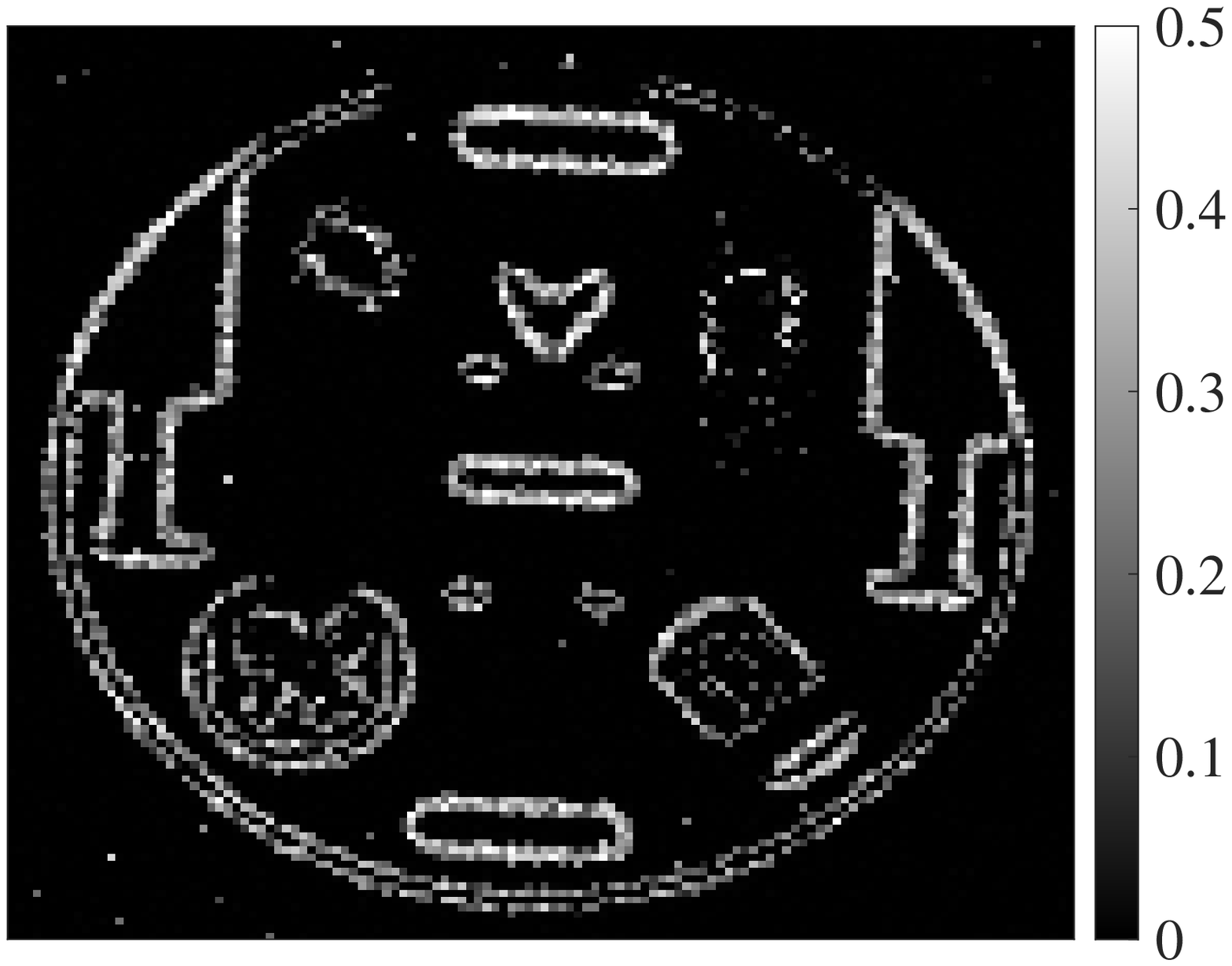}
    \caption{${\vect{x}_3}^{\beta,\boldsymbol q}_{\JHBL}$}
    \end{subfigure}
    \begin{subfigure}[b]{.23\textwidth}
    \includegraphics[width=\textwidth]{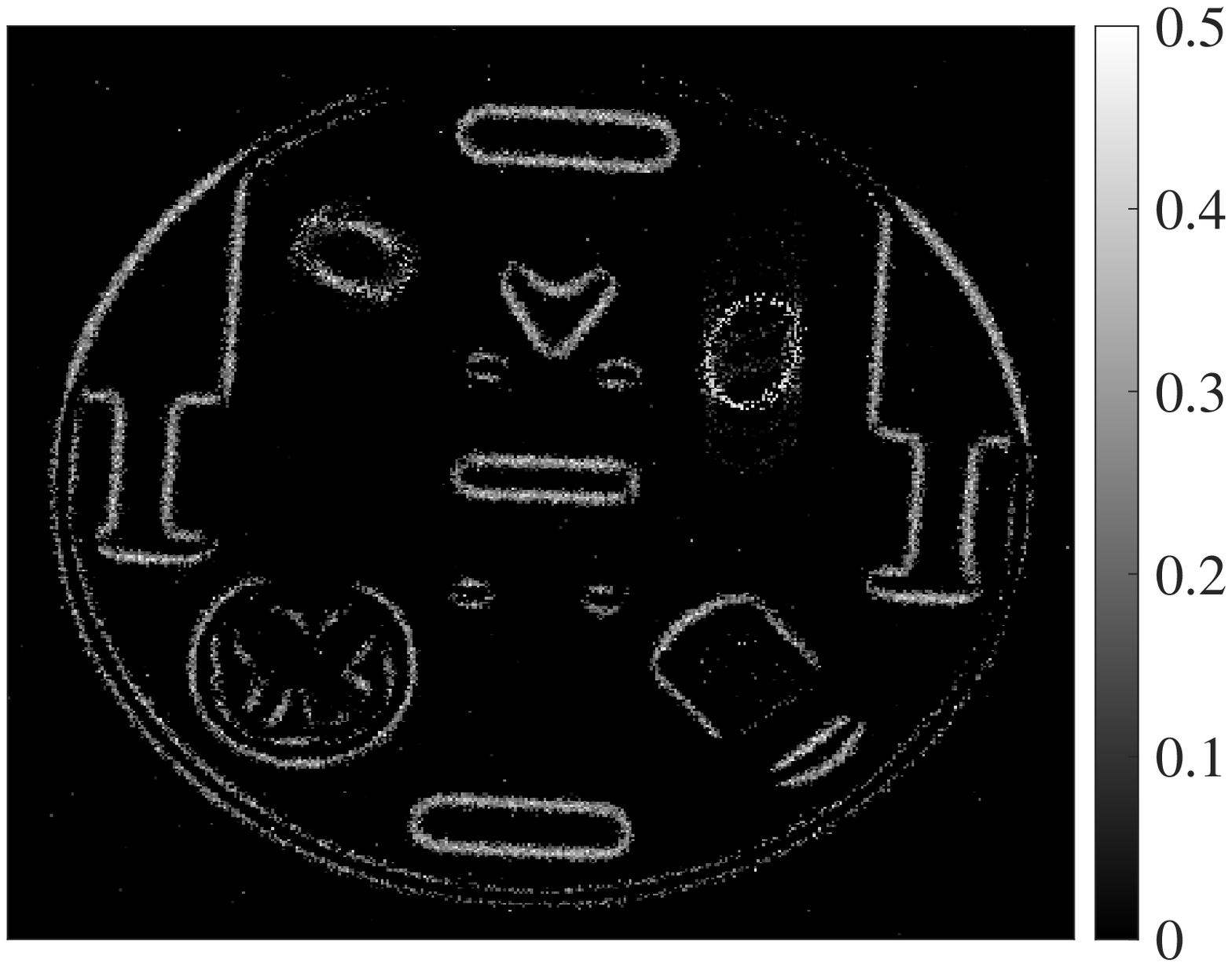}
    \caption{${\vect{x}_4}^{\beta,\boldsymbol q}_{\JHBL}$}
    \end{subfigure}
    
    \caption{(top row) Edge map recovery using Algorithm \ref{alg:JHBL-fixed-beta}.
    (middle row) Edge map recovery using Algorithm \ref{alg:JHBL}.
    (bottom row) Edge map recovery using Algorithm \ref{alg:JHBL-refine-2D}.}
    \label{fig:edge_SBL_2d}
\end{figure}

\begin{figure}[h!]
    \centering
    \begin{subfigure}[b]{.32\textwidth}
    \includegraphics[width=\textwidth]{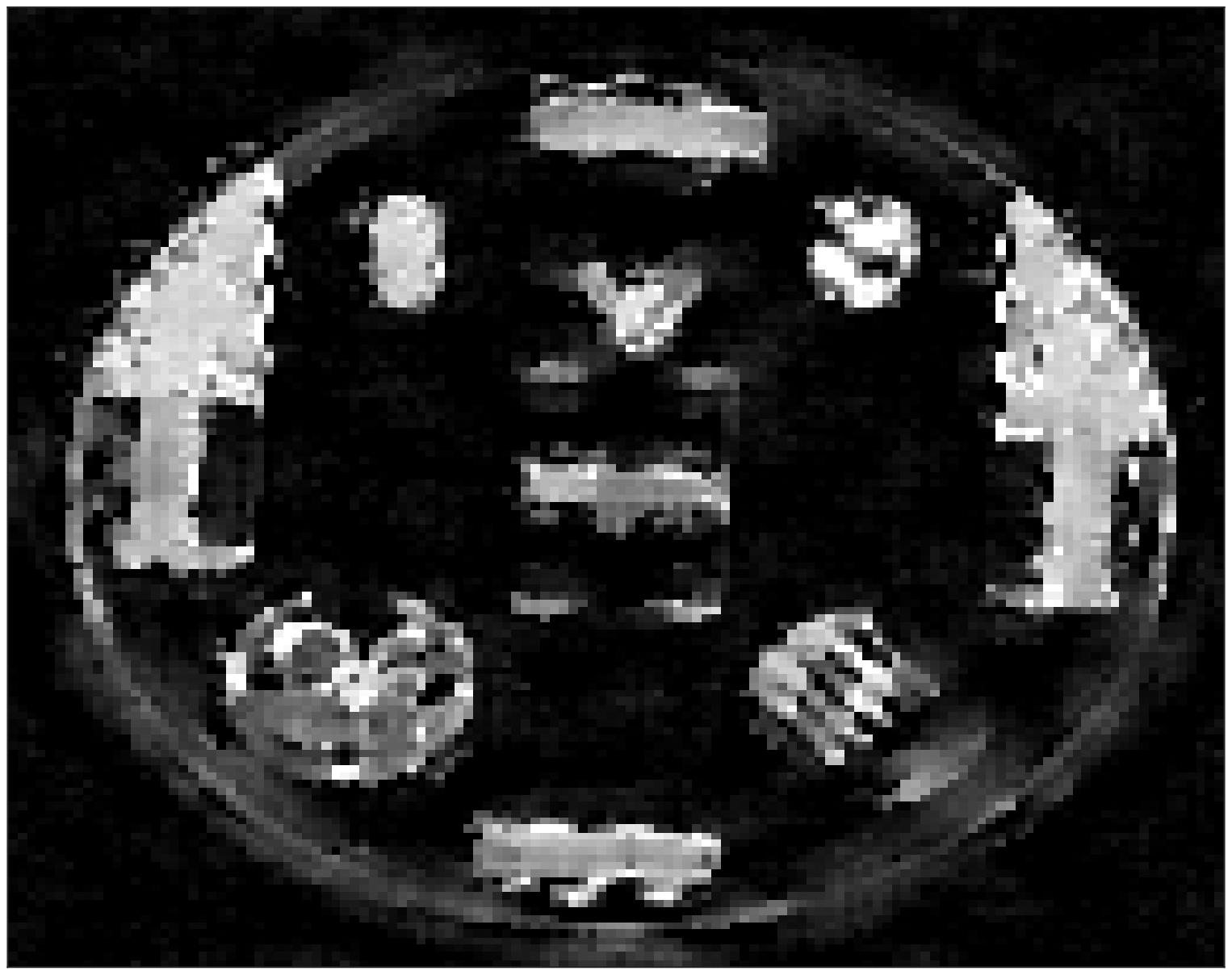}
    \caption{${\vect{x}_1}_{\JHBL}$}
    \end{subfigure}
    \begin{subfigure}[b]{.32\textwidth}
    \includegraphics[width=\textwidth]{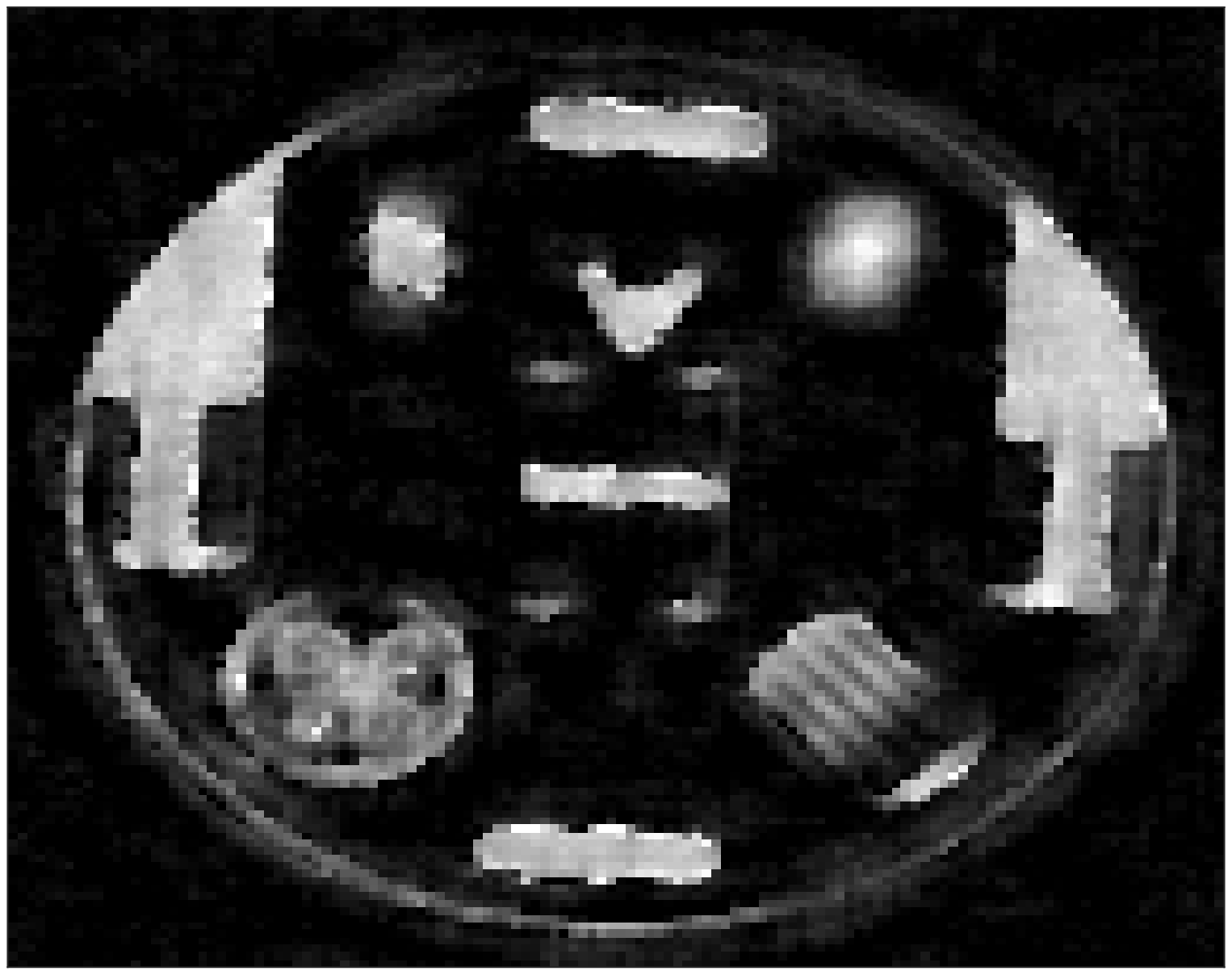}
    \caption{${\vect{x}_1}^{\beta}_{\JHBL}$}
    \end{subfigure}
    \begin{subfigure}[b]{.32\textwidth}
    \includegraphics[width=\textwidth]{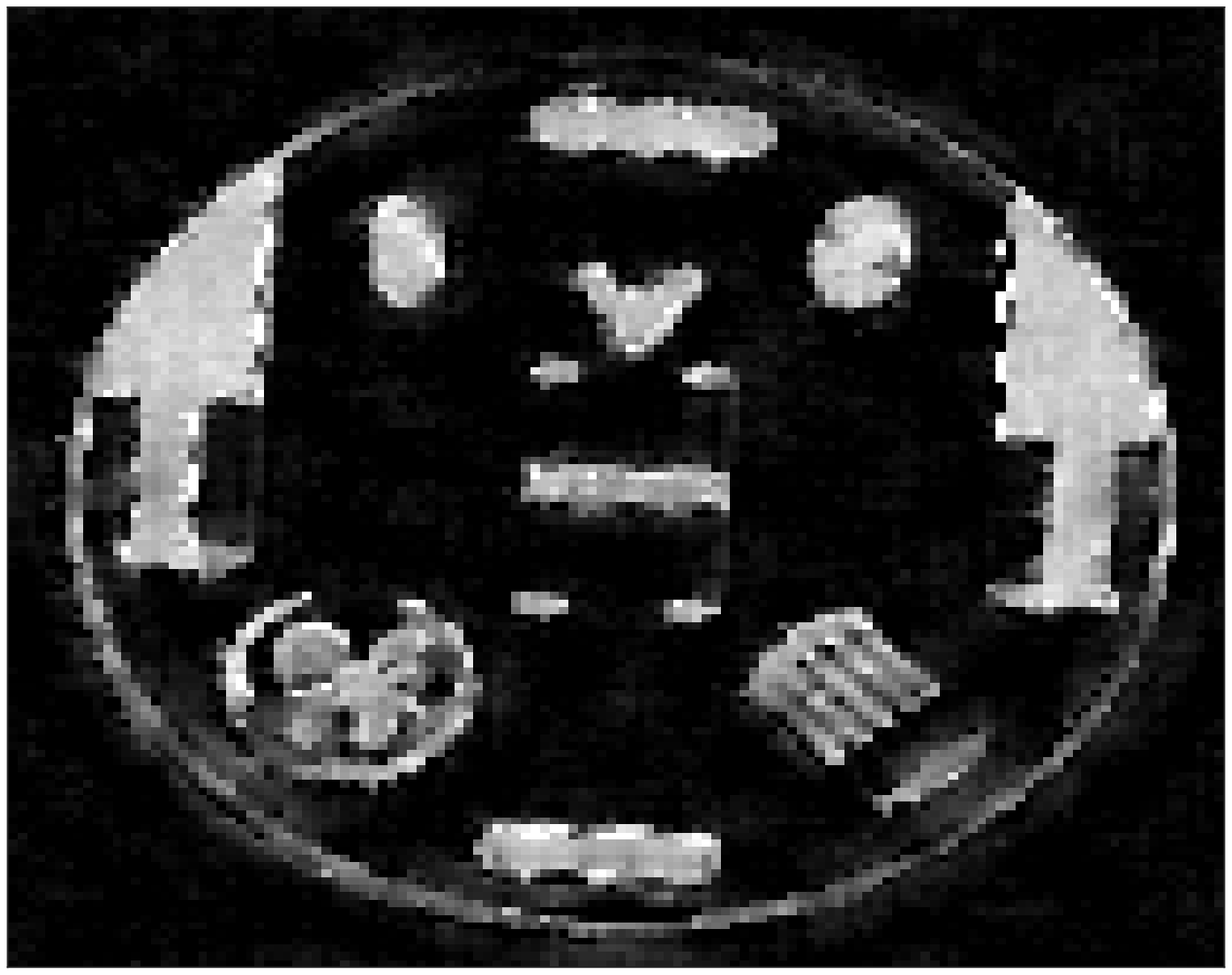}
    \caption{${\vect{x}_1}^{\beta,\boldsymbol q}_{\JHBL}$}
    \end{subfigure}
    \\
    \begin{subfigure}[b]{.325\textwidth}
    \includegraphics[width=\textwidth]{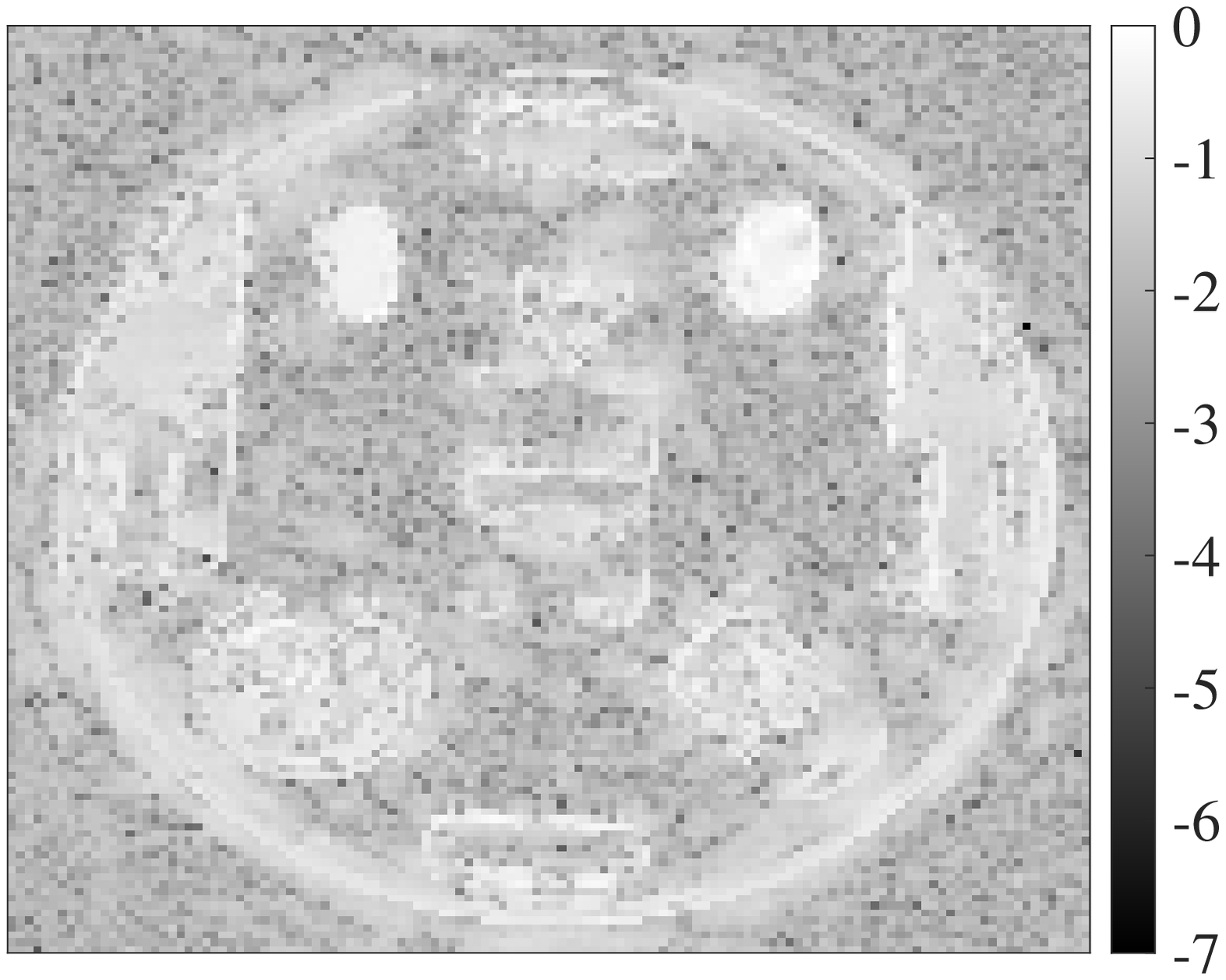}
    \caption{${\vect{x}_1}_{\JHBL}$}
    \end{subfigure}
    \begin{subfigure}[b]{.325\textwidth}
    \includegraphics[width=\textwidth]{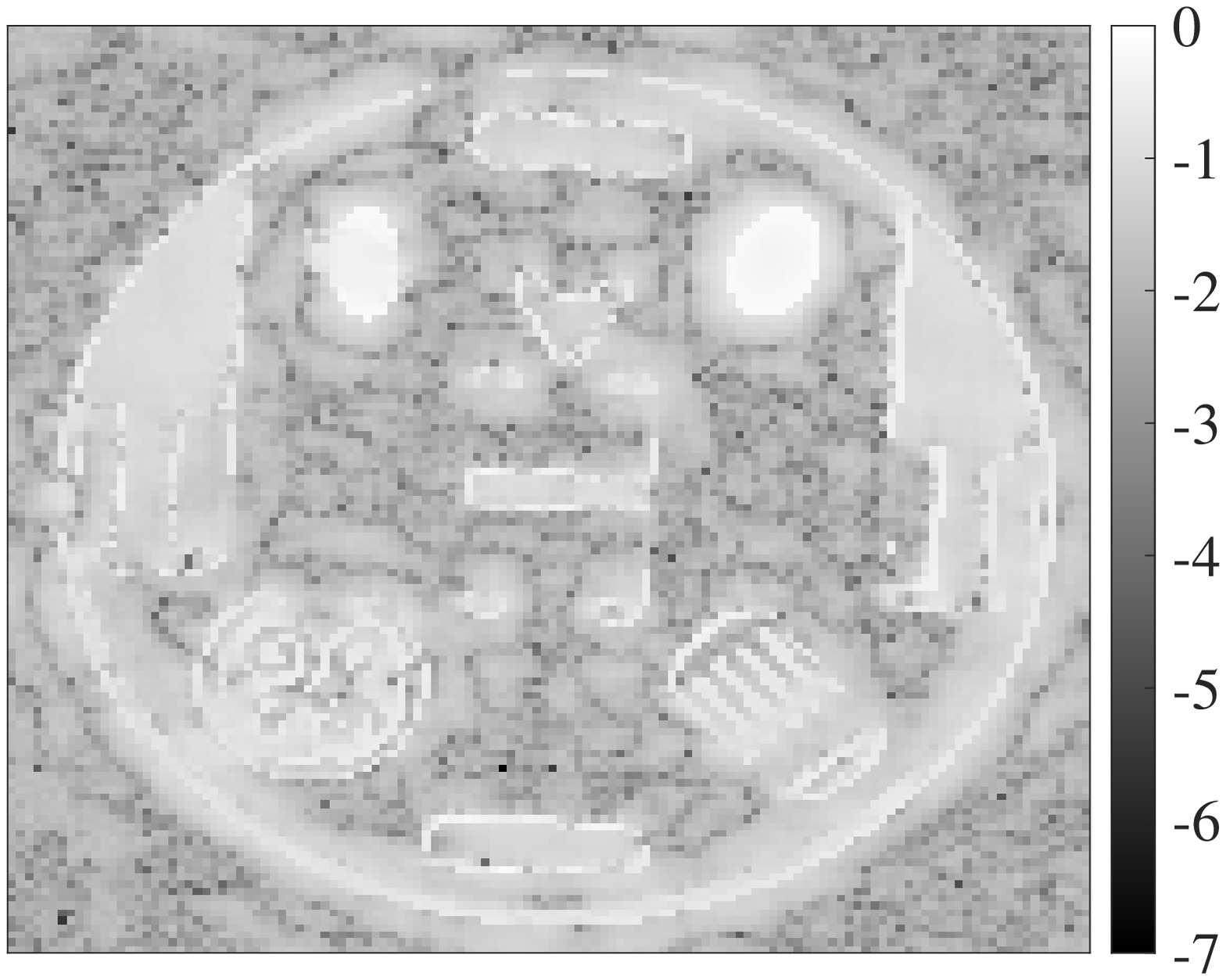}
    \caption{${\vect{x}_1}^{\beta}_{\JHBL}$}
    \end{subfigure}
    \begin{subfigure}[b]{.325\textwidth}
    \includegraphics[width=\textwidth]{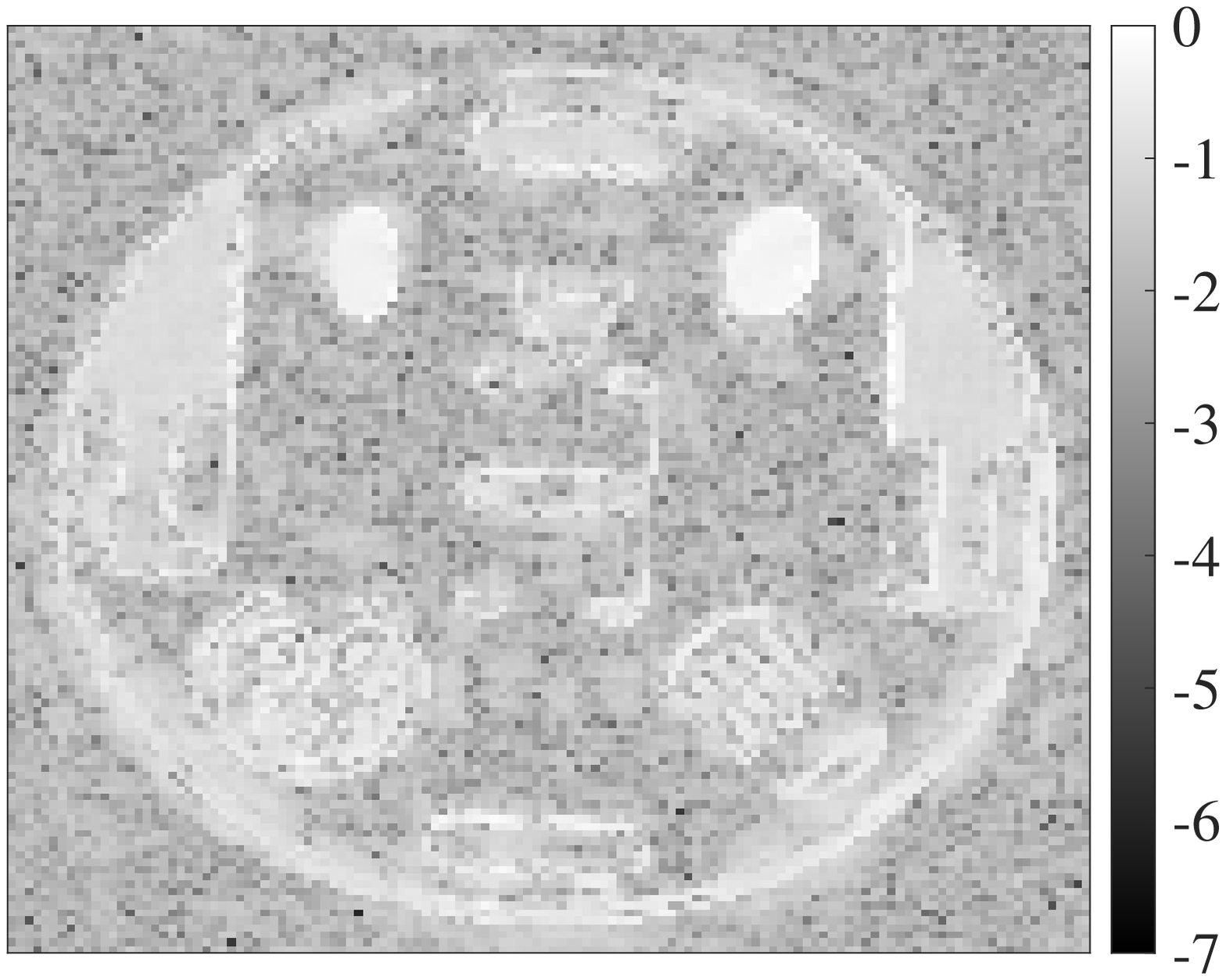}
    \caption{${\vect{x}_1}^{\beta,\boldsymbol q}_{\JHBL}$}
    \end{subfigure}
    
    \caption{(top) Image recovery of the first image in the sequence via \eqref{eq:wl1_model} with weights informed by (left) Algorithm \ref{alg:JHBL-fixed-beta};
    (middle) Algorithm \ref{alg:JHBL};
    and (right) Algorithm \ref{alg:JHBL-refine-2D}.
    (bottom) The corresponding log-scale pointwise error.}
    \label{fig:rec_SBL_2d}
\end{figure}

 Figure \ref{fig:edge_SBL_2d}  compares edge recovery using the JHBL method given by Algorithm \ref{alg:JHBL-fixed-beta},  which assumes information regarding $\beta$ is known a-priori, the standard JHBL method, given by Algorithm \ref{alg:JHBL}, which learns $\beta$ but does not refine the parameters based on inter-signal information, and  Algorithm \ref{alg:JHBL-refine-2D}, which refines the parameter selection by accounting for both intra- and inter-image information at each of the four time instances, again based on $J = 6$ original data sets. It is evident that which band is missing \eqref{eq:bandzero} plays an important role in how well each method is able to resolve the edges. Using a priori information regarding $\beta$ so that  $\beta=\frac{1}{\sigma^2}$, (top row) captures the internal structures but appears to result in additional clutter. 
Learning $\beta$ {\em without} refining the hyperparameters according to inter-signal information results in loss of moving edge information in each recovered edge map (second row).  In all cases, refining ${\boldsymbol q}$ improves resolution while mitigating the effects of both the corrupted data and the change of support locations in the edge domain (bottom row). For this example we chose $\vartheta=0.3$ in \eqref{eq:SBL-2D-s-refine} and note that some additional tuning may improve the results.

We observe in particular the poor edge map recovery quality using the JHBL approach with fixed \(\beta \)  (Algorithm \ref{alg:JHBL-fixed-beta}) in the first column of the first row in Figure \ref{fig:edge_SBL_2d}, which is likely due to the zeroed out low frequencies in \eqref{eq:bandzero}.  In the middle row, we see that while the {\em stationary} edge map structures are recovered more accurately, the {\em changed}  regions are not retrieved in the update process used by Algorithm \ref{alg:JHBL}.  By contrast,  our new approach in Algorithm \ref{alg:JHBL-refine-2D} is able both to enhance the quality observed when using Algorithm \ref{alg:JHBL-fixed-beta} while also recovering the changed regions in the edge maps (bottom row).

Figure \ref{fig:rec_SBL_2d} demonstrates how the edge maps from each recovery can be used to inform the weights in \eqref{eq:weights_scaled} which is in turn  used in the weighted $\ell_1$ regularization method  given by \eqref{eq:wl1_model} for image recovery. While only the first image in the sequence is displayed, the results for the rest of the sequence are comparable.  In particular we observe that using Algorithm \ref{alg:JHBL-refine-2D} consistently performs as well as or better than the other two edge recovery methods which do not properly account for change information.   It is also once again evident that which band is missing \eqref{eq:bandzero} plays an important role in how well each method is able to resolve the features in each image.

\subsubsection*{Sequential SAR images}

\begin{figure}[h!]
    \centering
    \begin{subfigure}[b]{.23\textwidth}
    \includegraphics[width=\textwidth]{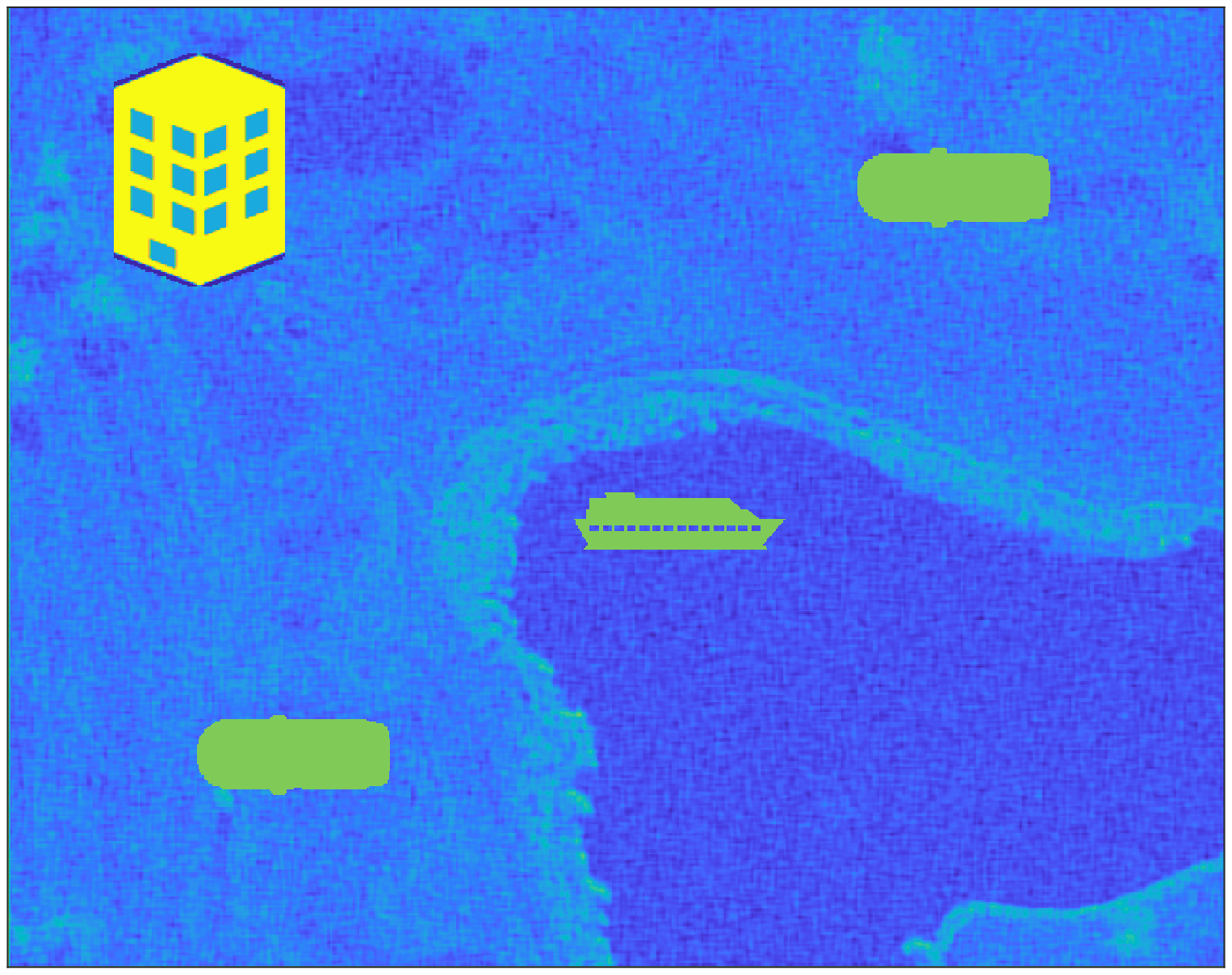}
    \caption{$\vect{f}_1$}
    \end{subfigure}
    \begin{subfigure}[b]{.23\textwidth}
    \includegraphics[width=\textwidth]{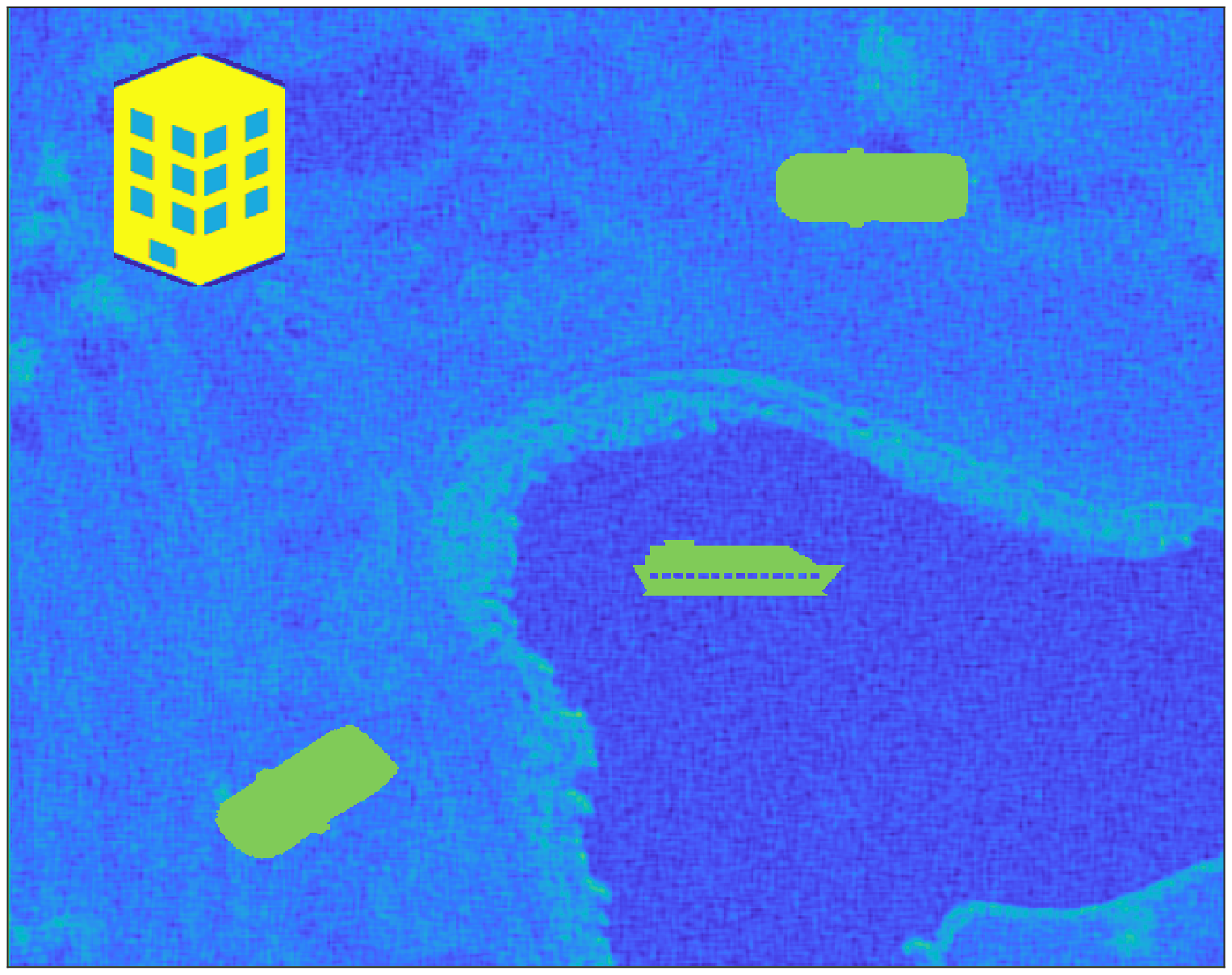}
    \caption{$\vect{f}_2$}
    \end{subfigure}
    \begin{subfigure}[b]{.23\textwidth}
    \includegraphics[width=\textwidth]{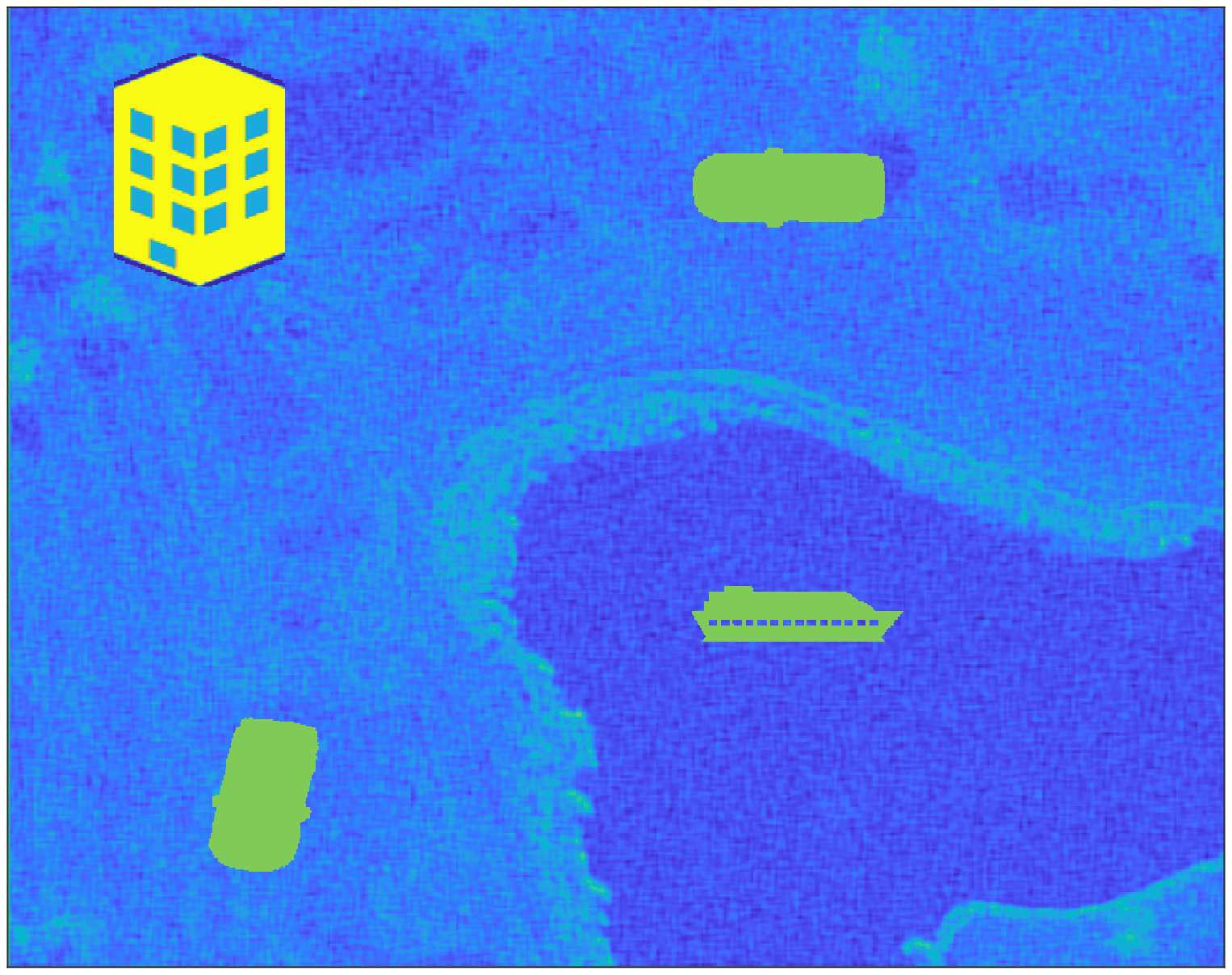}
    \caption{$\vect{f}_3$}
    \end{subfigure}
    \begin{subfigure}[b]{.23\textwidth}
    \includegraphics[width=\textwidth]{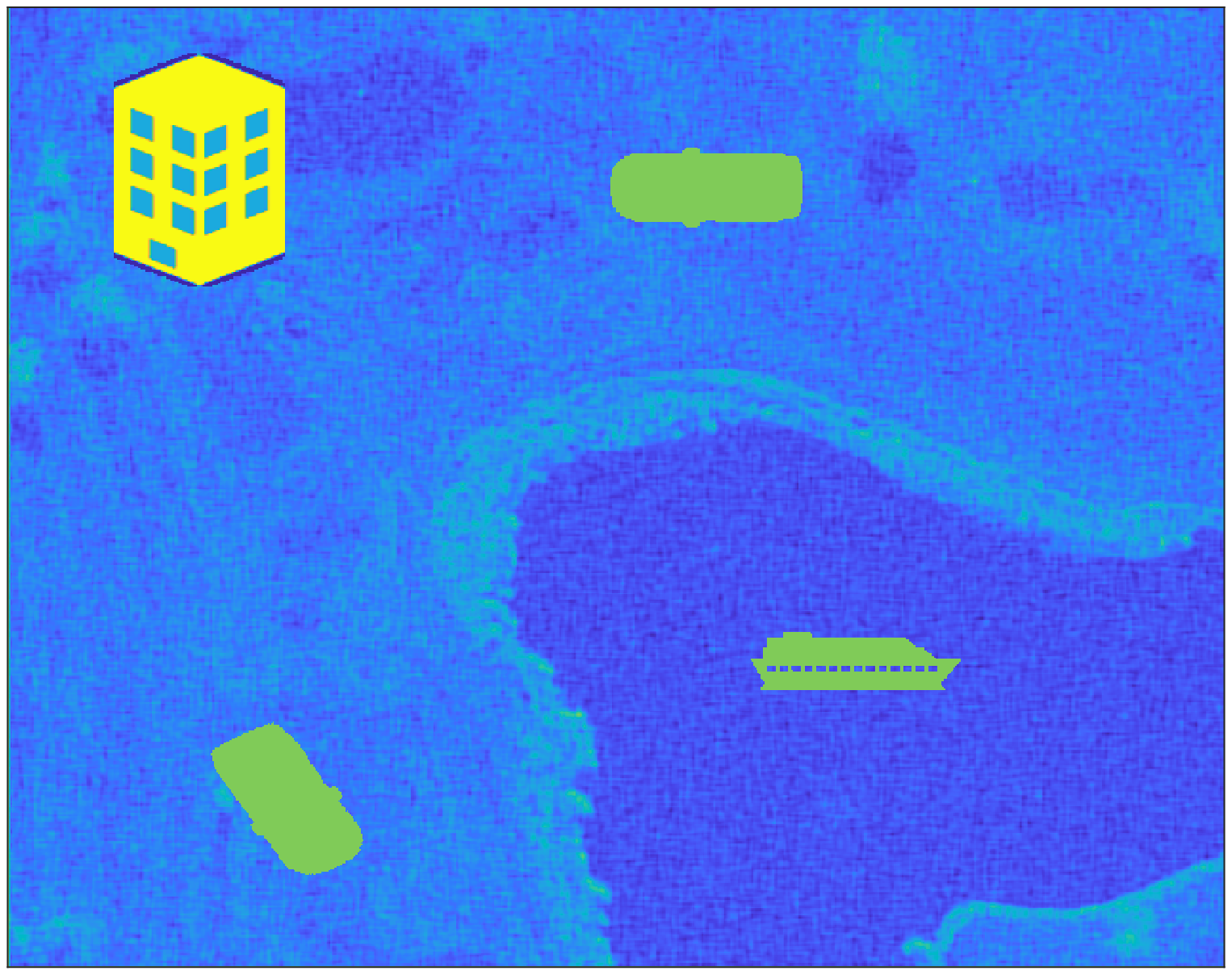}
    \caption{$\vect{f}_4$}
    \end{subfigure}

    \caption{The underlying scene is the SAR image of a golf course, \cite{SAR_Image_ref}, with a superimposed (static) hotel, and translating/rotating cars and boats.}
    \label{fig:rec_golf}
\end{figure}

For the second experiment we consider a temporal sequence of six SAR images of a golf course, \cite{SAR_Image_ref}, four of which are displayed in Figure \ref{fig:rec_golf}. Observe there is no ``ground truth'' in this case.  As in the MRI case, we again use the discrete Fourier transform of the $128\times128$ pixelated image to obtain the measurement data.  
We once again assume a symmetric band of measurements, $\{\mathcal{K}_j\}_{j=1}^J$ given in \eqref{eq:bandzero}, is for some reason not available for use.  Finally, we add noise with SNR $=2$. 

To respectively simulate moving and background objects, we impose cars and boats on the scene, each of magnitude $1$, and a building of magnitude $1.5$.  

\begin{figure}[h!]
    \centering
    \begin{subfigure}[b]{.23\textwidth}
    \includegraphics[width=\textwidth]{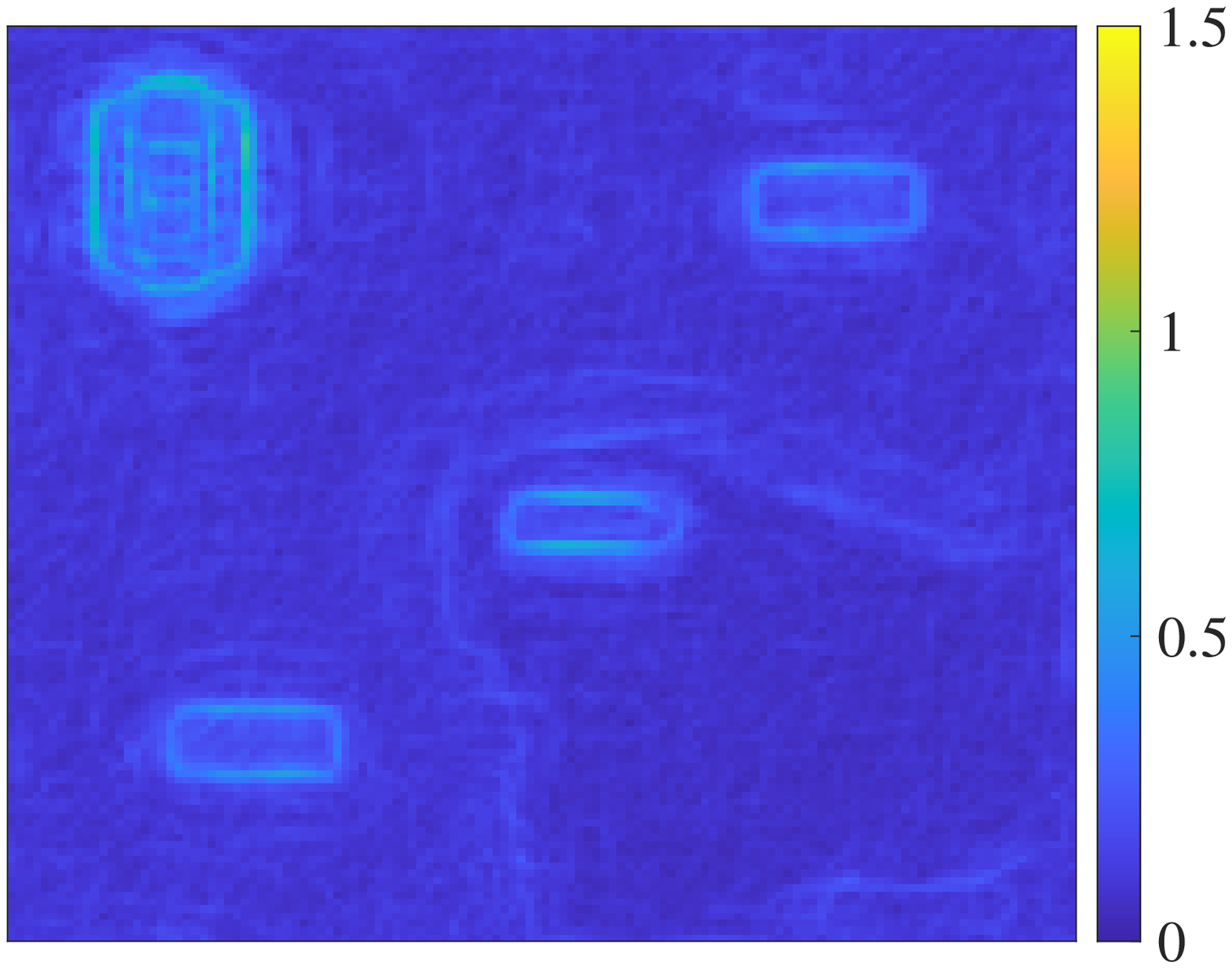}
    \caption{${\vect{x}_1}_{\JHBL}$}
    \end{subfigure}
    \begin{subfigure}[b]{.23\textwidth}
    \includegraphics[width=\textwidth]{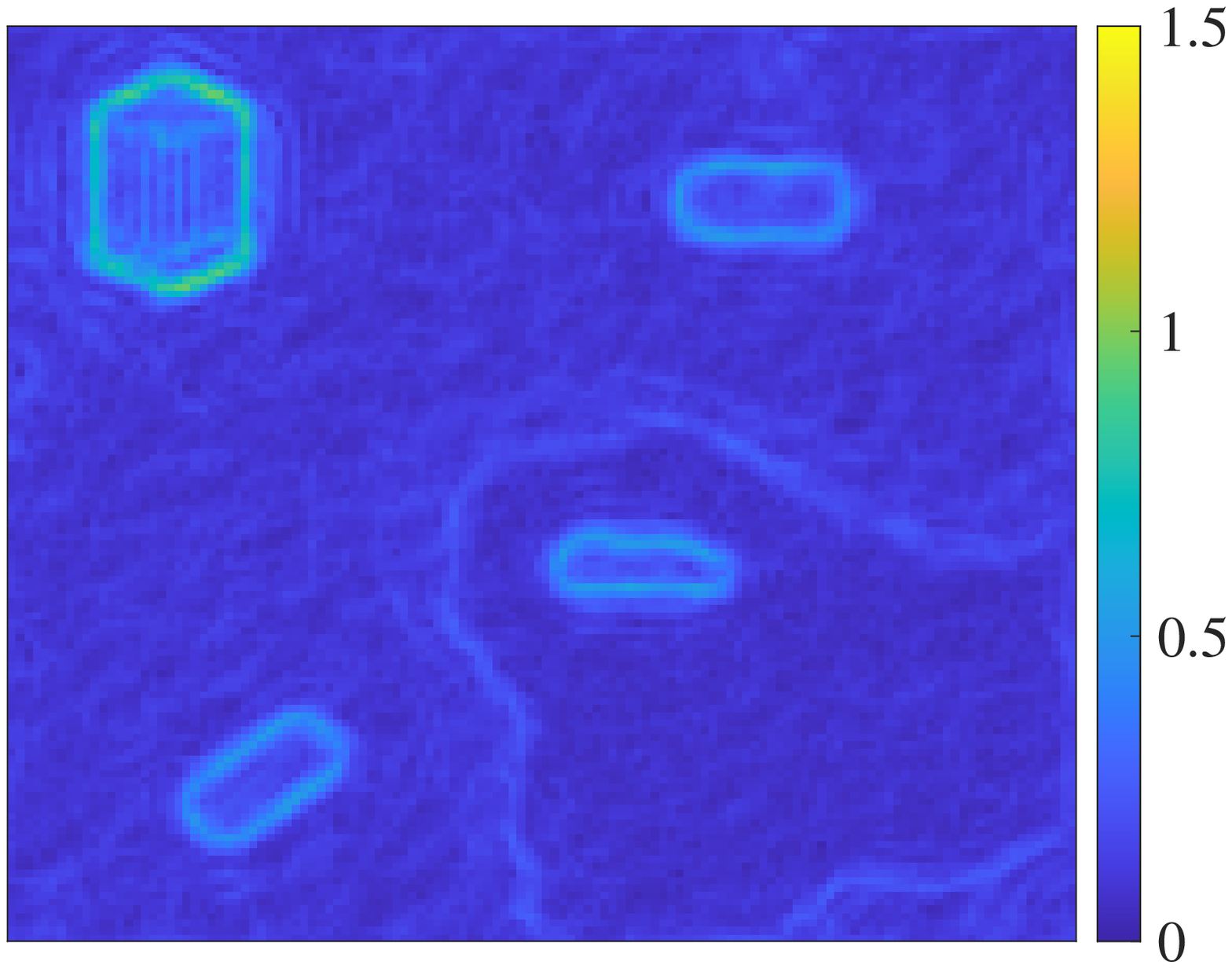}
    \caption{${\vect{x}_2}_{\JHBL}$}
    \end{subfigure}
    \begin{subfigure}[b]{.23\textwidth}
    \includegraphics[width=\textwidth]{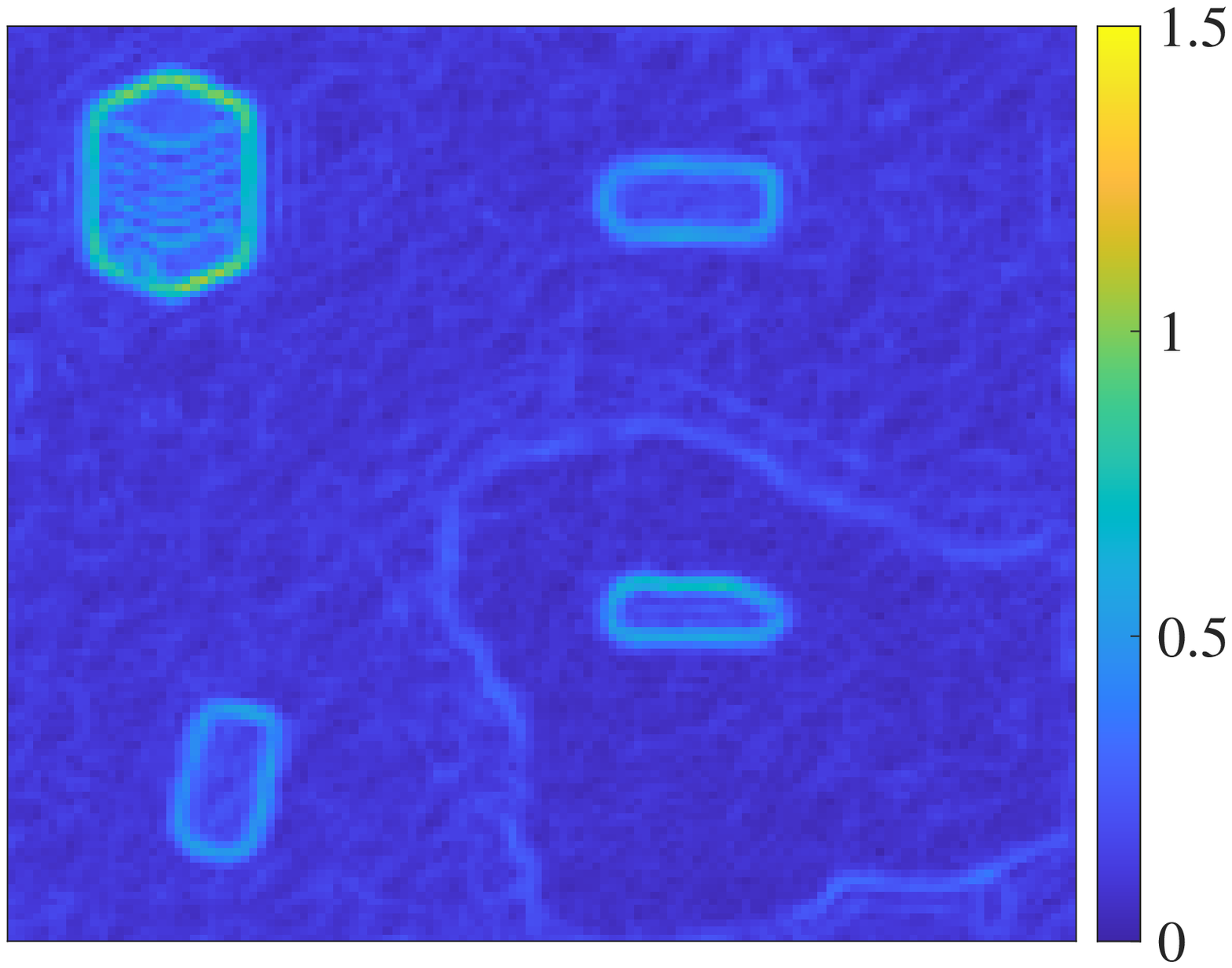}
    \caption{${\vect{x}_3}_{\JHBL}$}
    \end{subfigure}
    \begin{subfigure}[b]{.23\textwidth}
    \includegraphics[width=\textwidth]{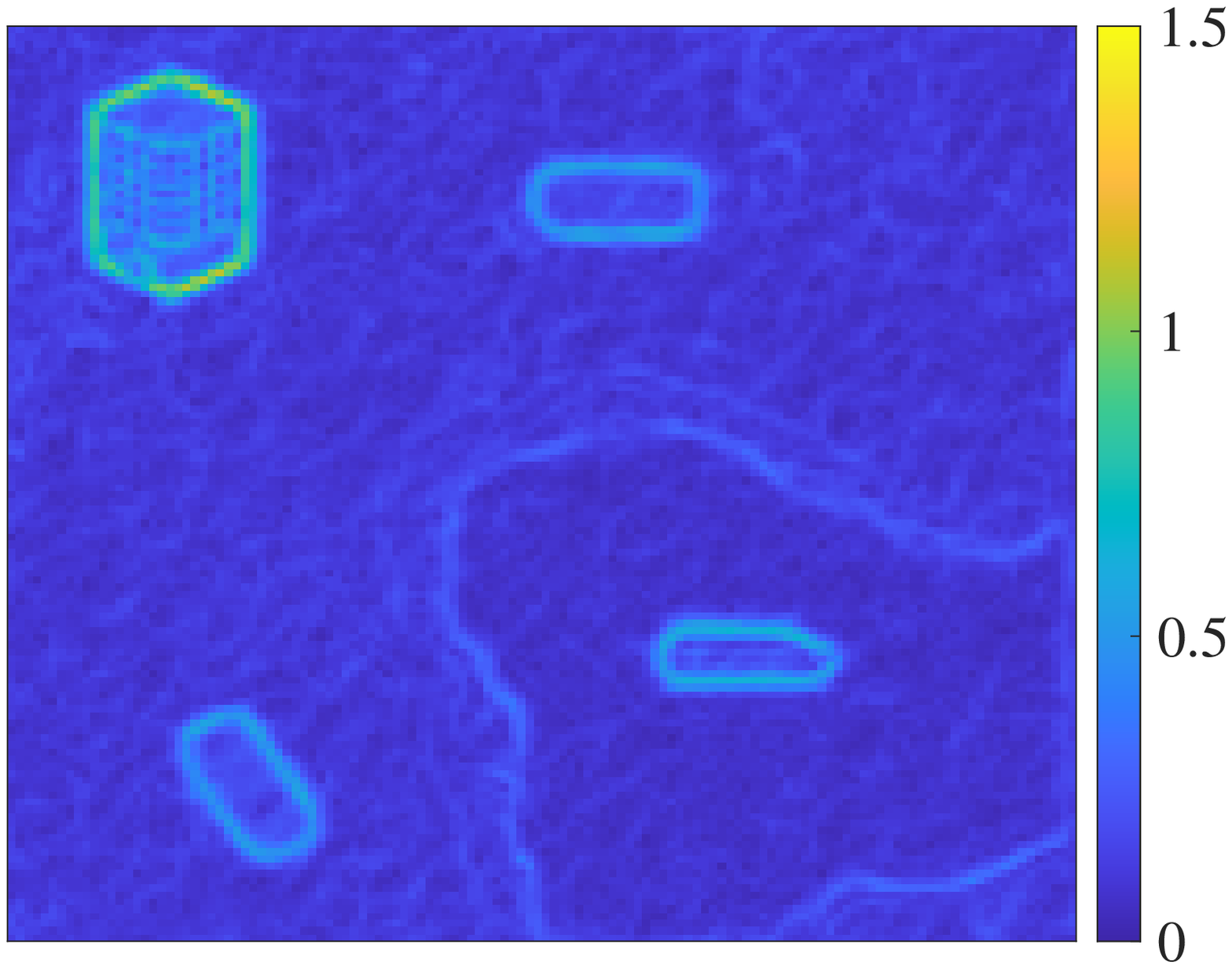}
    \caption{${\vect{x}_4}_{\JHBL}$}
    \end{subfigure}
    \\
    \begin{subfigure}[b]{.23\textwidth}
    \includegraphics[width=\textwidth]{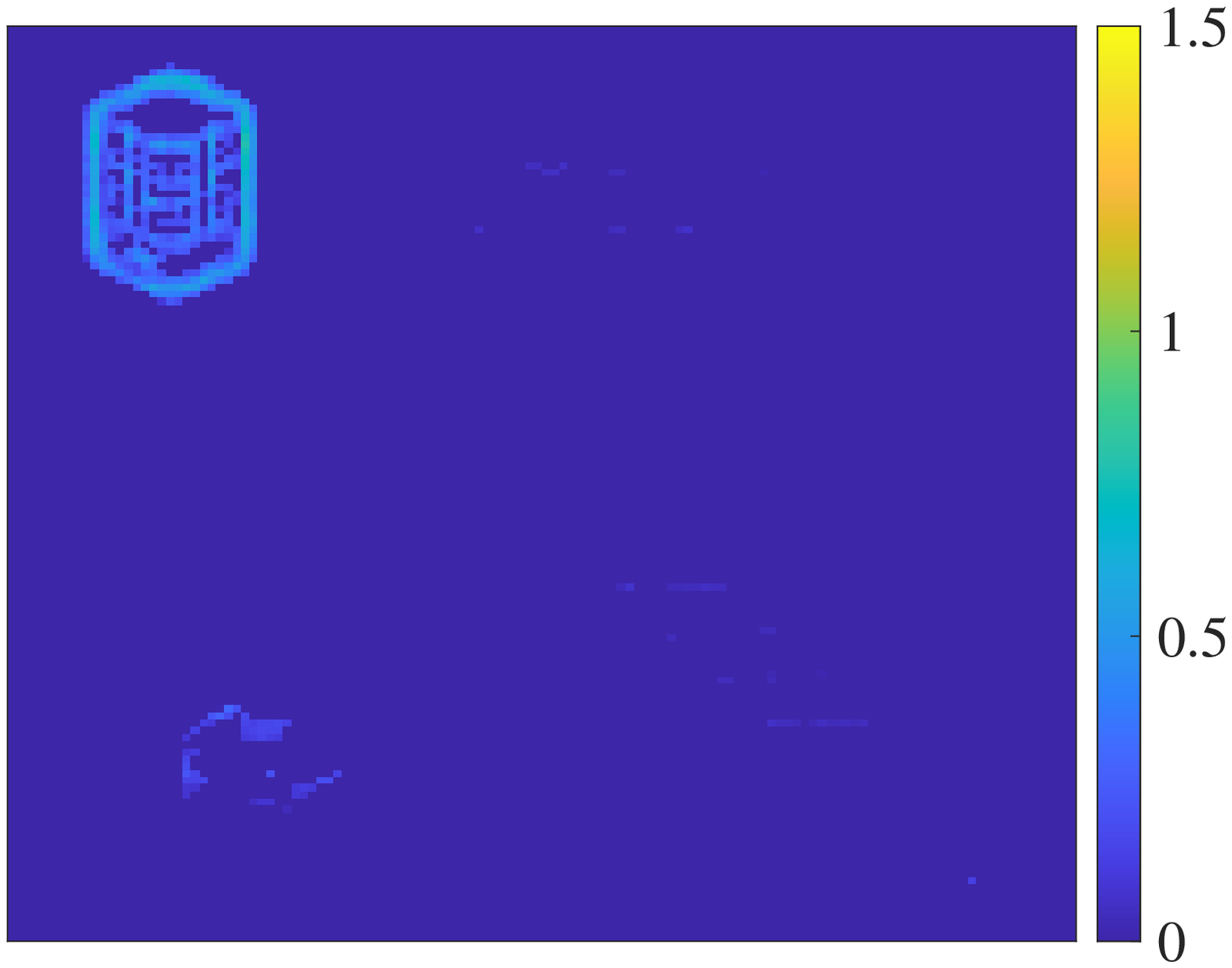}
    \caption{${\vect{x}_1}^{\beta}_{\JHBL}$}
    \end{subfigure}
    \begin{subfigure}[b]{.23\textwidth}
    \includegraphics[width=\textwidth]{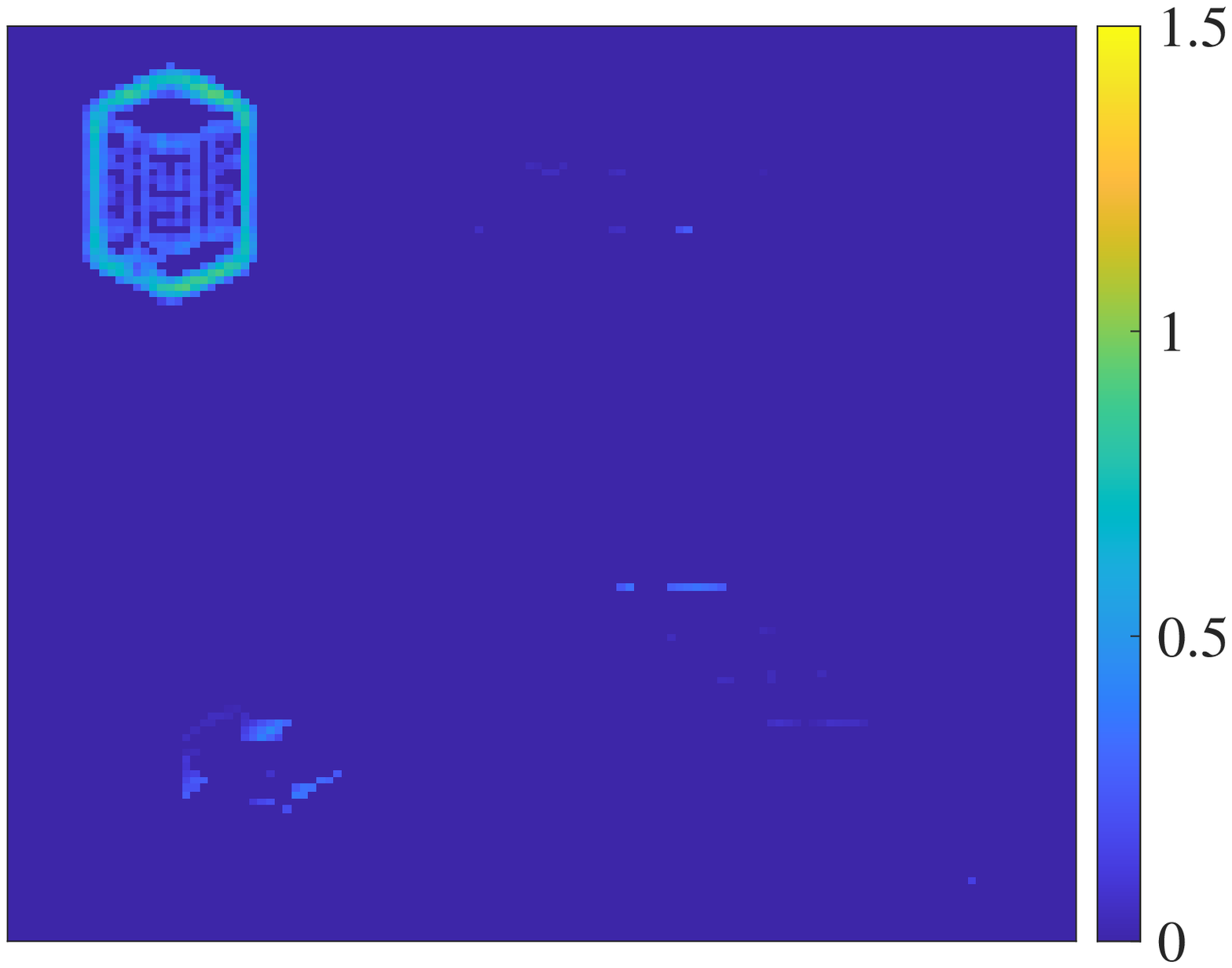}
    \caption{${\vect{x}_2}^{\beta}_{\JHBL}$}
    \end{subfigure}
    \begin{subfigure}[b]{.23\textwidth}
    \includegraphics[width=\textwidth]{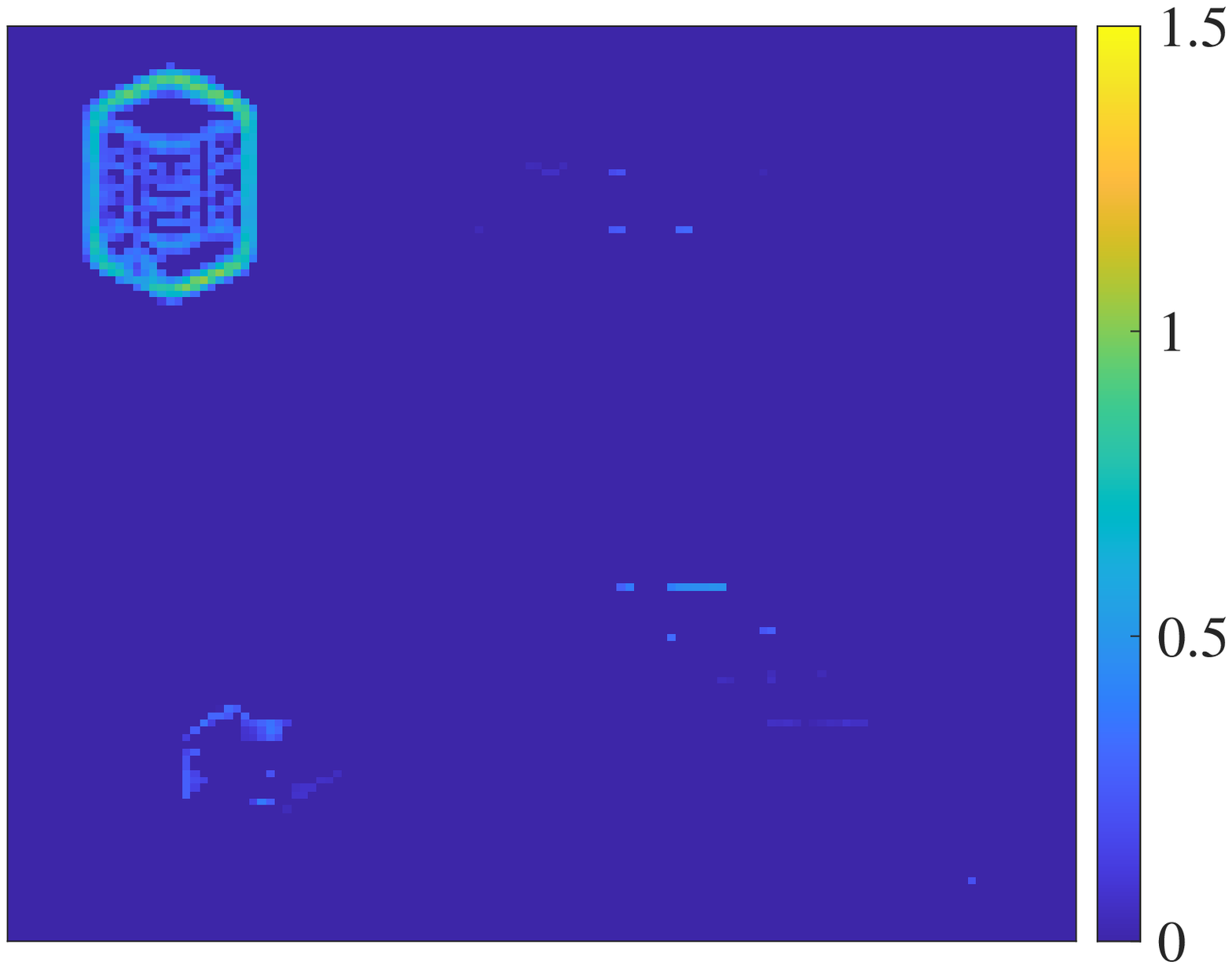}
    \caption{${\vect{x}_3}^{\beta}_{\JHBL}$}
    \end{subfigure}
    \begin{subfigure}[b]{.23\textwidth}
    \includegraphics[width=\textwidth]{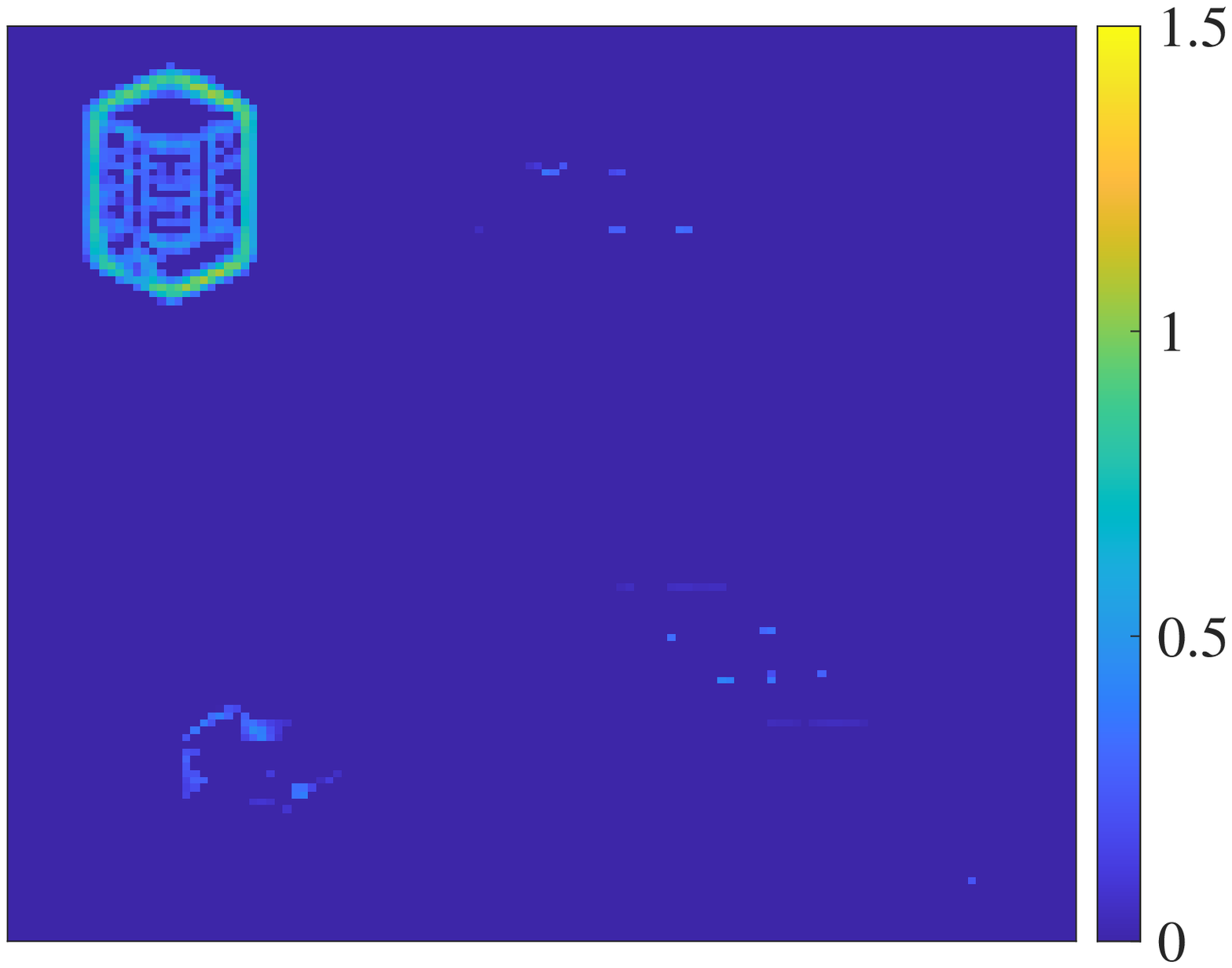}
    \caption{${\vect{x}_4}^{\beta}_{\JHBL}$}
    \end{subfigure}
    \\
    \begin{subfigure}[b]{.23\textwidth}
    \includegraphics[width=\textwidth]{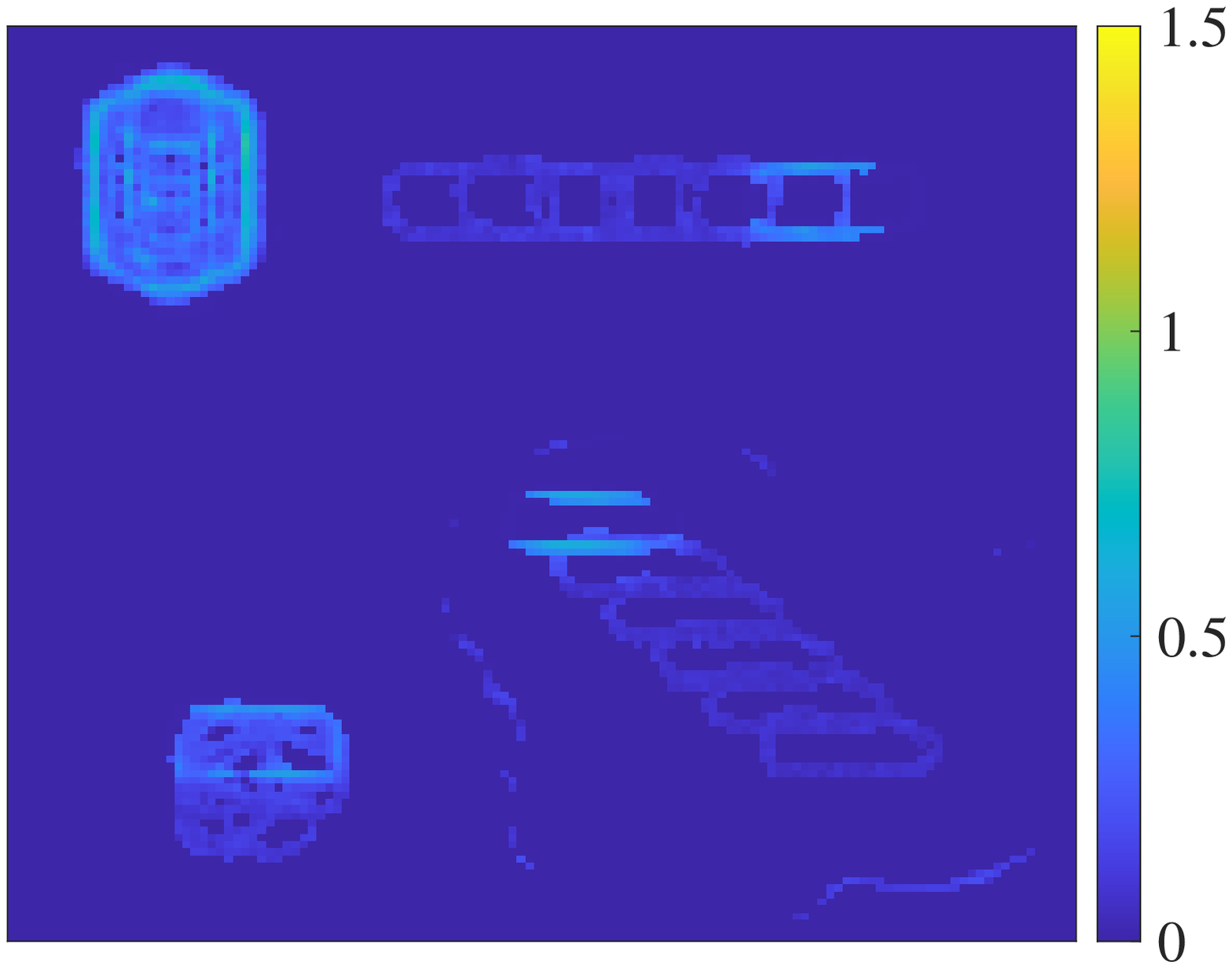}
    \caption{${\vect{x}_1}^{\beta,\boldsymbol q}_{\JHBL}$}
    \end{subfigure}
    \begin{subfigure}[b]{.23\textwidth}
    \includegraphics[width=\textwidth]{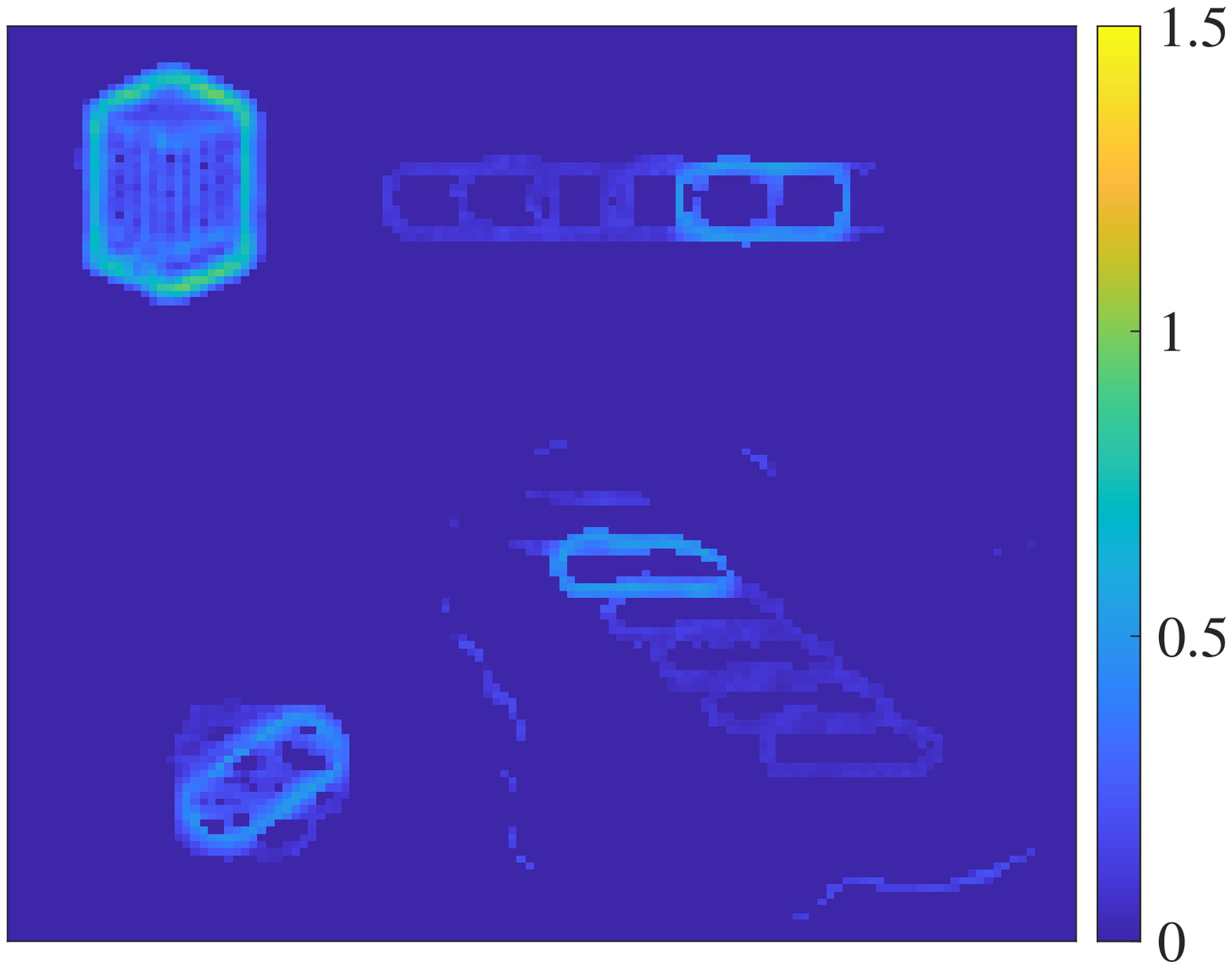}
    \caption{${\vect{x}_2}^{\beta,\boldsymbol q}_{\JHBL}$}
    \end{subfigure}
    \begin{subfigure}[b]{.23\textwidth}
    \includegraphics[width=\textwidth]{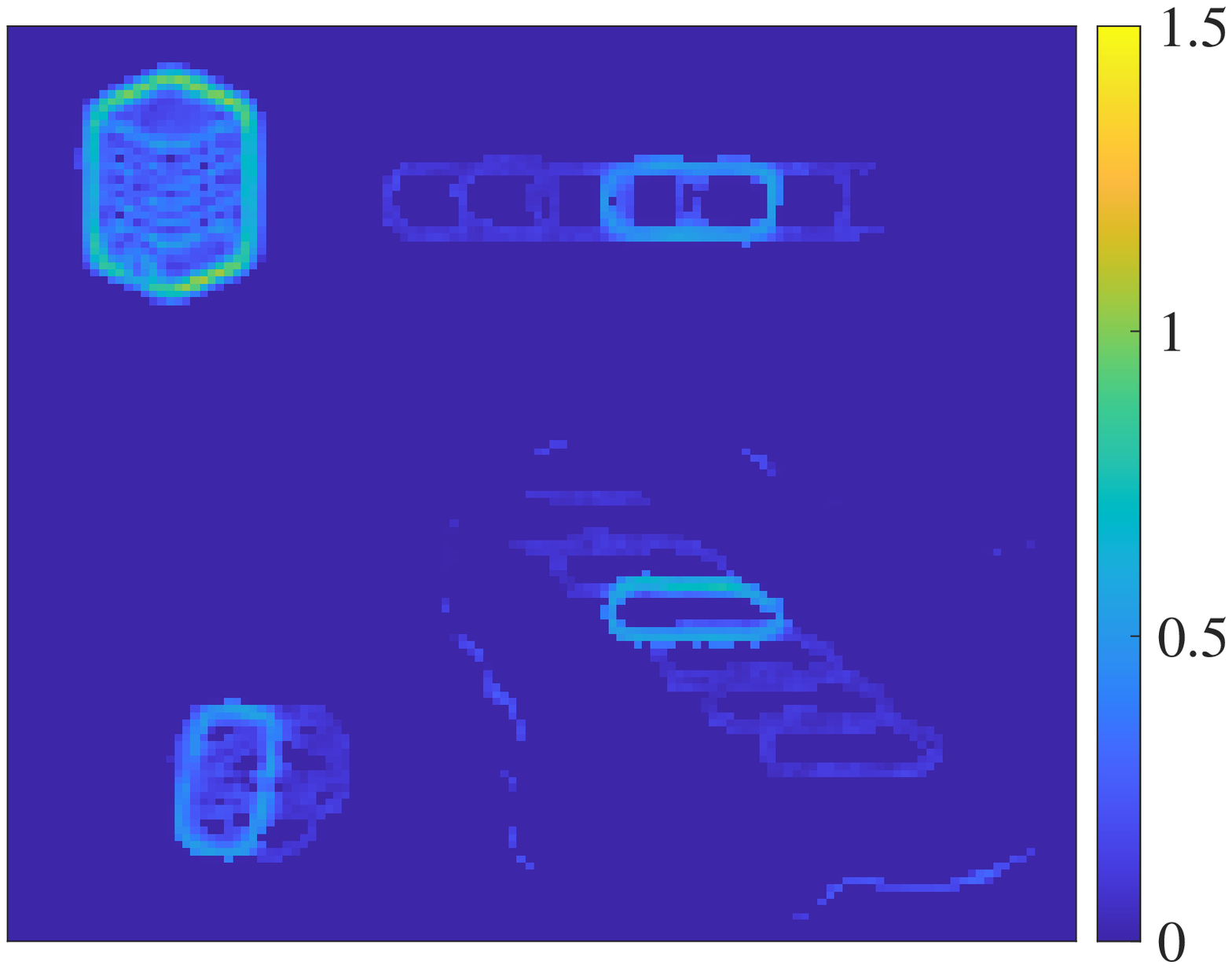}
    \caption{${\vect{x}_3}^{\beta,\boldsymbol q}_{\JHBL}$}
    \end{subfigure}
    \begin{subfigure}[b]{.23\textwidth}
    \includegraphics[width=\textwidth]{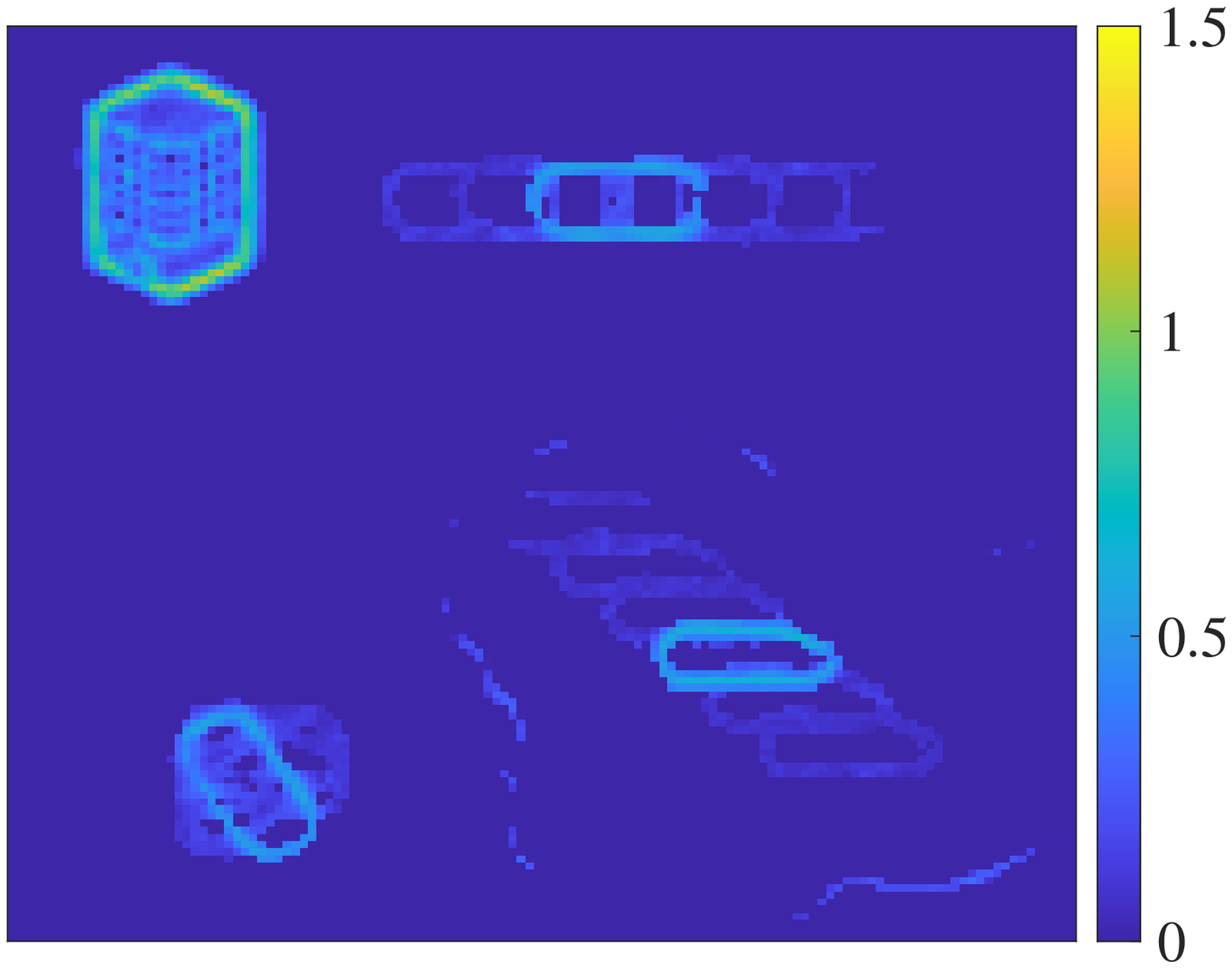}
    \caption{${\vect{x}_4}^{\beta,\boldsymbol q}_{\JHBL}$}
    \end{subfigure}
    
    \caption{Sequential edge map recovery shown at four time stamps  using (top) Algorithm \ref{alg:JHBL-fixed-beta};
    (middle) Algorithm \ref{alg:JHBL}; and (bottom) Algorithm \ref{alg:JHBL-refine-2D}.}
    \label{fig:edge_golf_2d}
\end{figure}

\begin{figure}
    \centering
    \begin{subfigure}[b]{.32\textwidth}
    \includegraphics[width=\textwidth]{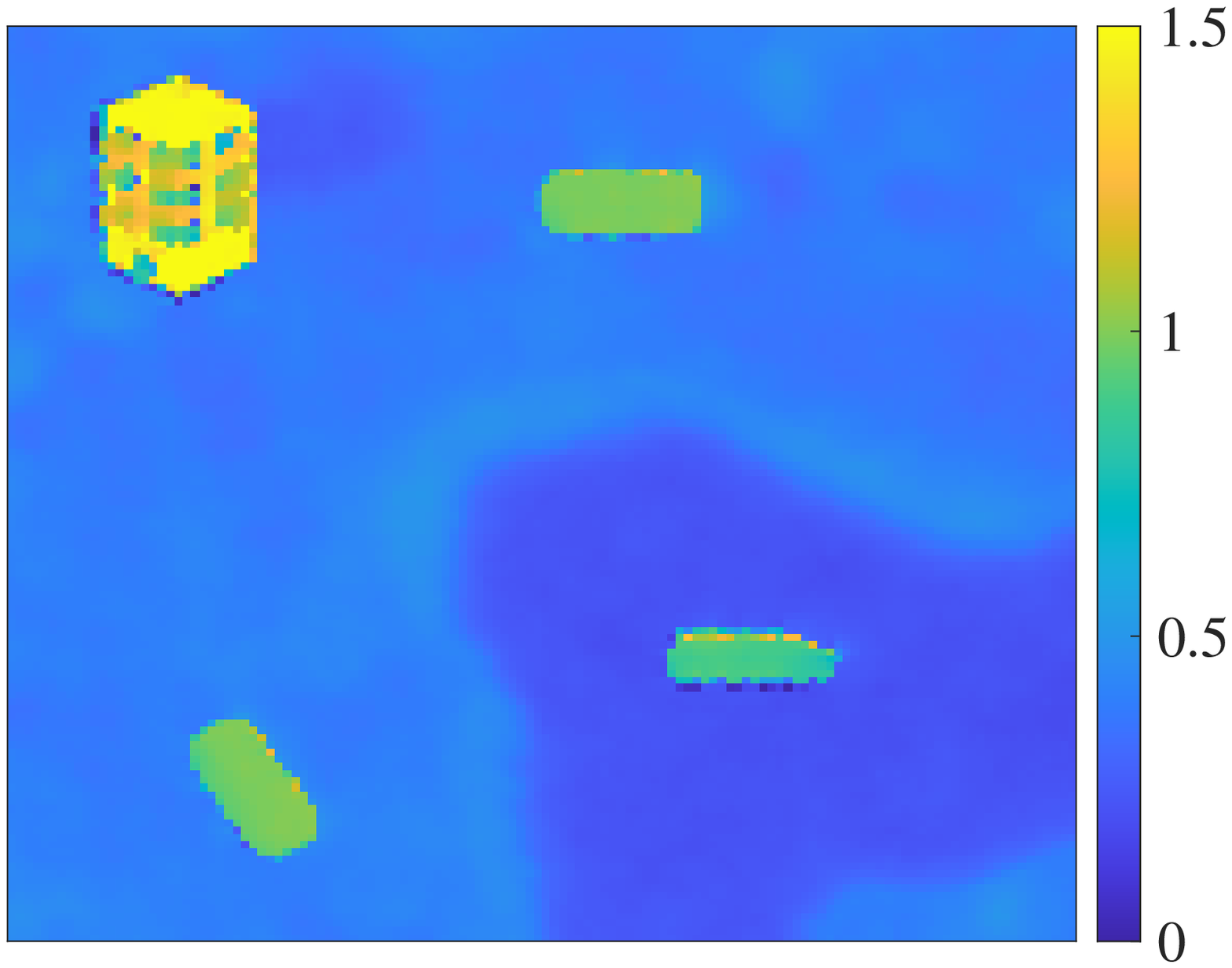}
    \caption{${\vect{x}_4}_{\JHBL}$}
    \end{subfigure}
    \begin{subfigure}[b]{.32\textwidth}
    \includegraphics[width=\textwidth]{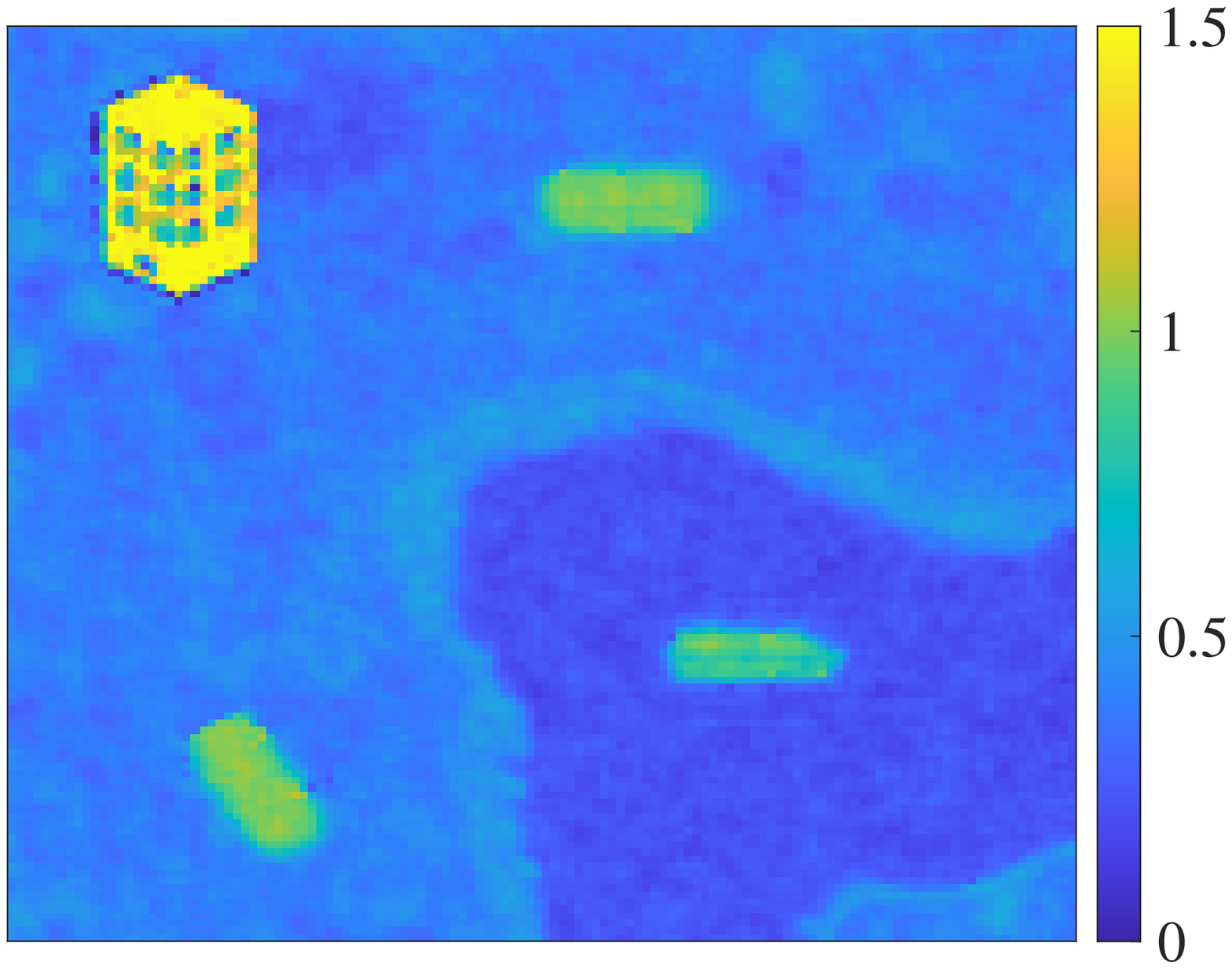}
    \caption{${\vect{x}_4}^{\beta}_{\JHBL}$}
    \end{subfigure}
    \begin{subfigure}[b]{.32\textwidth}
    \includegraphics[width=\textwidth]{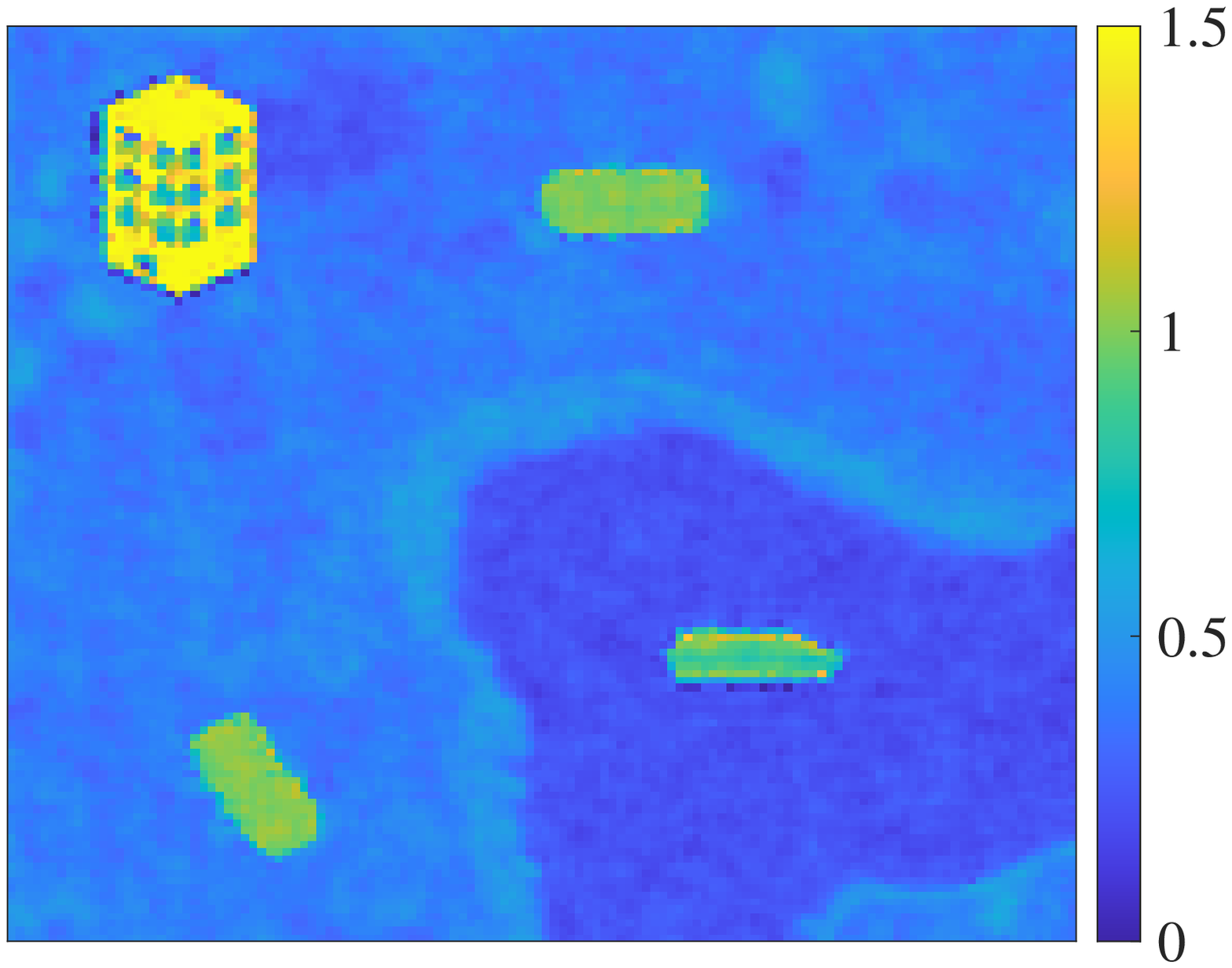}
    \caption{${\vect{x}_4}^{\beta,\boldsymbol q}_{\JHBL}$}
    \end{subfigure}
    \\
    \begin{subfigure}[b]{.32\textwidth}
    \includegraphics[width=\textwidth]{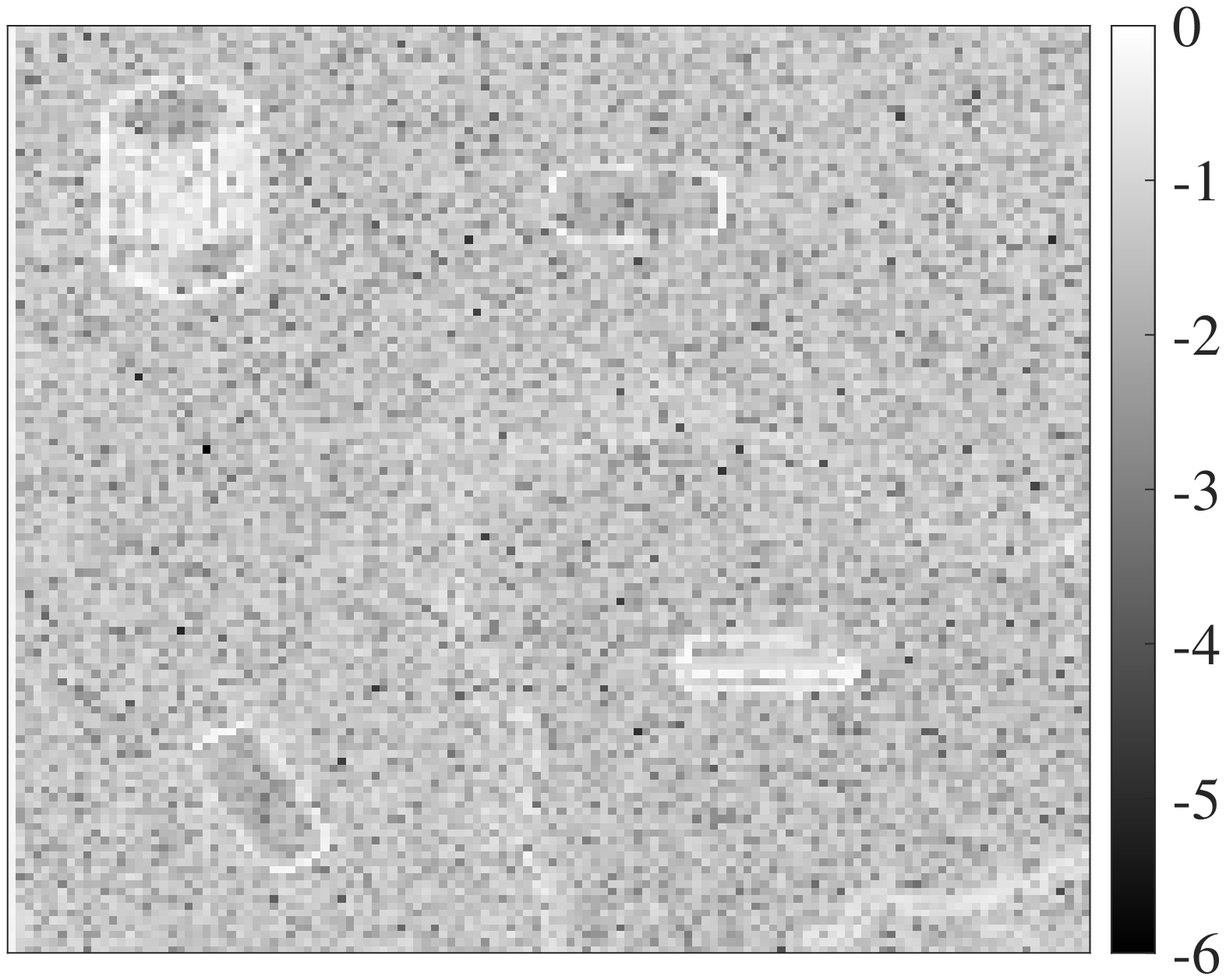}
    \caption{${\vect{x}_4}_{\JHBL}$}
    \end{subfigure}
    \begin{subfigure}[b]{.32\textwidth}
    \includegraphics[width=\textwidth]{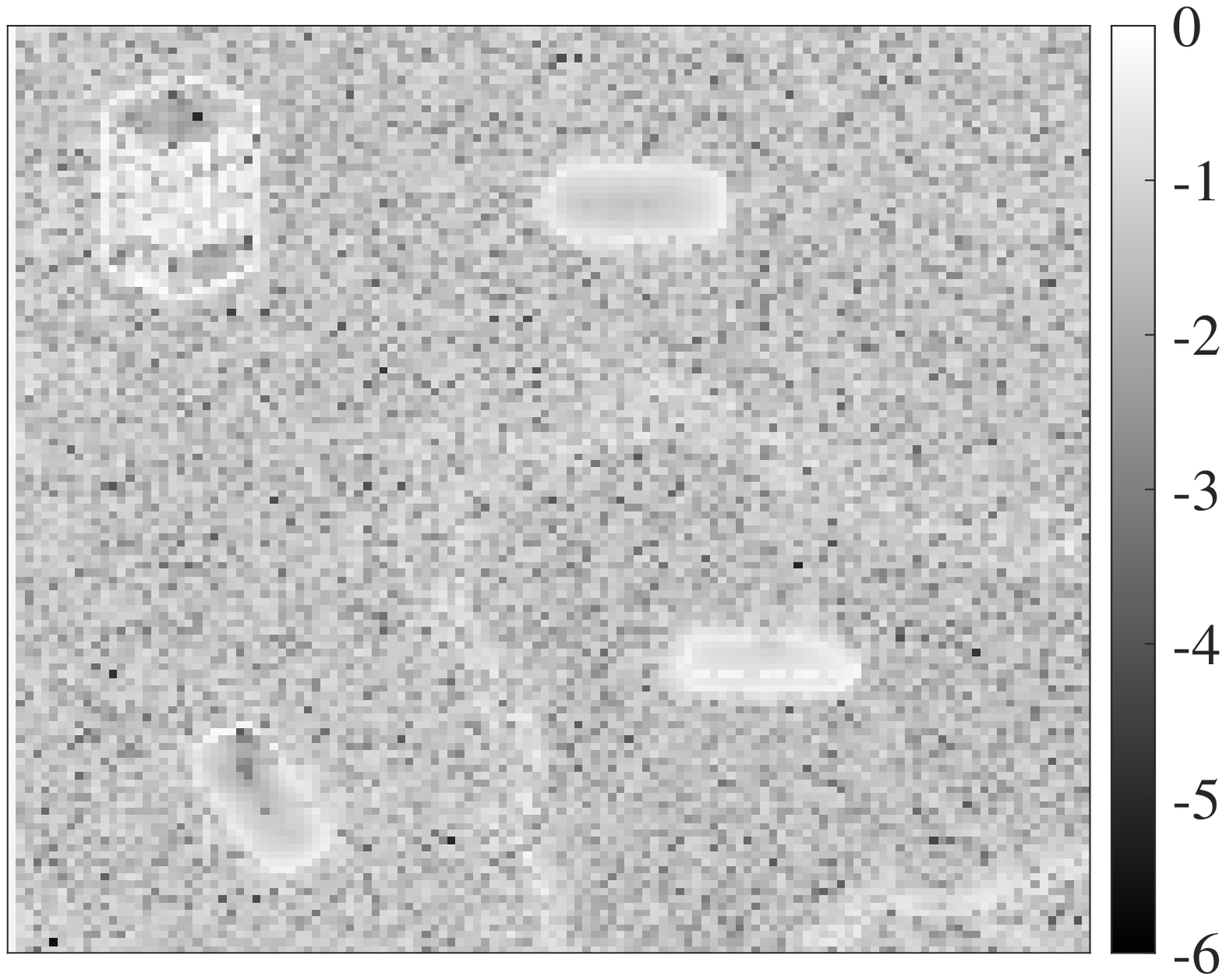}
    \caption{${\vect{x}_4}^{\beta}_{\JHBL}$}
    \end{subfigure}
    \begin{subfigure}[b]{.32\textwidth}
    \includegraphics[width=\textwidth]{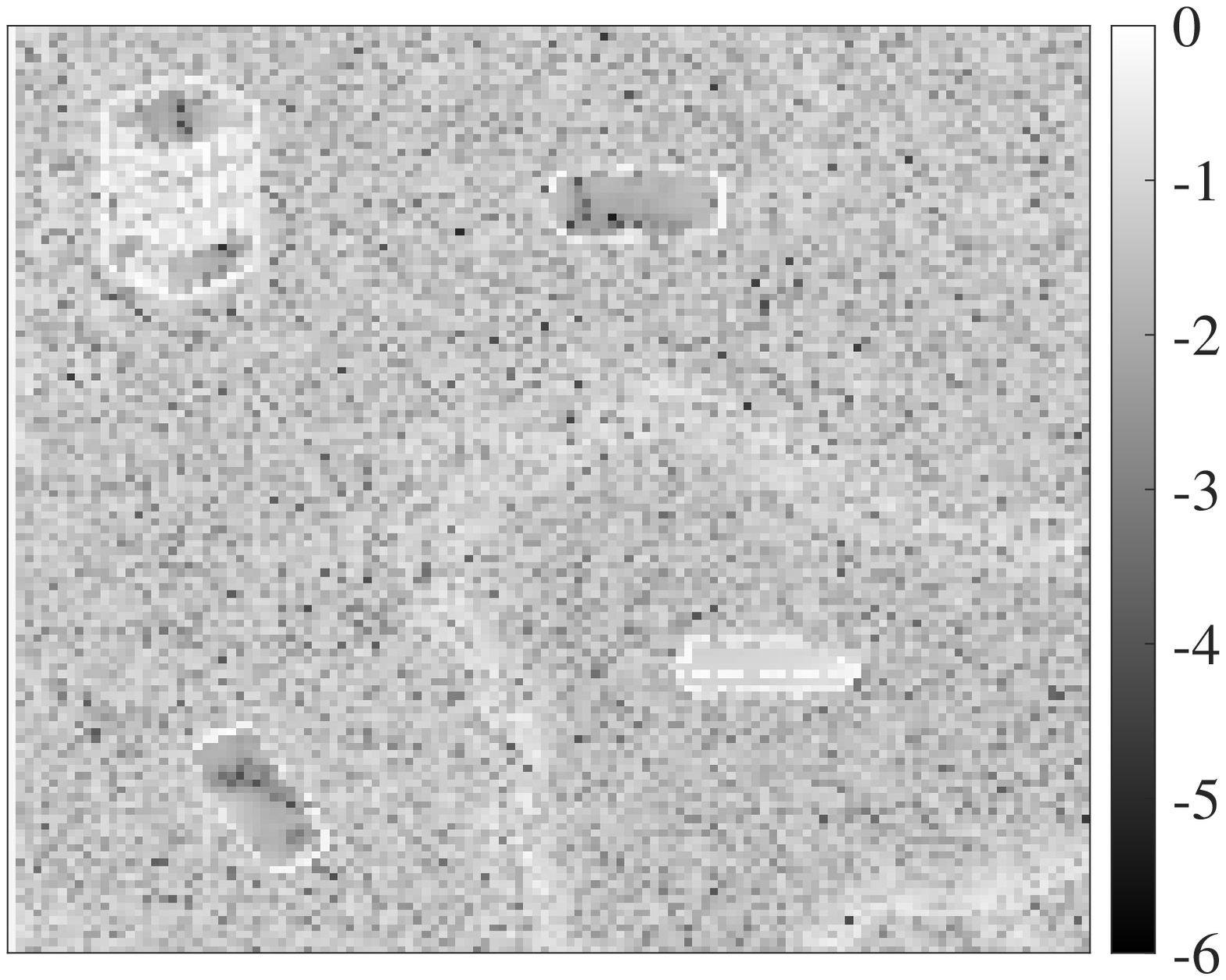}
    \caption{${\vect{x}_4}^{\beta,\boldsymbol q}_{\JHBL}$}
    \end{subfigure}
    
    \caption{(top row) Image recovery using \eqref{eq:wl1_model} where the weights are informed by (left) Algorithm \ref{alg:JHBL-fixed-beta}
    (middle) Algorithm \ref{alg:JHBL} and (right) Algorithm \ref{alg:JHBL-refine-2D}.
    (bottom row) The corresponding log-scale pointwise error of the images in the top row.}
    \label{fig:rec_golf_2d}
\end{figure}

As in the sequential MRI example, we first compare edge map recoveries using the different algorithms in Figure \ref{fig:edge_golf_2d}. The edges are then used to inform the weights in \eqref{eq:weights_scaled} for the weighted $\ell_1$ regularization method \eqref{eq:wl1_model}.  Figure \ref{fig:rec_golf_2d} displays the results for the fourth image in the sequence. The results for the rest of the sequence are similar. Overall, it is clear that by using Algorithm \ref{alg:JHBL-refine-2D} we are more able to capture the structural details in the underlying images without ``oversmoothing'' the background. We note that using either Algorithm \ref{alg:JHBL} or Algorithm \ref{alg:JHBL-fixed-beta} is sufficient for obtaining weights in \eqref{eq:wl1_model} for some images in the sequence, however the results are not consistent.  

More distinction between the results of each algorithm are observed in the sequential edge map recovery.   Figure \ref{fig:edge_golf_2d} shows cluttering in the edge maps when using Algorithm \ref{alg:JHBL-fixed-beta}.  Learning the hyperparameters without the benefit of correlating temporal information (Algorithm \ref{alg:JHBL}) reduces the clutter, but as was the case for the MRI example, the rotating and translating structures are lost.  Only Algorithm\ref{alg:JHBL-refine-2D}, which refines the hyperparameters to consider inter-signal correlations is able to capture moving objects.    

\section{Conclusion}
\label{sec:discussion_ongoing}
In this paper we proposed a new sparse Bayesian learning algorithm  to jointly recover a temporal sequence of edge maps from under-sampled and noisy Fourier data of piecewise smooth functions and images. Since each data set in the temporal sequence only acquires partial information regarding the underlying scene, it is not possible to {\em individually} recover each edge map in the sequence. Our new method incorporates both {\em inter-} and {\em intra-}image information into the design of the prior distribution. This is in contrast to standard multiple-measurement SBL approaches, which require stationary support across the temporal sequence. Moreover, unlike the deterministic method developed  in \cite{xiao2022sequential},  our approach does {\em not} require the explicit construction of sequential change masks, which adds numerical cost as well as introduces more parameters.

Our new algorithm compares favorably to more standard SBL approaches in regions where the underlying function is smooth (correspondingly zero in the jump function).
In particular we observed in our one-dimensional  example that fewer non-zero values were returned in smooth regions where the data sequence share the joint sparsity profile. Furthermore, our new method does not suffer as much magnitude loss at the true (non-zero) jump locations.
The magnitudes at varied jump locations (across the sequence) are also much more accurate using our method, although there are some oscillations in the surrounding neighborhoods. By contrast, we observed that jump locations were completely missed when the hyperparameters were not temporally correlated.

While these initial results are promising, our method should be refined for more complicated sequences of images.  For example, an empirical method can be introduced to determine the shape and rate of hyper-parameters for the partial joint support of the temporal data stream. This would  potentially involve another layer of hyper-hyper-parameters in the hierarchical Bayesian structure.   Similarly, the proposed framework can be applied to complex-valued signals (or images), where there is sparsity in the magnitude of the signal.  This would be important in SAR or ultrasound images.  We anticipate using sampling techniques such as Markov chain Monte Carlo (MCMC) and Gibbs sampling  in these cases which will also allow uncertainty quantification of the solution posterior.

\bmhead{Data Availability}
All underlying data sets can be made available upon request or are publicly available.  MATLAB codes used for obtaining results is available upon request to the authors.  MRI GE images were originally publicly provided by researchers at the Barrow Neurological Institute for the purpose of algorithmic development.   The SAR golf course image is provided in \cite{SAR_Image_ref}.


\bmhead{Acknowledgments}

This work is partially supported by the NSF grants DMS \#1912685 (AG), DMS \#1939203 (GS), AFOSR grant \#FA9550-22-1-0411 (AG), DOE ASCR \#DE-ACO5-000R22725 (AG), and ONR MURI grant \#N00014-20-1-2595 (AG).


\bibliographystyle{spmpsci}
\bibliography{literature}


\end{document}